\documentclass[a4paper,twocolumn,times,10pt]{article}
\usepackage{fullpage}
\AtBeginDocument{%
	}

\usepackage{tikz}
\usepackage{amsmath}

\usepackage{epsfig}
\usepackage{amsfonts}
\usepackage{graphicx}
\usepackage{balance}  
\usepackage{cleveref}

\usepackage{algpseudocode}
\usepackage{bm}
\usepackage{graphicx}
\usepackage{comment}
\usepackage{array}
\usepackage{float}
\floatstyle{plaintop}
\restylefloat{table}
\usepackage{amsthm}
\usepackage{amsmath}
\usepackage{algorithm}
\usepackage{subfig}
\usepackage{titlesec}
\usepackage{blindtext}
\usepackage{multirow}
\usepackage{yfonts}
\usepackage{xcolor}
\usepackage{url}
\usepackage{balance}
\usepackage{color, colortbl}
\usepackage{booktabs}
\usepackage{caption}

\definecolor{Gray}{gray}{0.9}

\begin{document}
\title{An Evaluation of Software Sketches}

\author{Roy Friedman}

\date{}

\maketitle

\begin{abstract}
This work presents a detailed evaluation of Rust (software) implementations of several popular sketching solutions, as well as recently proposed optimizations.
We compare these solutions in terms of computational speed, memory consumption, and several approximation error metrics.
Overall, we find a simple hashing based solution employed with the Nitro sampling technique~\cite{NitroSketch} gives the best trade-off between memory, error and speed.
Our findings also include some novel insights about how to best combine sampling with Counting Cuckoo filters depending on the application.
\end{abstract}

\section{Introduction}
\label{sec:intro}

Sketches are approximate compact synopsis that represent statistical features of large data sets and streams of events.
Particularly, in this work we focus on software implementation of frequency estimation sketches.
That is, sketches that answer queries about a given item's frequency in the dataset, or stream, and whose reply may include an estimation error bounded by some accuracy parameter $\epsilon$.
In some cases, the error bound only holds with a given probability $1-\delta$.

Historically, sketches were designed to give the minimal memory to error trade-offs.
This makes sense for hardware implementations of sketches, where computation speed is often a secondary issue due to hardware's inherent parallelism.
In resource constrained environments, being as memory frugal is also very important.
Finally, the claim is that since SRAM memory is much smaller than DRAM, memory frugality is also an important characteristic of software implementations to obtain fast execution in practice.
That is, given the orders of magnitude difference in access times between SRAM and DRAM, it is important for a data structure to fit into the hardware cache and present a hardware cache-friendly access pattern.

Yet, several works have shown that on modern processors, simple implementations often outperform complex data structures, and the cost of calculating multiple hash functions can be detrimental to performance~\cite{AMM15,AMZYM18,SketchVisor,NitroSketch}.
In particular, NitroSketch~\cite{NitroSketch} was offered as generic approach to expediting sketches based on multiple counter arrays updated with independent hash functions, which include, e.g., Count-Min Sketch~\cite{CountMinSketch}, Count Sketch~\cite{CountSketch}, UnivMon~\cite{UnivMon}, and K-ary Sketch~\cite{k-ary}.
This is by sampling, through geometric distribution, how many sketch' counters (and items) should be skipped before any counter should be~updated.

Another interesting optimization that targets the Space Saving algorithm~\cite{SpaceSavings} is RAP~\cite{RAP}.
Space Saving is an algorithm for heavy-hitter detection as well as frequency estimation.
Space Saving keeps track of a fixed number of items, and whenever an un-tracked item arrives, it gets the entry of an item whose frequency count is minimal at that point.
RAP improves on Space Saving's accuracy and performance by only replacing an un-tracked item with a minimally frequent item $x$ with a probability that is inversely proportional to $x$'s estimated frequency.

In this work, we measure the performance of several known algorithms, with and without the RAP and Nitro optimizations described above (depending on their applicability).
Unlike previous studies we are aware of, we have implemented all algorithms in Rust, which is a modern memory safe, strongly typed, platform neutral and highly efficient programming language (claimed to be as fast as C/C++).
Specifically, we have implemented and measured the performance of a simple Hash based solution, a hashing plus Nitro solution nicknamed NitroHash, Count-Min Sketch (CMS)~\cite{CountMinSketch} and NitroCMS, counting Cuckoo filter~\cite{CuckooFilter}, NitroCuckoo, Space Saving~\cite{SpaceSavings} and Space Saving with the RAP optimization~\cite{RAP}.
Notice that NitroHash and NitroCuckoo were never explored before, especially in terms of the throughput impact of the geometric probabilistic process used for sampling in Nitro.

Many of our findings echo previous studies, e.g.,~\cite{AMM15,AMZYM18,SketchVisor,NitroSketch}.
For example, speed wise, a simple hash based solution gives the fastest result when Nitro is not applied (and of course the most accurate), with Nitro giving it another significant performance boost with a very small accuracy degradation.
Our new insights are related to the questionable benefit of the minimum increment (also known as conservative update) optimization of CMS, and guidelines for the use of sampling, e.g., Nitro, with finger-print based hash tables like Cuckoo depending on whether the application is update intensive or query intensive.
Finally, we make all our code available in open source~\cite{FiltersSketchesCode}.

\section{Background}
\label{sec:background}

\subsection{Hash Tables}
\label{sec:hash}
\begin{figure}[t]
	\centering
	\includegraphics[width=0.8\columnwidth]{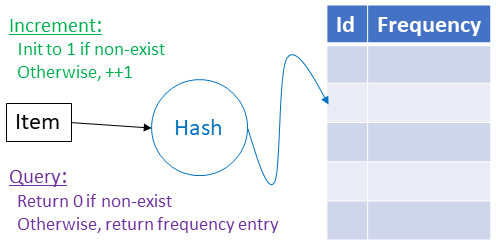}
	\caption{Hash Table Based Frequency Counting}
	\label{fig:hashtbl}
\end{figure}

The simplest approach to track the frequency of items in a data stream is to maintain a counter for each encountered item in a hash table, as shown in Figure~\ref{fig:hashtbl}.
The first time the item appears, it is inserted into the hash table with a count of $1$.
Each subsequent occurrence simply increment the counter.
A frequency count query for an item $x$ simply returns $x$ counter from the hash table if it exists, or $0$ otherwise.

The main benefit of this approach is its simplicity.
The drawbacks include a linear memory consumption.
This is because we need to allocated a counter for each unique item in the stream, and in the worst case it is on the same order as the stream length.
Moreover, the hash table needs to store the exact identifier, which can be quite large.
For example, a typical 4-tuple consisting of source IP, source port, destination IP, and destination port consumes $128$ on IPv4 and $196$ bits on IPv6.
In case of URIs, the size is much larger.
Further, there is the need to calculate a hash function on each element in the stream, which also involves non-negligible computational overhead.

\subsection{Count-Min sketch}
\begin{figure}[t]
	\centering
	\includegraphics[width=0.85\columnwidth]{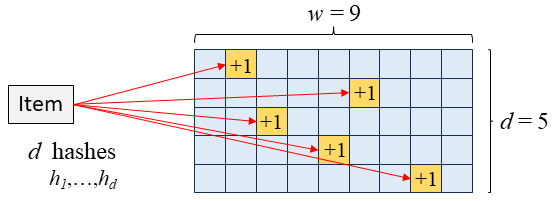}
	\caption{Count-Min Sketch}
	\label{fig:cms}
\end{figure}

A Count-Min sketch (CMS)~\cite{CountMinSketch} data structure consists of $d$ arrays of $w$ counters each, denoted CMS[$d,w$], and $d$ pairwise independent hash functions $\{h_1,\ldots,h_d\}$, as illustrated in Figure~\ref{fig:cms}.
In order to increment the frequency of an item $x$, all counters CMS[$i,h_i(x)$] are incremented, for all $i\in [1,\ldots,d]$.
In order to estimate the frequency of $x$, CMS returns $\min_{i\in[1,\ldots,d]}$ CMS[$i,h_i(x)$].

Denote the true frequency of $x$ after processing $N$ items by $f(x)$.
Assigning $w=e\epsilon^{-1}$ and $d=\log(\delta^{-1})$ ensures that a frequency the estimate $0 \leq \hat{f}(x) - f(x) \leq \epsilon N$ with probability $1-\delta$.
Notice that the memory requirement of CMS is independent of the number of items that are inserted $N$ (discounting the $O(\log N)$ bits per counter).
However, the absolute error value does grow linearly with $N$.
Also, CMS does not save any identifiers, which reduces its memory overhead, especially when identifiers are large.

The \emph{conservative update}~\cite{ConservativeUpdate} or \emph{minimum increment}~\cite{SBF} optimization mandates that only the minimal counter(s) would be incremented on each update.
The motivation for this is that non-minimal counters have reached their current value due to hash collisions, so incrementing them further increases the error for the colliding flows, without impacting the estimation of the current flow.
Thus, this optimization helps reduce the expected error~\cite{CUAnalysis}.
The downside of this optimization is that it prevents handling negative values (or decrements) and is not always applicable.
It also requires additional computational steps and memory accesses.

The main drawbacks of CMS include the fact that it requires computing $k$ hash functions and also accessing $k$ random memory locations on each query and update.
This implies both heavy computational load and hardware-cache unfriendly access~pattern.

CMS is used here as a representative of several additional counter matrices sketches, such as Count Sketch~\cite{CountSketch}, UnivMon~\cite{UnivMon}, K-ary~\cite{k-ary}, and Spectral Bloom Filters~\cite{SBF}.
Each has a slightly different update rule and guarantees, and all suffer from the same shortcomings.

\subsection{NitroSketch}

Nitrosketch~\cite{NitroSketch} solves the above mentioned inherent shortcomings of Count-Min sketch and similar counter array based sketches mentioned above for software implementation.
This is by updating the respective counter of each array with probability $p$ (a parameter), where $p$ can be even smaller than $1/d$ (where $d$ is the number of arrays) to enable skipping entire events.
Further, since invoking the PRNG has non-negligible costs, the actual process used is that at the beginning and after every counter update, the number of counters updates to skip is drawn from geometric distribution with an average of $1/p$.
This gives an equivalent behavior, at a much reduced computational cost as now the PRNG is only invoked on average every $1/p$ potential counter updates (once in every $d/p$ events).
As been shown in~\cite{NitroSketch}, the error rapidly converges to the same one as provided by the original sketch.
A downside of sampling in general and Nitro in particular is that it does not support~decrementing.

\subsection{Space Saving}
\begin{figure}[t]
	\centering
	\includegraphics[width=\columnwidth]{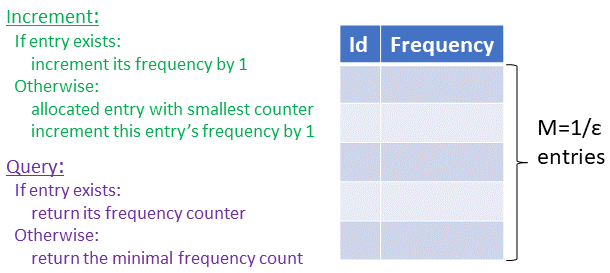}
	\caption{Space Saving}
	\label{fig:SS}
\end{figure}

Space Saving (SS) is an efficient mechanism for detecting heavy-hitters, but it also provides an effective method for estimating specific items' frequency~\cite{SpaceSavings}.
Specifically, Space Saving includes an array of $M$ pairs of IDs and tuples, as illustrated in Figure~\ref{fig:SS}.
When an item with id $x$ arrives, if $x$ has an allocated entry, it associated counter is increment.
Otherwise, $x$ replaces the item whose counter $C_m$ is minimal, and $C_m$ is incremented (without being reset first).
For frequency estimation, if $x$ has an entry, then the value of the associated counter is returned.
Otherwise, the value of the minimal counter is returned.

Space Saving guarantees that $0 \leq \hat{f}(x) - f(x) \leq N/M$.
Hence, when $M = \epsilon^{-1}$, the error is (one side) bounded by $\epsilon N$.

The memory access pattern of Space Saving is more hardware-cache friendly than CMS.
Space Saving can be implemented so that its computational complexity is $O(1)$ with a single hash function computation, but the details are not trivial~\cite{ben2016heavy}.
Unlike CMS, the error guarantee is deterministic and it holds only $1/\epsilon$ counters.
Yet, Space Saving needs to save $1/\epsilon$ identifiers, so its memory efficiency when compared to CMS depends on the identifiers' size\footnote{
	Space Saving can be augmented to store $O(\log \delta^{-1})$-bit fingerprints instead of identifiers, bringing its size down to below CMS, in which case the error guarantee of Space Saving becomes accordingly probabilistic.}.
Also, Space Saving as is can detect heavy-hitters, while CMS needs to be complemented with a heap structure in order to do the same\footnote{
	It is also possible to employ reversible CMS for heavy-hitter detection~\cite{ReversibleSketch}.
	While its space and computational overheads are even worse, it does support deletions.}.
Finally, the number of non-zero counters in CMS can be used to estimate the number of unique flows (also known as cardinality estimation and count distinct), which is not the case with Space~Saving.
Yet, given the significant computation cost and poor accuracy of unique flow estimation from CMS vs. the negligible cost of HLL~\cite{flajolet2007hyperloglog} and its high accuracy, in practice it is extremely rare to employ CMS for this. 

Space Saving is a representative of several similar algorithms, including Misra-Gries~\cite{MG82}, Frequent~\cite{Frequent,BatchDecrement}, and Lossy Counting~\cite{LossyCounting}.
We have chosen it here since Space Saving has been claimed to be the fastest and most accurate among them~\cite{SpaceSavingIsTheBest}.

\subsection{RAP}

RAP improves on Space Saving's ability to detect heavy hitters and the frequency estimation accuracy, especially with heavy-tailed workloads~\cite{RAP}.
With RAP, when an item from a flow $x$ that does not have an allocated counter arrives, $x$ is only allocated the entry of the minimal counter $C_m$ with probability $1/(C_m+1)$ rather than unconditionally as done in Space Saving.
The idea here is that since most untracked items belong to the tail, always giving them a counter hurts the ability of Space Saving to track and heavy hitters and reduces its estimation accuracy.
By using the above mentioned probabilistic insertion filter, RAP avoids giving an entry to most tail items, while still admitting heavy-hitters as the latter will most probably arrive often enough to obtain a counter.

\subsection{Counting Cuckoo Filter}

A cuckoo hash table~\cite{Cuckoo} is composed of multiple buckets, each consisting of a (small) fixed number of slots that can hold hashed items, as well as two independent hash functions $h_1$ and $h_2$.
An item with id $x$ may be placed in one of the buckets $h_1(x)$ and $h_2(x)$.
Hence, looking up an item requires checking at most these two buckets.
As for insertion, if the bucket $h_1(x)$ has empty slots, then $x$ is inserted there.
Otherwise, $x$ is inserted into bucket $h_2(x)$.
In case this bucket was already full, then a random item $y$ is removed from bucket $h_2(x)$, to make room for $x$, and $y$ is moved to its alternate bucket.
This may start a chain reaction, in case the alternate bucket of $y$ was also full, and the process could continue until a maximal number of attempts, after which failure is declared.
It has been shown that when the bucket size is $3$, the insertion process will succeed with high probability as long as the number of items is less than $80\%$ of the total hash table~capacity~\cite{Cuckoo-Survey}.

Cuckoo filters extends the above idea to implement a Bloom filter (approximate set-membership) like functionality~\cite{CuckooFilter}.
Rather than storing an exact item, a Cuckoo filter only stores a fingerprint of size $L$ of the item.
Also, the hash functions are defined such that $h_2(x)=h_1(x) \oplus h_1(\mathrm{fingerprint}(x))$.
Notice that due to the symmetry of $\oplus$, $h2(x)$ can be computed from $h_1(x)$ and vice versa.
The latter is needed since Cuckoo filters do not store identifiers.
This way, a Cuckoo filter can answer approximate set membership queries without false negatives and having a false positive ratio bounded by $2^{-L}$.
Counting Cuckoo filters add to each slot a counter, which is logically initialized to zero and is incremented for each add/insert operation.

\section{Our Novel Combinations}
\label{sec:ours}

\subsection{NitroHash}
\label{sec:nitrohash}

In NitroHash, we add a NitroSketch like optimization to the hash based solution described in Section~\ref{sec:hash}.
Specifically, rather than adding every item to the hash table, and incrementing its value if already there, we do so only with probability $p$.
Further, to reduce the number of PRNG invocations, at the beginning and after each update, we draw the number of item to skip before the next actual update from a geometric distribution with an average of $1/p$.
When required to estimate the frequency of an item, we multiply its counter in the table, if exist, by $1/p$.
In case there is no counter, we return~$0$.

This optimization has three complementing runtime benefits:
We invoke the hash function and perform the associated random memory access only once every $1/p$ items in expectation.
We invoke the PRNG only once every $1/p$ items in expectation.
We expect to insert much fewer items into the hash table, so its size can be much smaller, as demonstrated in the evaluation section.

Let us note that this is not the same as Sticky Sampling~\cite{LossyCounting}.
In the latter, on each item arrival, a non-monitored item is allocated a counter with probability $p$, whereas an item that already has a counter is always incremented (deterministically).
This means performing a hash table lookup for each item arrival plus a PRNG invocation for each occurrence of a non-monitored item.
With NitroHash, each item is always sampled with probability $p$, where we employ the geometric based sampling to reduce the number of PRNG invocations.

\subsection{NitroCuckoo}
\label{sec:nitrocuckoo}

Similarly to above, in NitroCuckoo augment a counting Cuckoo filter with a NitroSketch like optimization.
That is, rather than performing an increment/add operation on the filter for every item, we do so only with probability $p$.
Also, we chose the number of items to skip from a geometric distribution with an average of $1/p$.
A frequency estimation query multiplies the associated counter by $1/p$ if exists, and returns $0$ otherwise.
The benefits here are similar to those mentioned for NitroHash.

\section{Implementation}
\label{sec:implementation}

We have implemented all data structures and associated algorithms in Rust, and the code is available in open source at~\cite{FiltersSketchesCode}.
Our code is single threaded.
For the hash table implementations, we have used Rust's default hash table.
For CMS, we have taken the open source implementation of~\cite{AmadeusStreaming}\footnote{This implements the conservative update optimization.}, and have also augmented it as needed to obtain our NirtoSketch implementation.
We have taken the Cuckoo filter implementation from~\cite{RustCuckoo} and augmented it to serve as a counting Cuckoo filter, and then further modified it to also serve as a NitroCuckoo filter.
We have implemented Space Saving ourselves, and maintain the counters sorted using Rust's default priority queue implementation~\cite{RustPriorityQueue}.
We have also realized the RAP optimization for it, and the exact behavior, with or without RAP, is controlled by a runtime parameter.
It is possible that a more sophisticated implementation along the lines of~\cite{ben2016heavy} would have yielded faster operation execution; this is left for future work.


\section{Evaluation}
\label{sec:evaluation}

\subsection{Method}

\subsubsection{Metrics}

\paragraph{Throughput}
We measure the throughput in terms of number of operations per second.
This is obtained by measuring the net elapsed time for invoking the relevant operations for each item in a given trace, in the order of which items appear in the trace, and then dividing the trace size by the measured execution time.

Specifically, the application first uploaded and parsed an entire trace file, storing its parsed lines as entries in an array.
Then, we iterated over the lines of the array, invoking the respective data structure method for each entry.
The memory, number of items, and various error metrics were measured in a separate run than the throughput measurement.
For throughput measurements, we measured the elapsed time during the iteration over the array.
In case of write only measurements, each iteration consisted of invoking only an add/increment operation for each encountered item.
In the case of write-read measurement, in each iteration we perform an add/increment operation followed immediately by a query for the same item.
For read-only, we perform two iterations: in the first, we add/increment all items, whereas in the second we issue queries for all items; here, we only measure the net execution time of the second iteration. 

\paragraph{Memory and Number of Entries}
We measure the memory consumed by the associated data structures of each algorithm.
For data-structures that store explicit item identifiers or fingerprints, we also measure the number of items in the data-structure.
This includes the hash table as well as the counting cuckoo filter, in both cases with and without the nitro optimization.

\paragraph{Approximation Error}
Denote $N$ the number of nodes in the trace, $M$ the number of unique flows in the trace, $\widehat{f(x)}$ the estimated frequency of $x$ at a given time and $f(x)$ the true frequency of flow $x$ at the same time.
We measure the following variants of approximation error\footnote{According to the findings of~\cite{SALSA}, the AVGERR measure becomes minimal when estimating all frequencies as~$0$.}:
\begin{description}
	\item[On Arrival:] For each encountered item $x_i$, we immediately query its frequency estimation and compute
		\small
		$$\text{OA MSRE} = \frac{\Sigma_i^N{\sqrt{\left(f(x_i)-\widehat{f(x_i)}\right)^2}}}{N}.$$
		$$\text{OA AVGERR} = \frac{\Sigma_i^N{|{f(x_i)-\widehat{f(x_i)}|}}}{N}.$$
		$$\text{OA AVGRELERR} = \frac{\Sigma_i^N{\left( |{f(x_i)-\widehat{f(x_i)}}|/f(x_i)\right )}}{N}.$$
		\normalsize
		
	\item[Per Flow:] After inserting all items in the trace, we scan all flow identifiers that appeared in the trace, query their frequency estimation and compute
		\small
		$$\text{Flow MSRE} = \frac{\Sigma_i^M{\sqrt{\left(f(x_i)-\widehat{f(x_i)}\right)^2}}}{M}.$$
		$$\text{Flow AVGERR} = \frac{\Sigma_i^M{|{f(x_i)-\widehat{f(x_i)}|}}}{M}.$$
		$$\text{Flow AVGRELERR} = \frac{\Sigma_i^M{\left( |{f(x_i)-\widehat{f(x_i)}}|/f(x_i)\right )}}{M}.$$
		\normalsize
		
	\item[Postmortem:] After inserting all items in the trace, we scan again all items in the trace in their appearance order and compute
		\small
		$$\text{PM MSRE} = \frac{\Sigma_i^N{\sqrt{\left(f(x_i)-\widehat{f(x_i)}\right)^2}}}{N}.$$
		$$\text{PM AVGERR} = \frac{\Sigma_i^N{|{f(x_i)-\widehat{f(x_i)}|}}}{N}.$$
		$$\text{PM AVGRELERR} = \frac{\Sigma_i^N{\left( |{f(x_i)-\widehat{f(x_i)}}|/f(x_i)\right )}}{N}.$$
		\normalsize
\end{description}

Notice that the formulas for the On Arrival errors and the Postmortem cases look the same, but they are computed differently, as explained in the textual description~above.

\subsubsection{Settings}
All measurements were performed on an Intel Core i9-13900 (8+16 Cores/36MB/32T/2.0GHz to 5.2GHz/65W) machine with 128GB (2X64GB) DDR5 DRAM memory, and an M.2 2280 2TB PCIe NVMe Class 40 SSD, running Microsoft Windows 11 Pro.
The L1 cache size is $2.1$MB, L2 is $32$MB, and L3 is $36$MB. 
The code was compiled for release mode using Rust's cargo application.

As mentioned before, the entire trace is first loaded into memory and parsed, so we measure the net time for processing the items in memory by the various algorithms and data structures.
The amount of DRAM memory in the machine used for testing is high enough so that there were no page swapping during the execution.

\subsubsection{Traces}

We have used several real-world traces taken from the CAIDA repository of backbound routers~\cite{caida-traces}.
These traces are summarized in Table~\ref{tbl:traces}.

\begin{table}[t]
\begin{center}
	\begin{tabular}{|l|r|r|}
		\hline
		Trace & \# items \ \ \ \ \ & uniques \ \ \ \ \\
		\hline
		\hline
		Chicago15 & $133,988,641$ & $3,495,149$ \\  
		\hline
		Chicago16 & $88,529,637$ & $1,650,097$ \\  
		\hline
		Chicago16Small & $1,000,000$ & $69,046$ \\
		\hline
		Chicago1610Mil & $10,000,000$ & $332,183$ \\  
		\hline
		NY19A & $30,098,745$ & $1,522,382$ \\
		\hline
		NY19B & $63,284,829$ & $2,968,038$ \\
		\hline
		SJ14 & $188,511,031$ & $2,922,904$ \\
		\hline
	\end{tabular}
\end{center}
\caption{Traces from~\cite{caida-traces} used in this work.}
\label{tbl:traces}
\end{table}

The traces Chicago16Small and Chicago1610Mil are prefixes of Chicago16.
They are used to exemplify the impact of trace length on the results when considering the same trace.
All displayed data points are the average of 13 runs, and we also plot the respective confidence~intervals.


\subsection{Throughput Results}
\label{sec:thput}

\subsubsection{Update Only}
\label{sec:thput-write}

\begin{figure*}[t]
	\centering
	\subfloat[][ny19A]{\includegraphics[width=0.32\textwidth]{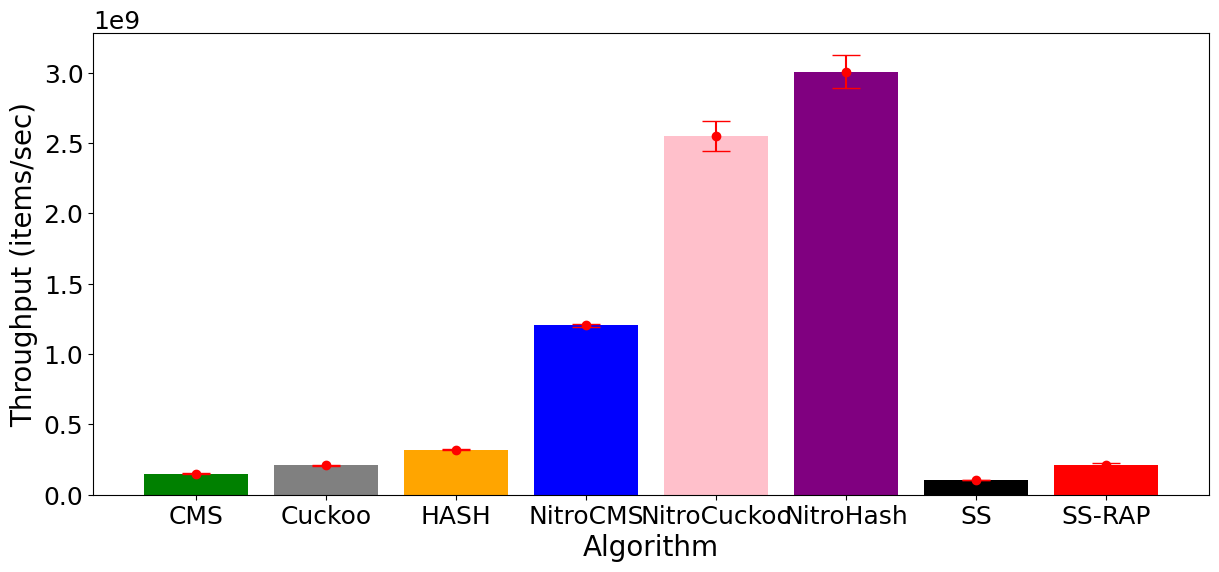}\label{fig:thpt-write:ny19A}}
	\subfloat[][SJ14]{\includegraphics[width=0.32\textwidth]{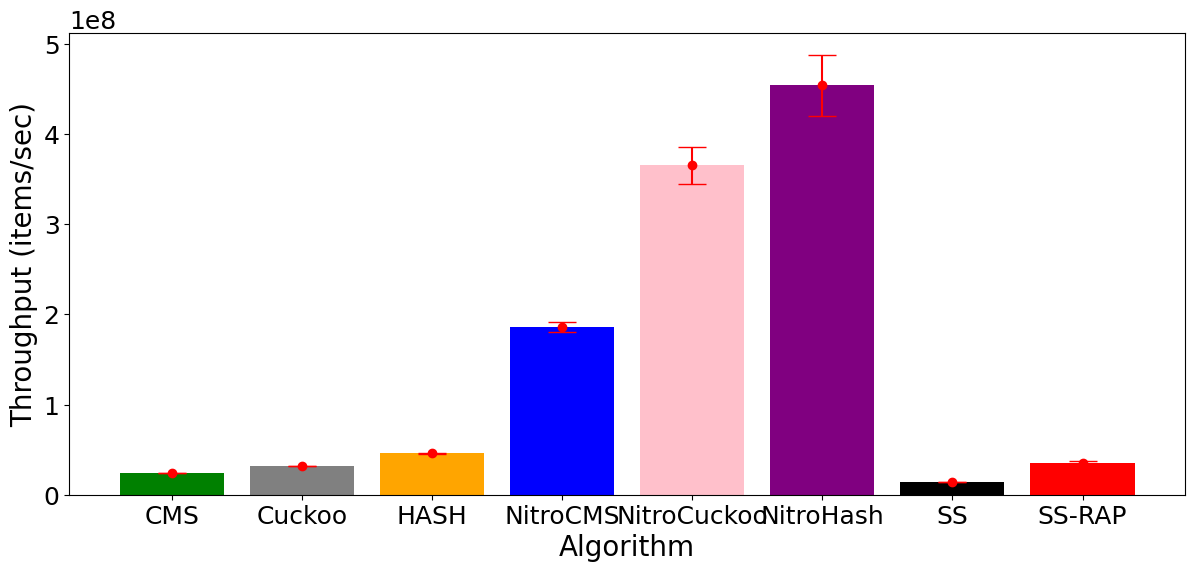}\label{fig:thpt-write:SJ14}}
	\subfloat[][Chicago15]{\includegraphics[width=0.32\textwidth]{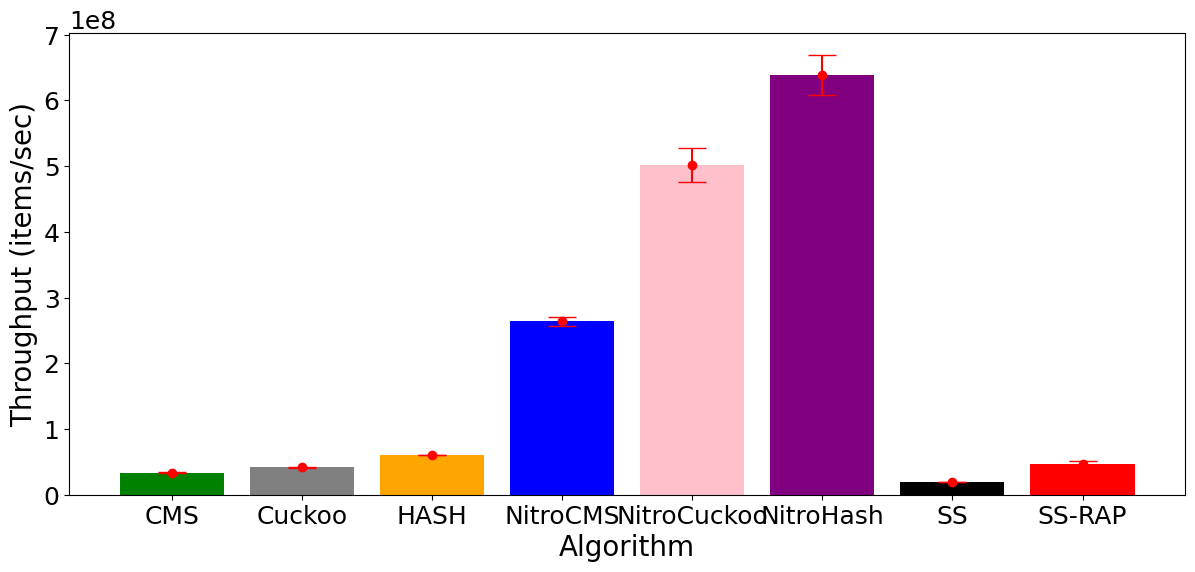}\label{fig:thpt-write:Chicago15}}\\
	\subfloat[][Chicago16Small]{\includegraphics[width=0.32\textwidth]{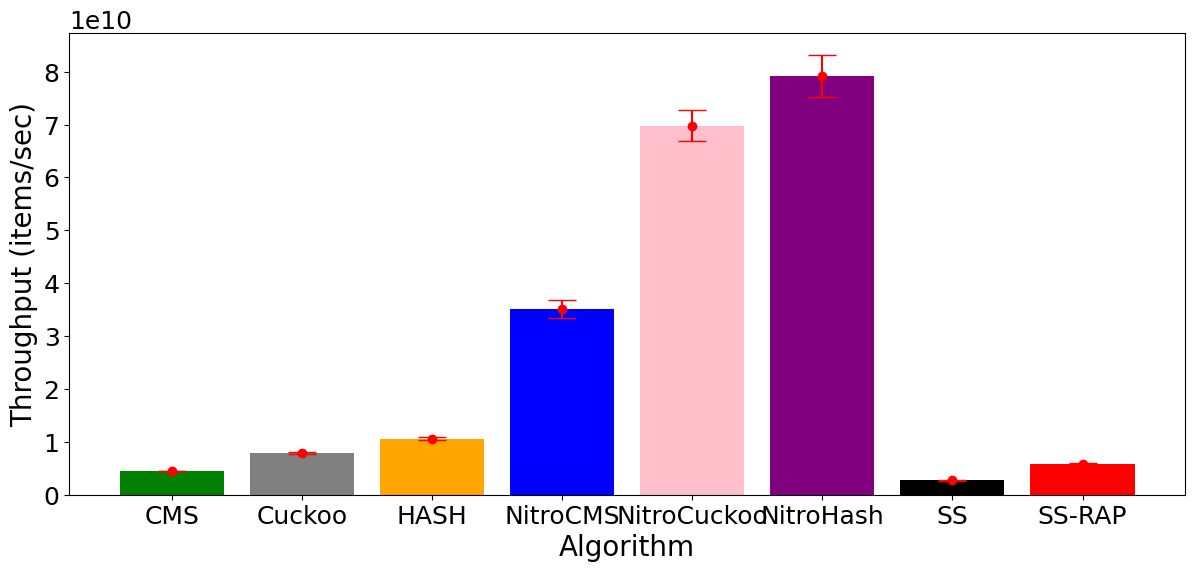}\label{fig:thpt-write:Chicago16Small}}
	\subfloat[][Chicago1610Mil]{\includegraphics[width=0.32\textwidth]{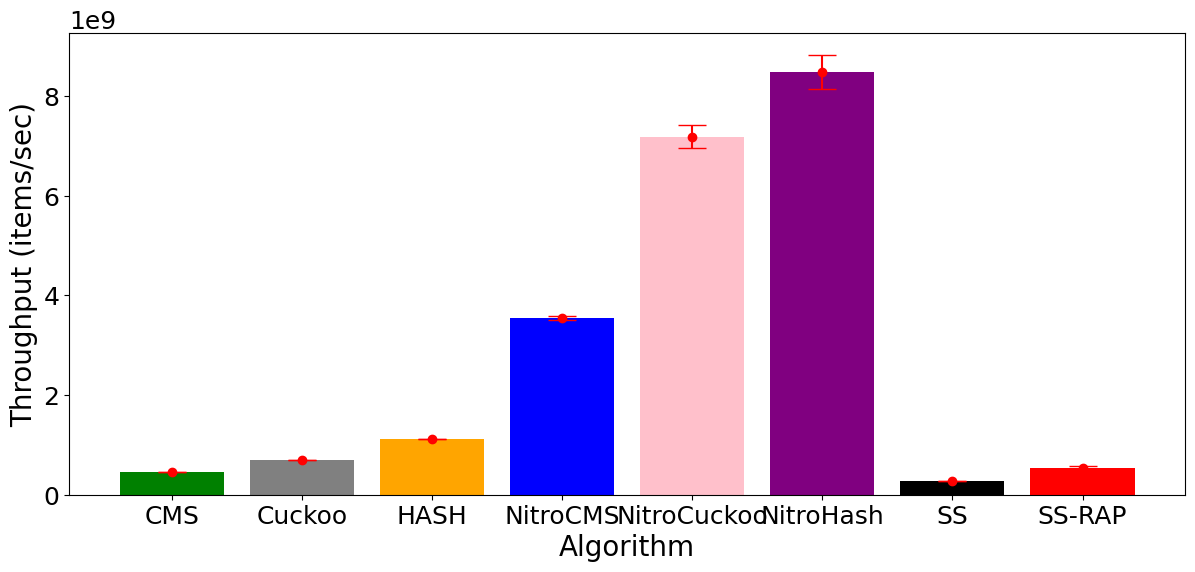}\label{fig:thpt-write:Chicago1610Mil}}
	\subfloat[][Chicago16]{\includegraphics[width=0.32\textwidth]{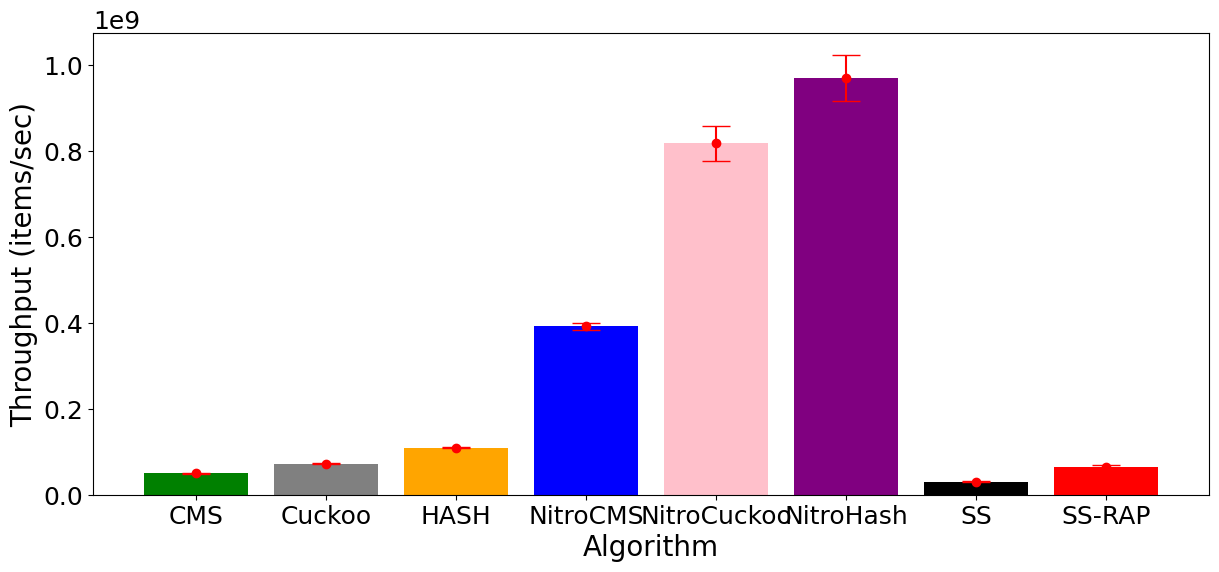}\label{fig:thpt-write:Chicago16}}
	\caption{Throughput results for the write only test}
	\label{fig:thpt-write}
\end{figure*}

The throughput results for the write-only measurements appear in Figures~\ref{fig:thpt-write}.
As can be seen, the Nitro optimization brings an order of magnitude improvement to the throughput for all three algorithms it was applied to: Hashing, Cuckoo, and CMS.
This is as expected, following the discussion above about this optimization.

Interestingly, among the three, Hash (NitroHash, respectively) came out the fastest policy, followed by Cuckoo (NitroCuckoo, respectively), with CMS (NitroCMS, respectively) being the slowest, due to its multiple hash computations and random memory accesses per update.
SpaceSaving is the slowest policy, with SpaceSaving-RAP being slightly better than CMS.
As expected, the RAP policy improves performance, since there are fewer inserts into the data structure.
Also, the implementation of SpaceSaving that we used is a bit naive, as it uses Rust's priority queue.
It is possible that a more sophisticated implementation along the lines of~\cite{ben2016heavy} would have resulted in better performance.
This is left for future work.

Another factor that slows down CMS is the conservative update optimization that was applied, by which only the minimal counter is incremented.
Our conjecture is that in software based implementation, it might be better to configure the structure with a lower theoretical error guarantee (i.e., more space) than applying this optimization.
This is explored below in Section~\ref{sec:cms-deepdive}.

When comparing the Chicago16Small trace (Figure~\ref{fig:thpt-write:Chicago16Small}) to the other traces in Figure~\ref{fig:thpt-write}, an interesting phenomenon is revealed:
We notice that the throughput of all algorithms is roughly an order of magnitude better with the Chicago16Small trace than the other traces, which are much longer.
For Hash and NitroHash, we can relate this to the fact that they require more memory as the trace grows.
One could argue that for Cuckoo and NitroCuckoo is that the fuller the Cuckoo table becomes, the more common relocation is needed, and the relocation chains become longer.
However, in all traces the load factor on the table is very low since the number of unique items is below 10\% of the trace length, and we allocate enough items for the entire trace.
Further, for CMS and NitroCMS, it is surprising, as both the size and amount of work performed by these algorithms are independent of the number of items.
Recall that we pre-load into an in-memory array the entire trace before the timing measurement begins.
Also, since the machine has plenty of memory, the reason is not related to swapping.
Rather, we speculate that when the trace is small, the array fits better into the various levels of the hardware cache.
Yet, when the trace size grows, we get more hardware cache misses when trying to fetch the next items from the in-memory array, which dominates the time measurement.
In that sense, the shorter trace (Chicago16Small) gives better indication of the maximal direct throughput bound of each algorithm.

\subsubsection{Write Read}
\label{sec:thput-rw}
\begin{figure*}[t]
	\centering
	\subfloat[][ny19A]{\includegraphics[width=0.32\textwidth]{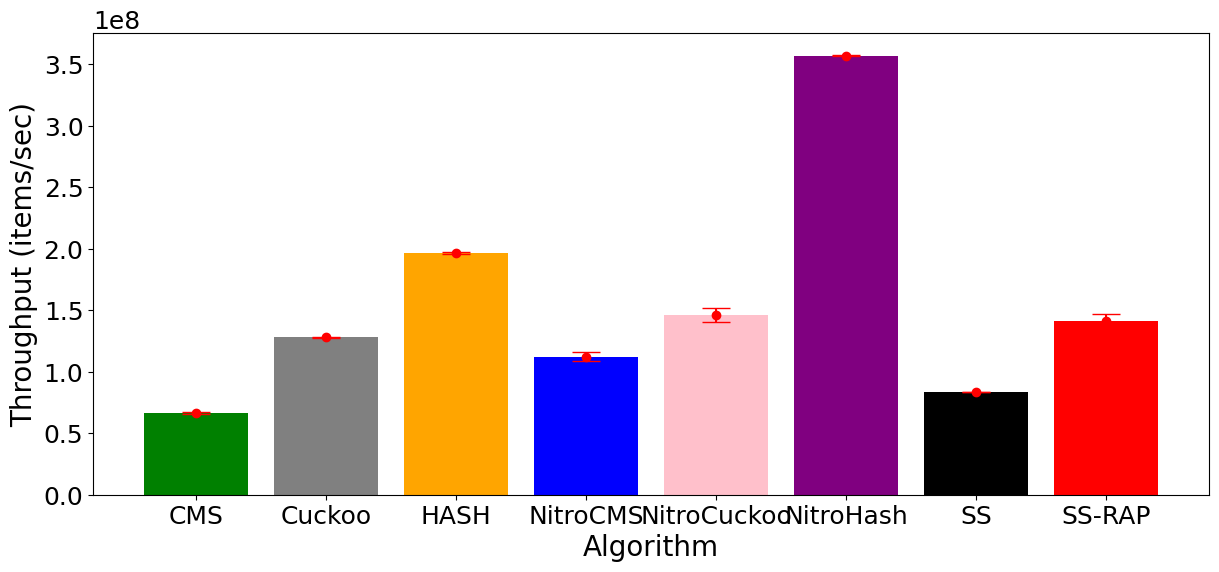}\label{fig:thpt-rw:ny19A}}
	\subfloat[][SJ14]{\includegraphics[width=0.32\textwidth]{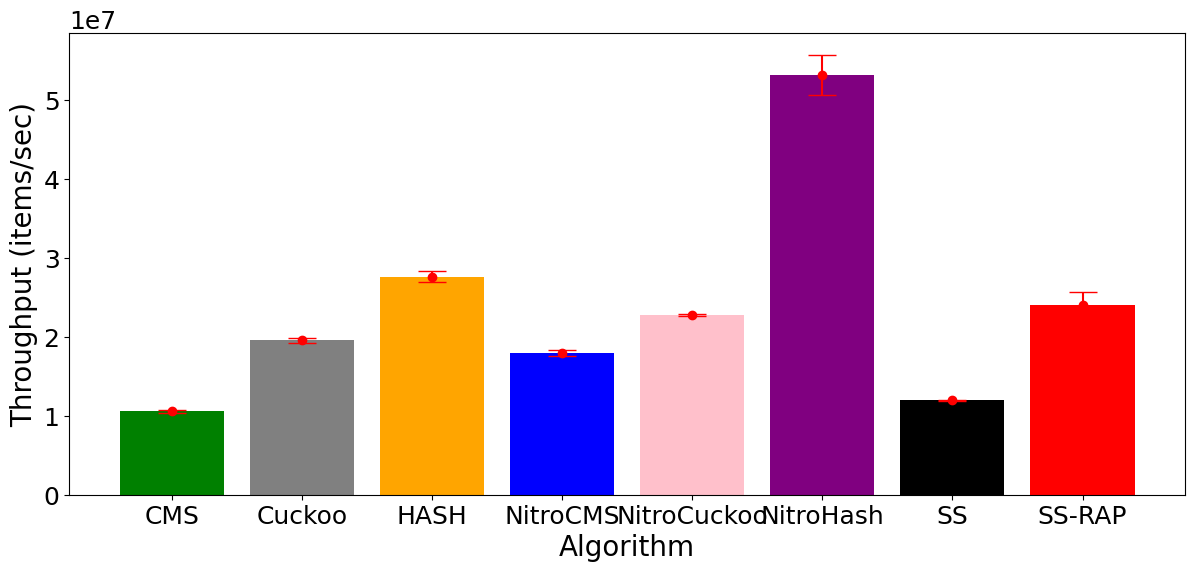}\label{fig:thpt-rw:SJ14}}
	\subfloat[][Chicago15]{\includegraphics[width=0.32\textwidth]{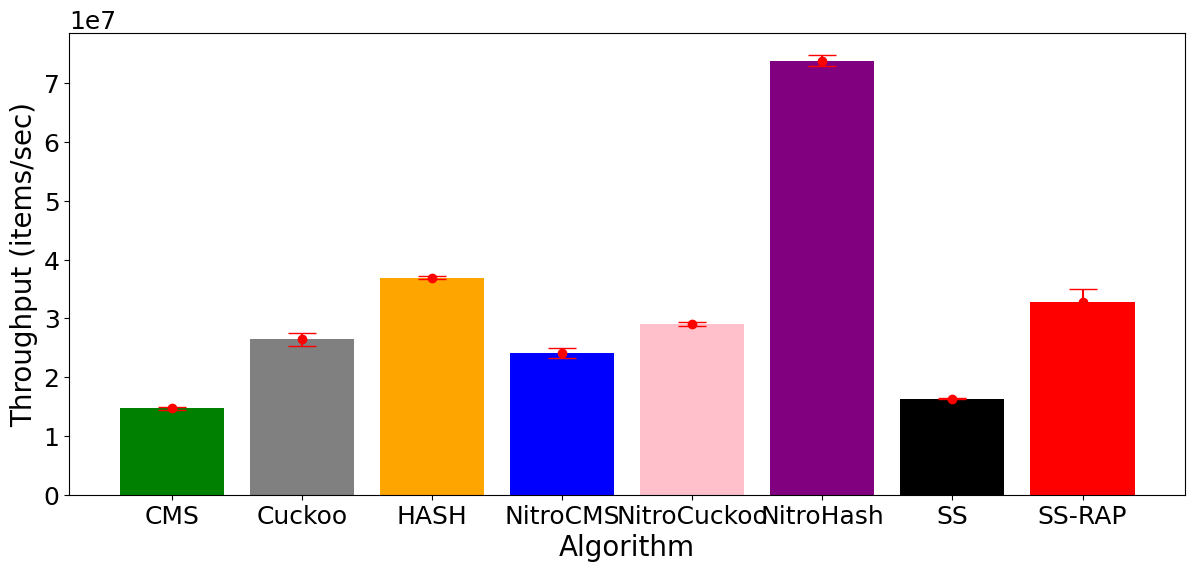}\label{fig:thpt-rw:Chicago15}}\\
	\subfloat[][Chicago16Small]{\includegraphics[width=0.32\textwidth]{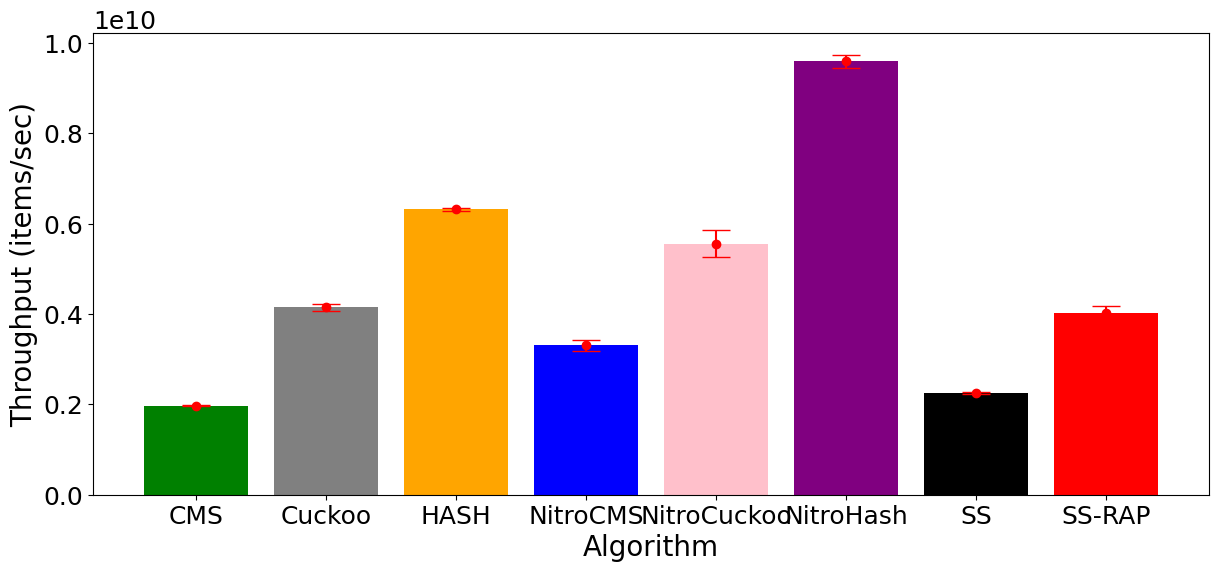}\label{fig:thpt-rw:Chicago16Small}}
	\subfloat[][Chicago1610Mil]{\includegraphics[width=0.32\textwidth]{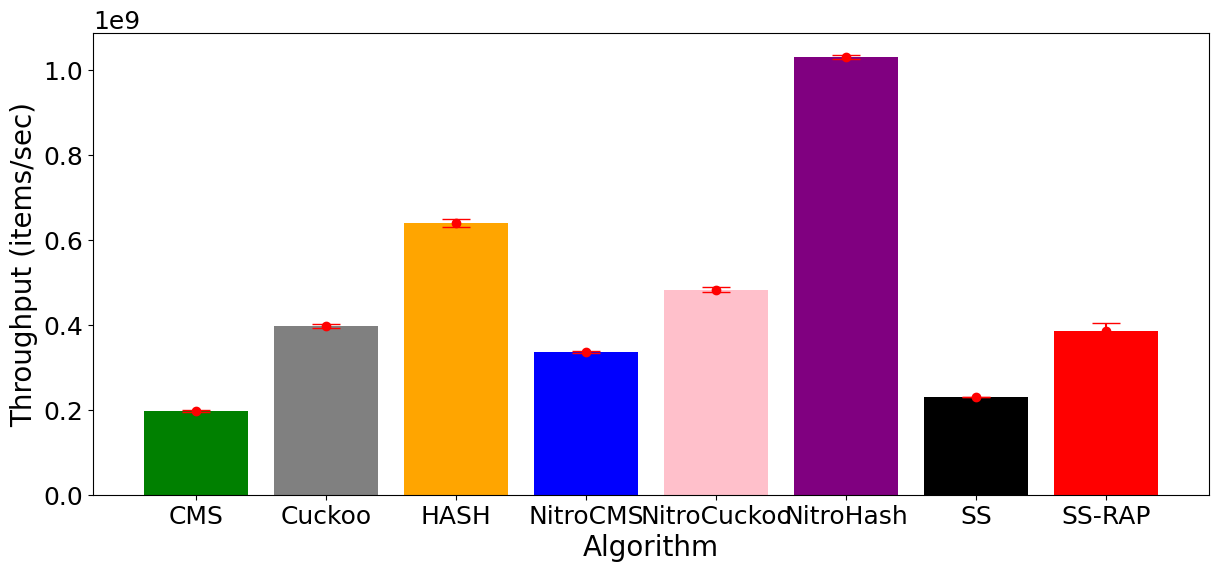}\label{fig:thpt-rw:Chicago1610Mil}}
	\subfloat[][Chicago16]{\includegraphics[width=0.32\textwidth]{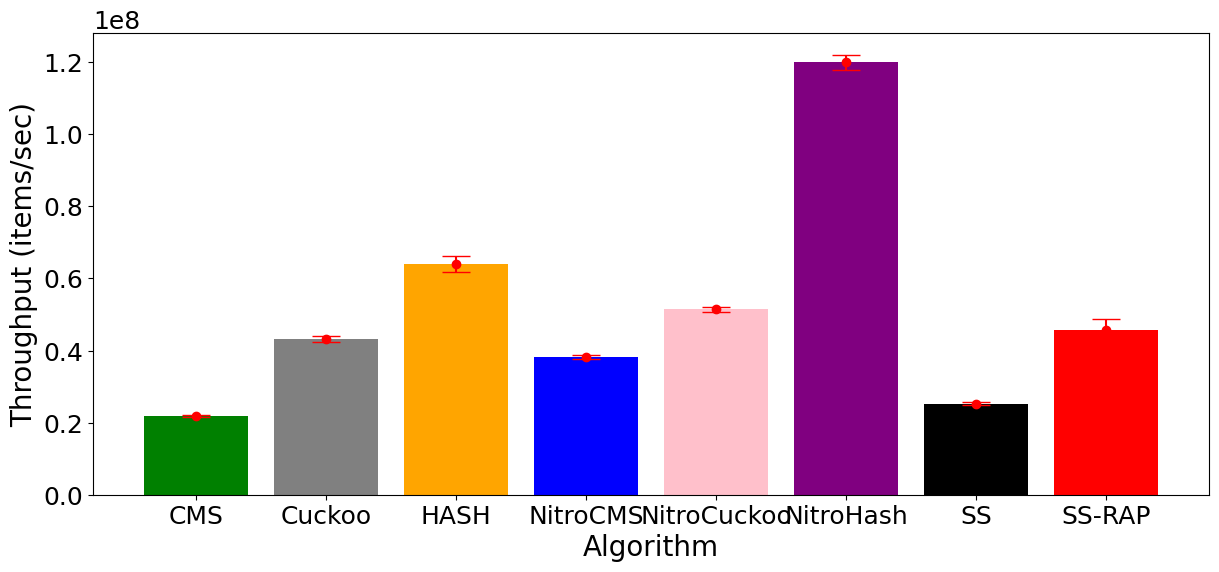}\label{fig:thpt-rw:Chicago16}}
	\caption{Throughput results for the write-read test}
	\label{fig:thpt-rw}
\end{figure*}

The throughput results for the write-read measurements appear in Figures~\ref{fig:thpt-rw}.
As can be seen, the Nitro optimization is less effective here, since Nitro only expedites the updating process, but has no impact on query execution.
Among the three Nitro enabled variants, NitroHash is the fastest, due to the fact that when using a plain hash table, each access only requires a single hash function calculation (in expectation).
NitroCMS is the slowest of the three since both its update and query processes involve multiple hash computation and random memory accesses.

Finally, Space Saving performed worse.
As mentioned above, this is due to the fact that our implementation only relies on Rust's priority queue, whose associative access time is slow.

\subsubsection{Read Only}
\label{sec:thput-read}

\begin{figure*}[t]
	\centering
	\subfloat[][ny19A]{\includegraphics[width=0.32\textwidth]{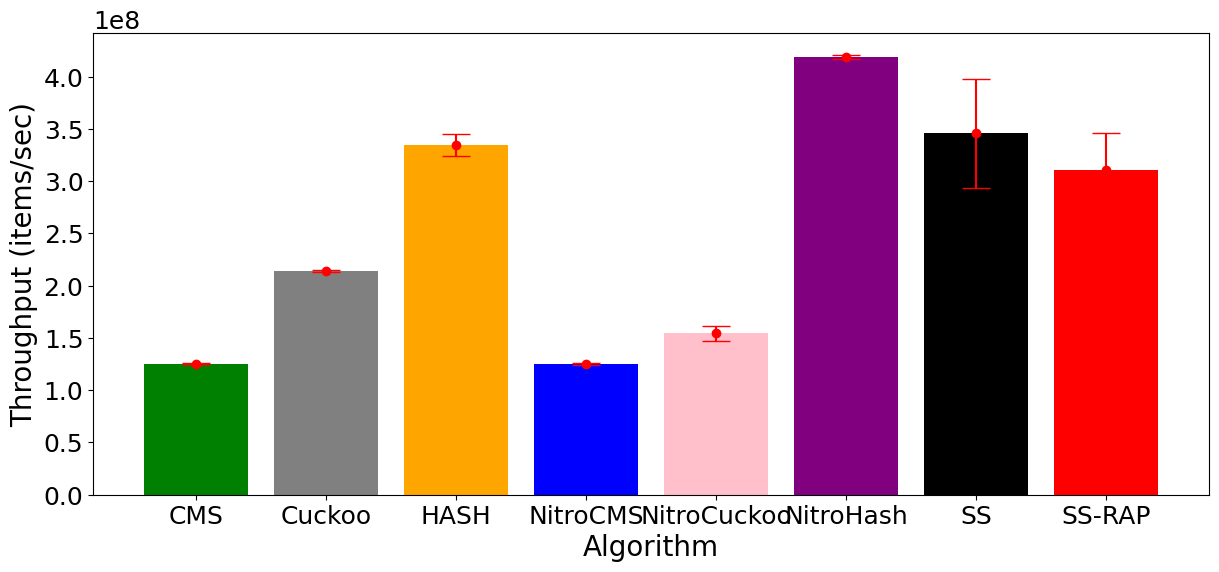}\label{fig:thpt-read:ny19A}}
	\subfloat[][SJ14]{\includegraphics[width=0.32\textwidth]{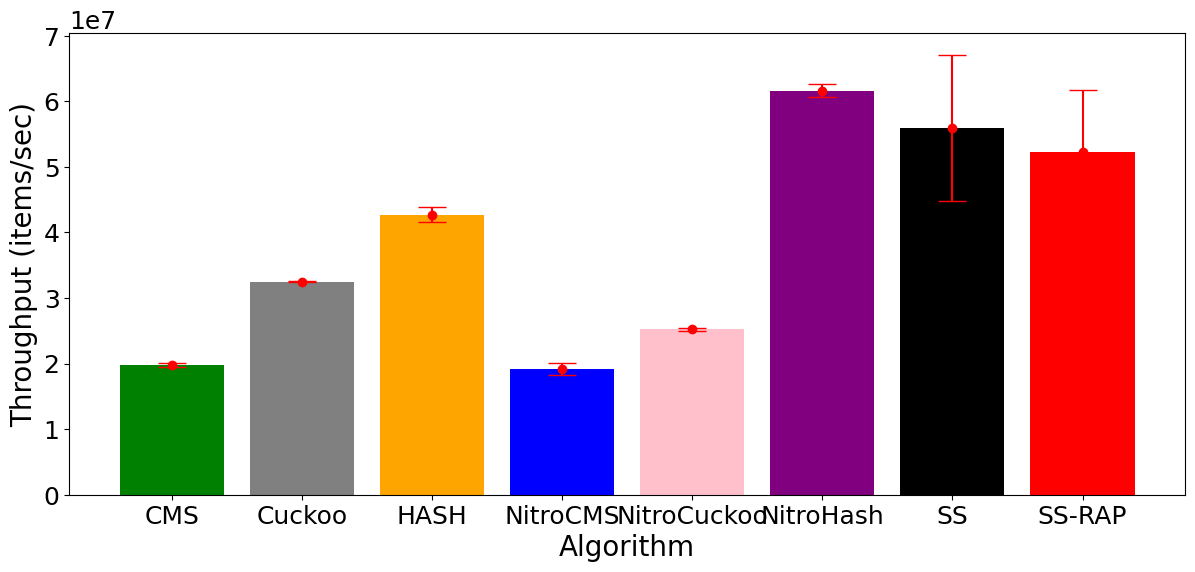}\label{fig:thpt-read:SJ14}}
	\subfloat[][Chicago15]{\includegraphics[width=0.32\textwidth]{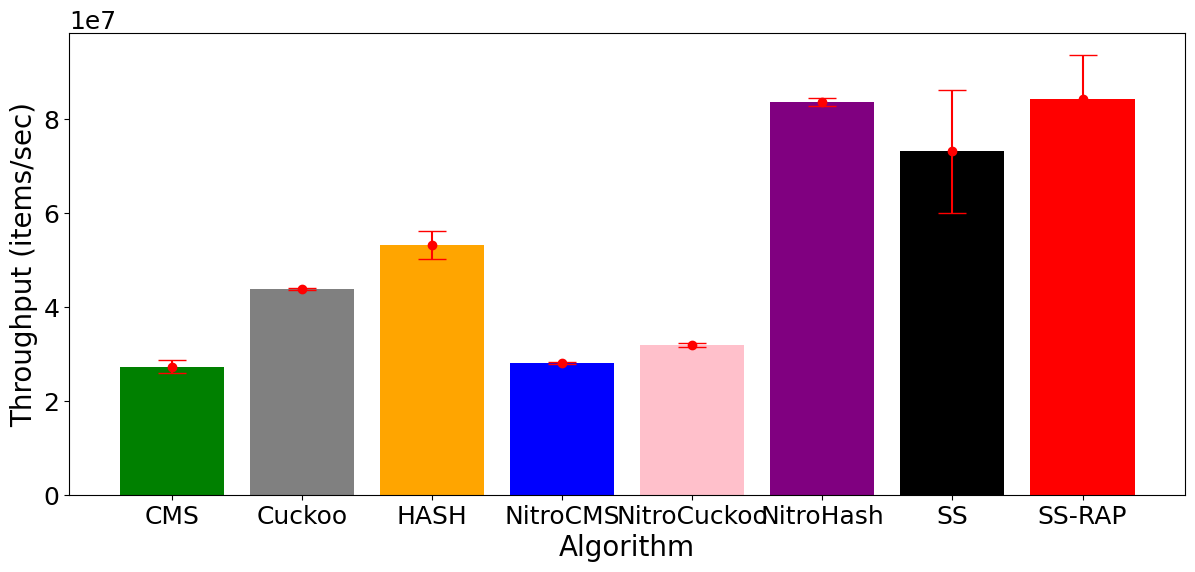}\label{fig:thpt-read:Chicago15}}\\
	\subfloat[][Chicago16Small]{\includegraphics[width=0.32\textwidth]{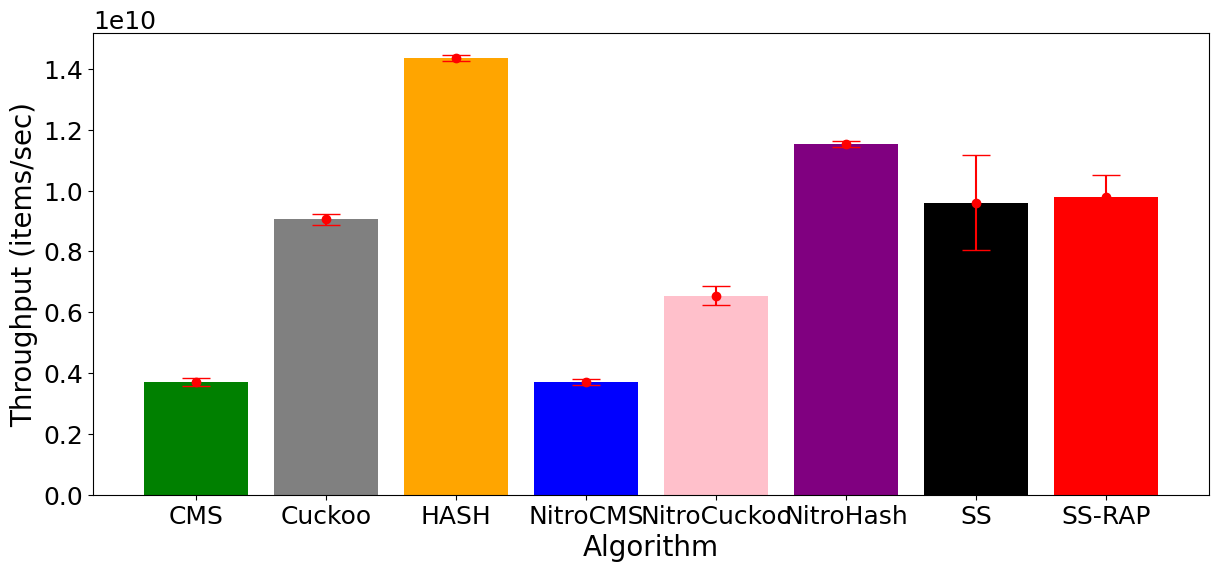}\label{fig:thpt-read:Chicago16Small}}
	\subfloat[][Chicago1610Mil]{\includegraphics[width=0.32\textwidth]{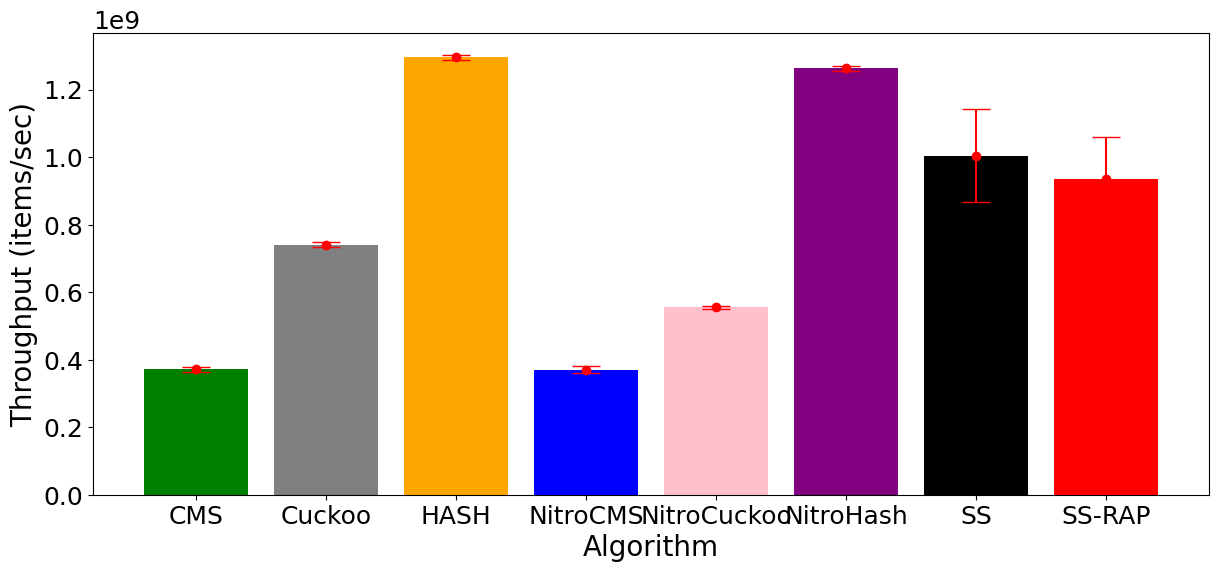}\label{fig:thpt-read:Chicago1610Mil}}
	\subfloat[][Chicago16]{\includegraphics[width=0.32\textwidth]{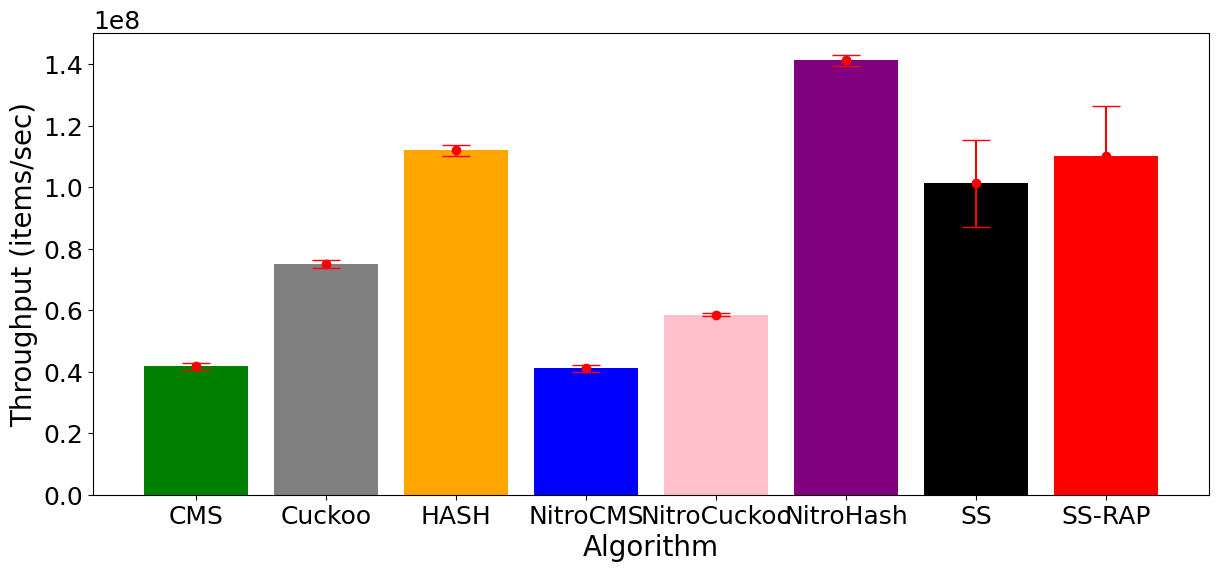}\label{fig:thpt-read:Chicago16}}
	\caption{Throughput results for the read only test}
	\label{fig:thpt-read}
\end{figure*}

The throughput results for the  read only measurements appear in Figures~\ref{fig:thpt-read}.
The most surprising result here is related to the impact of the Nitro optimization on the Hash algorithm.
Nitro improved performance here, despite not being part of the query process at all.
The reason is that fewer items have been inserted into the hash table.
This translates into a smaller final size and fewer items, thereby faster completion times.

In contrast, Cuckoo is faster than NitroCuckoo.
The reason is that as fewer items are inserted in the Cuckoo table with the Nitro optimization, additional lookup invocations require computing both hash functions and accessing two random memory locations than when the table is much fuller, which happens when Nitro is not used.
This indicates that when the read performance is more important, the use of sampling when creating the table is actually counter-productive, unless the table is pre-configured to accept fewer items.

In fact, for finger-printing hash table based solutions, be it counting Cuckoo filters, TinyTable~\cite{TinyTable}, Counting Quotient Filter~\cite{CountingQuotientFilter}, DASH~\cite{DASH}, etc., sampling reduces the load factor for a given table size and stream length.
Hence, the decision of whether to use sampling or not and how to configure the table in return depends on the type of application.
For network monitoring applications, which are update dominant and it is paramount to keep up with the line rate, the use of sampling while keeping a relatively large hash table, e.g., one that can accommodate all possible items even without sampling, is the best way to go.
On the other hand, in data-base applications, when the sketch (or filter) is used to represent a large but mostly static relation, the use of sampling can be beneficial when the size of the hash table is reduced roughly proportionately to the sampling ratio.
This way, both the space consumption is significantly lowered, and the query time improves.
We explore this below for NitroCuckoo in Section~\ref{sec:cuckoo-deepdive}.

As for Space Saving, we see here too the RAP slightly hurts the performance.
Once again, this is due to the fact that we used a priority queue, so searching for a non-existing item takes longer.
In general, the relative performance of Space Saving, with and without RAP, compared to the other algorithms is much better.
This is because the update process of Space Saving is much more expensive than query, at least with the priority queue based realization we used.

\subsection{Approximation Errors Results}
\label{sec:approx-err}

\begin{figure*}[t]
	\centering
	\subfloat[][ny19A]{\includegraphics[width=0.32\textwidth]{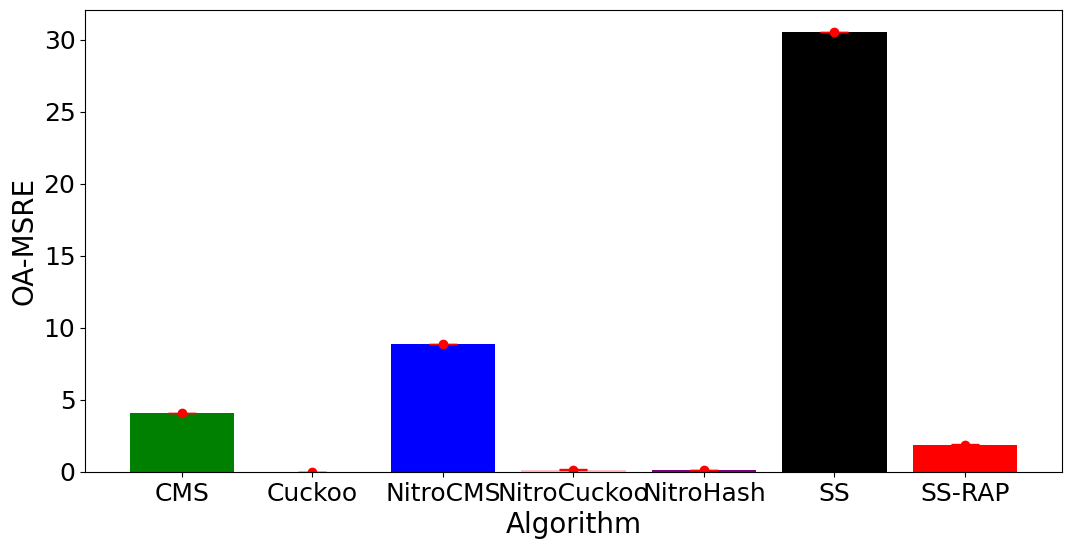}\label{fig:oa-msre:ny19A}}
	\subfloat[][SJ14]{\includegraphics[width=0.32\textwidth]{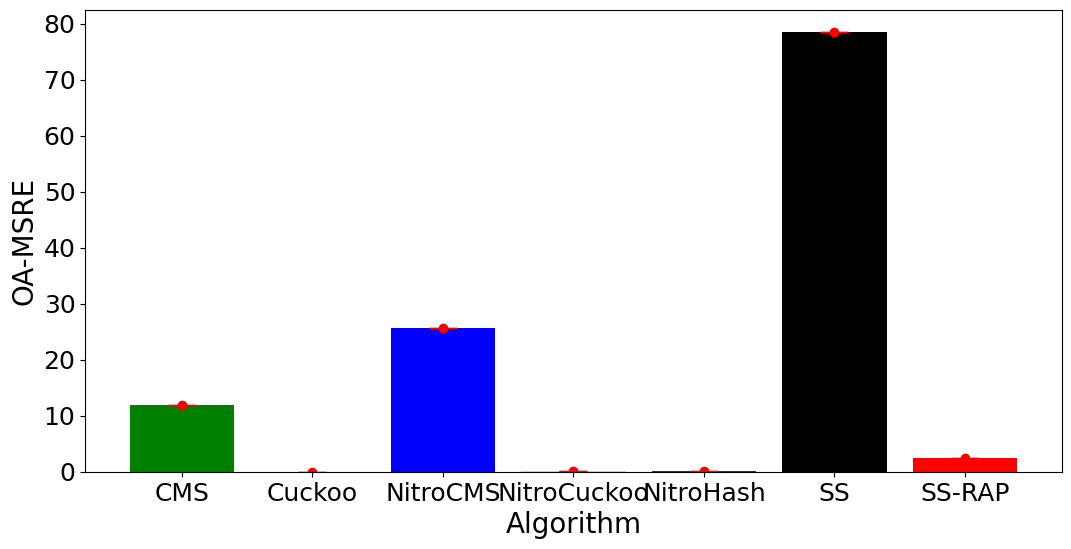}\label{fig:oa-msre:SJ14}}
	\subfloat[][Chicago15]{\includegraphics[width=0.32\textwidth]{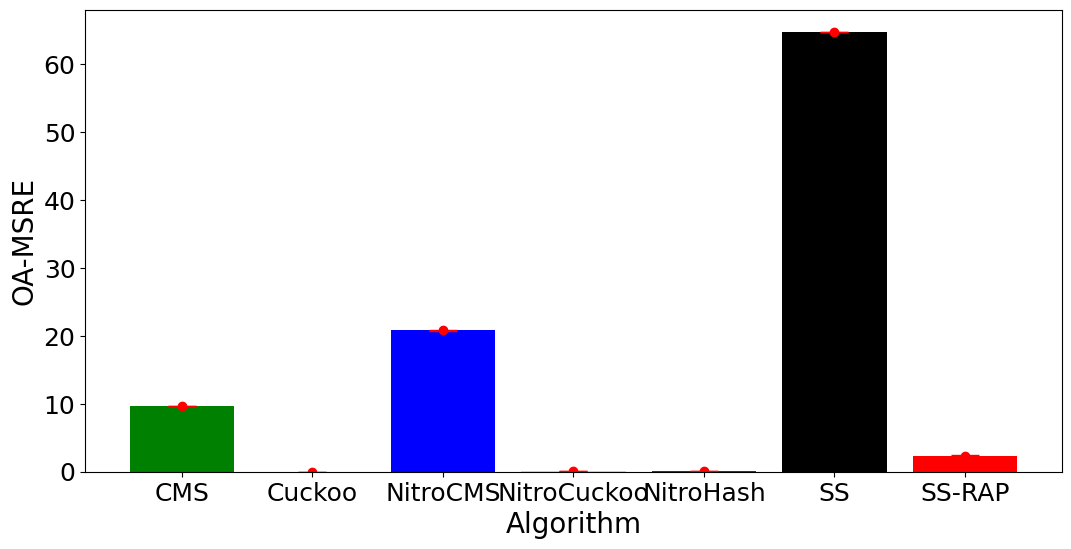}\label{fig:oa-msre:Chicago15}}\\
	\subfloat[][Chicago16Small]{\includegraphics[width=0.32\textwidth]{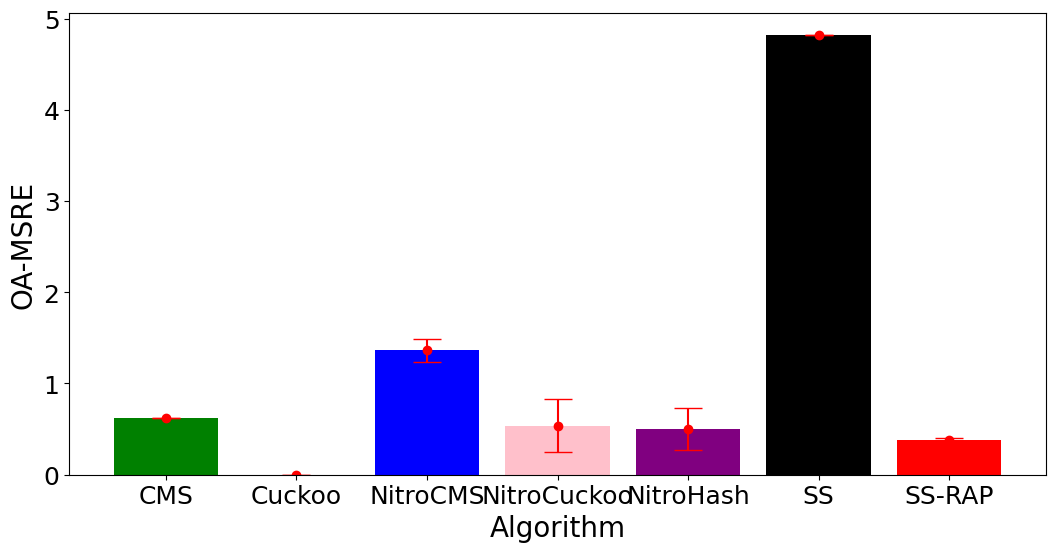}\label{fig:oa-msre:Chicago16Small}}
	\subfloat[][Chicago1610Mil]{\includegraphics[width=0.32\textwidth]{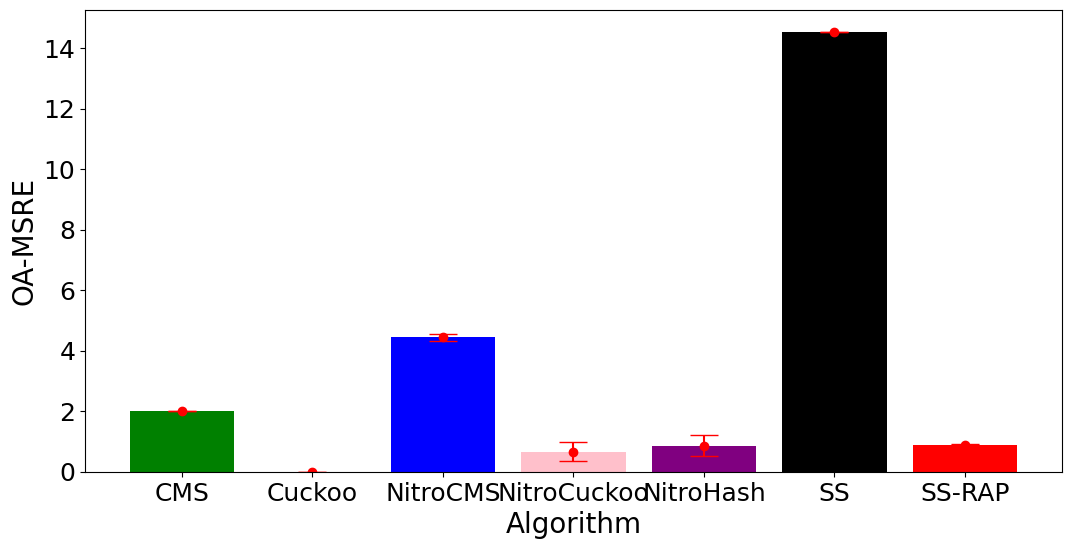}\label{fig:oa-msre:Chicago1610Mil}}
	\subfloat[][Chicago16]{\includegraphics[width=0.32\textwidth]{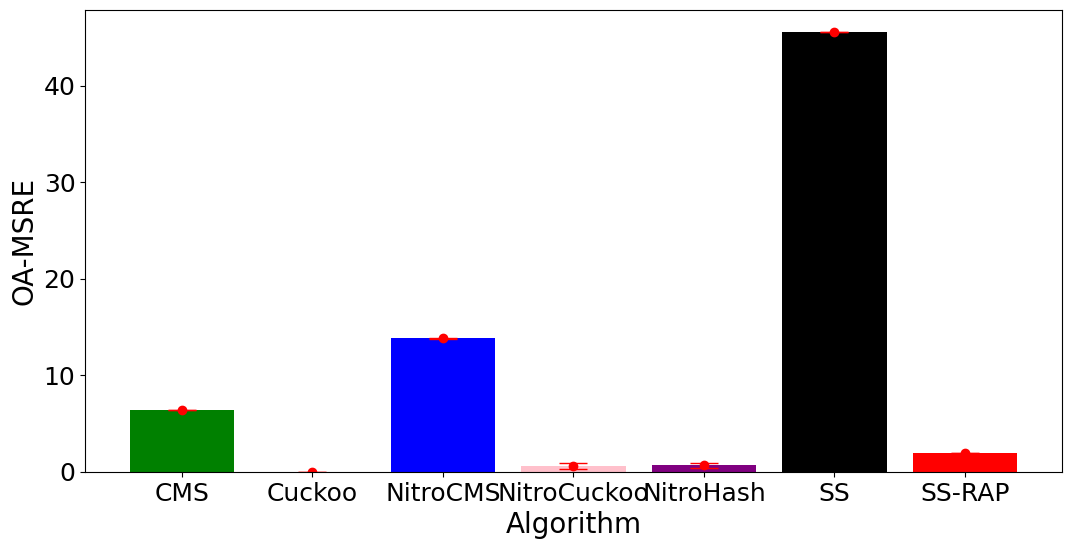}\label{fig:oa-msre:Chicago16}}
	\caption{OA Arrival MSRE}
	\label{fig:oa-msre}
%
	\subfloat[][ny19A]{\includegraphics[width=0.32\textwidth]{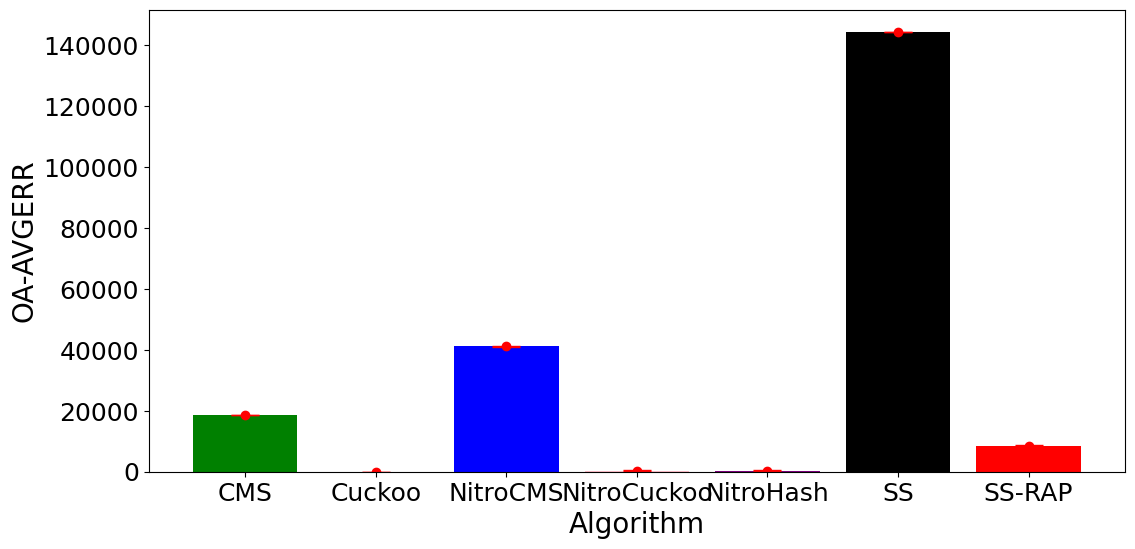}\label{fig:oa-avgerr:ny19A}}
	\subfloat[][SJ14]{\includegraphics[width=0.32\textwidth]{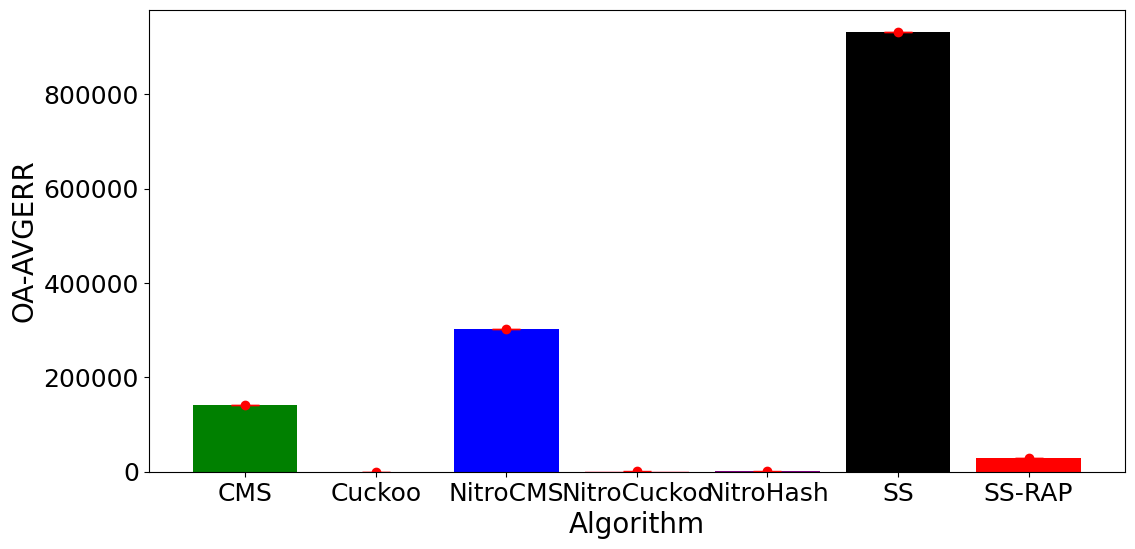}\label{fig:oa-avgerr:SJ14}}
	\subfloat[][Chicago15]{\includegraphics[width=0.32\textwidth]{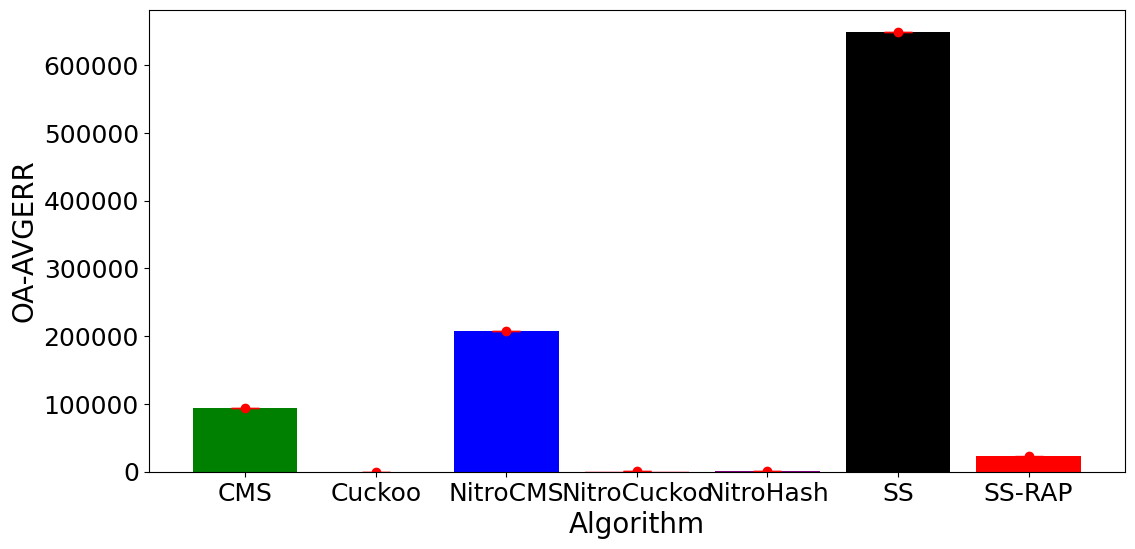}\label{fig:oa-avgerr:Chicago15}}\\
	\subfloat[][Chicago16Small]{\includegraphics[width=0.32\textwidth]{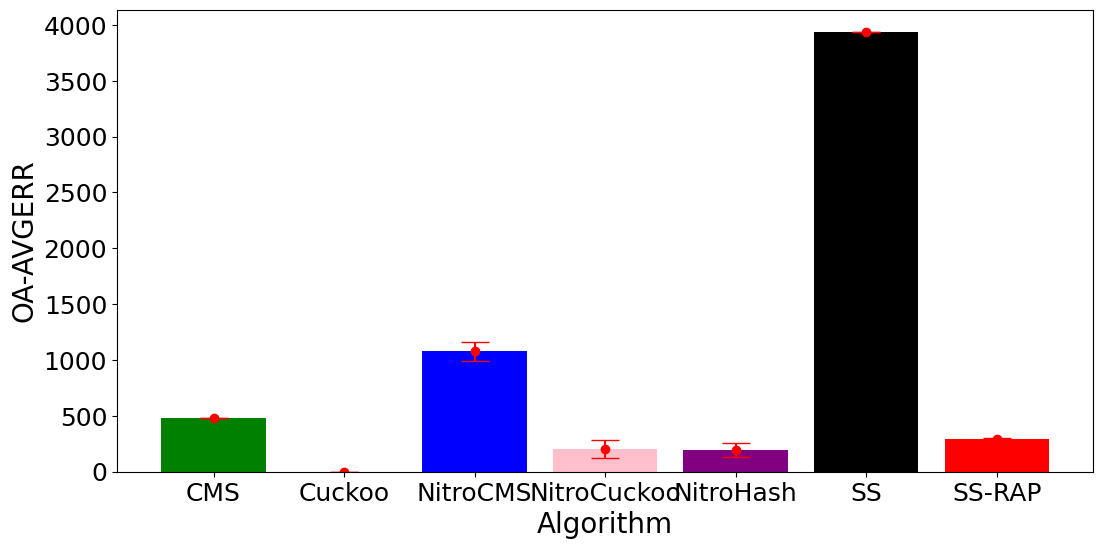}\label{fig:oa-avgerr:Chicago16Small}}
	\subfloat[][Chicago1610Mil]{\includegraphics[width=0.32\textwidth]{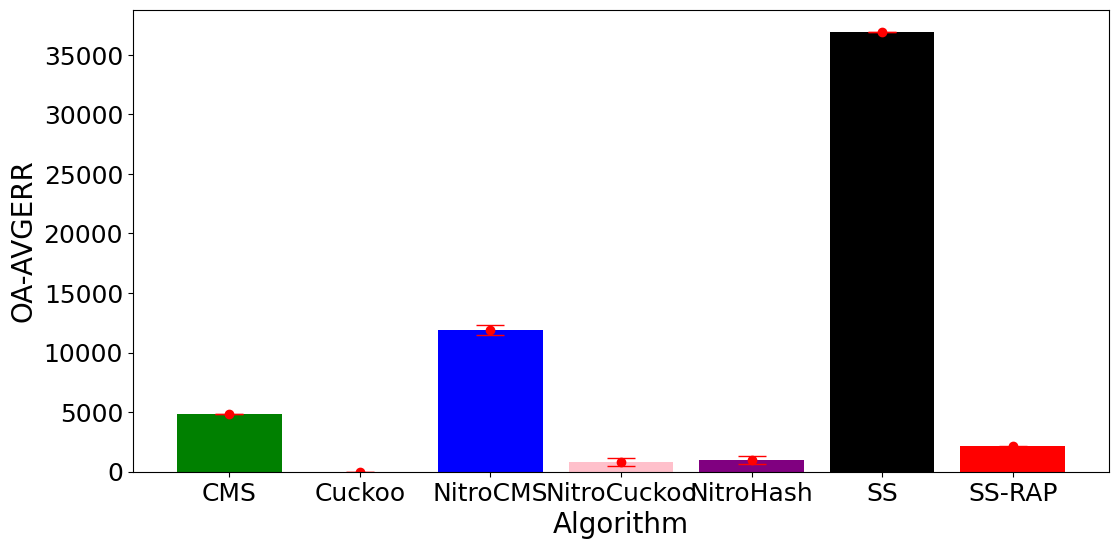}\label{fig:oa-avgerr:Chicago1610Mil}}
	\subfloat[][Chicago16]{\includegraphics[width=0.32\textwidth]{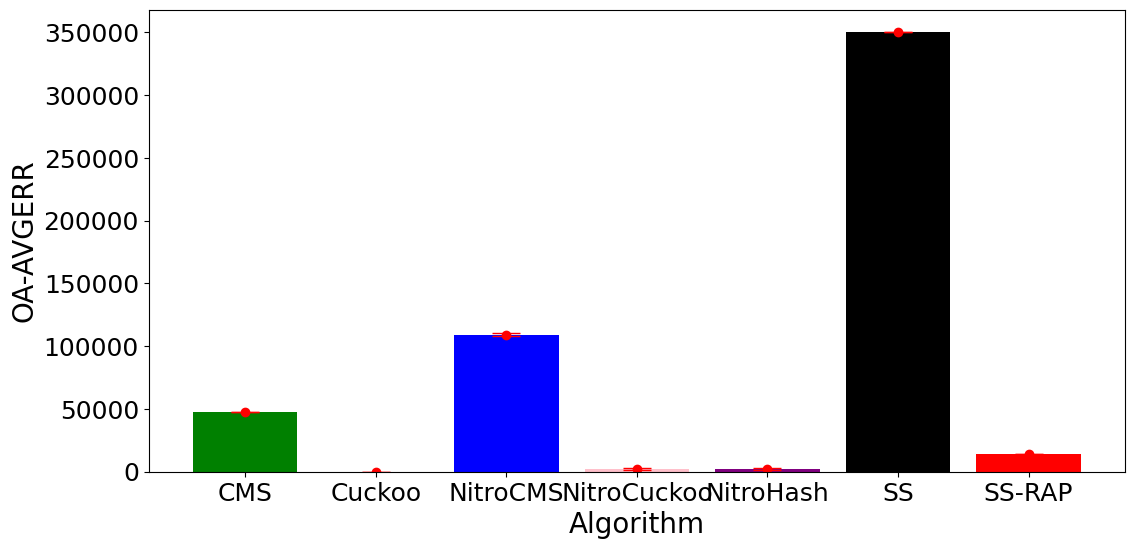}\label{fig:oa-avgerr:Chicago16}}
	\caption{OA Arrival AVGERR}
	\label{fig:oa-avgerr}
%
	\subfloat[][ny19A]{\includegraphics[width=0.32\textwidth]{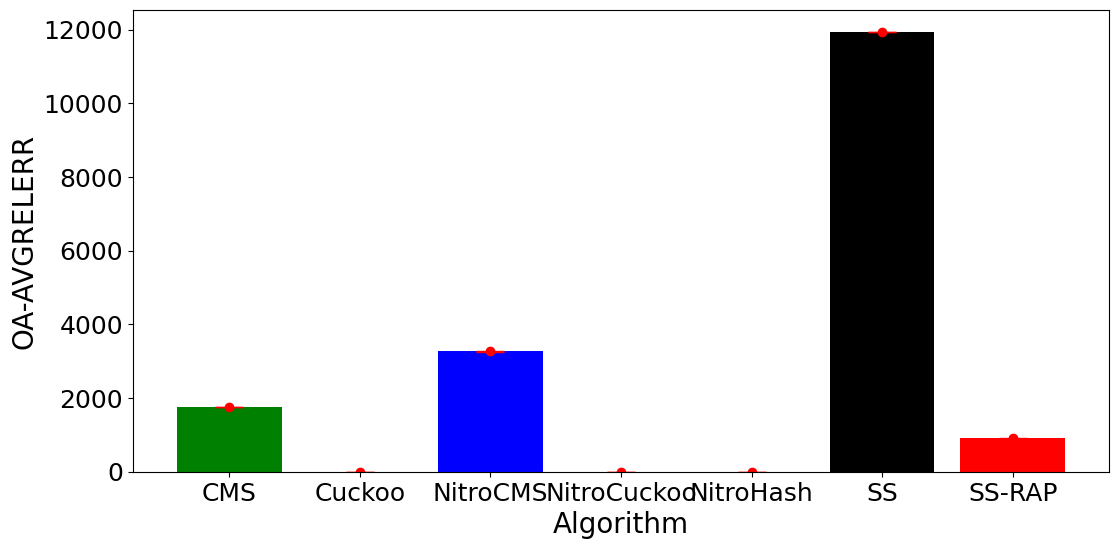}\label{fig:oa-avgrelerr:ny19A}}
	\subfloat[][SJ14]{\includegraphics[width=0.32\textwidth]{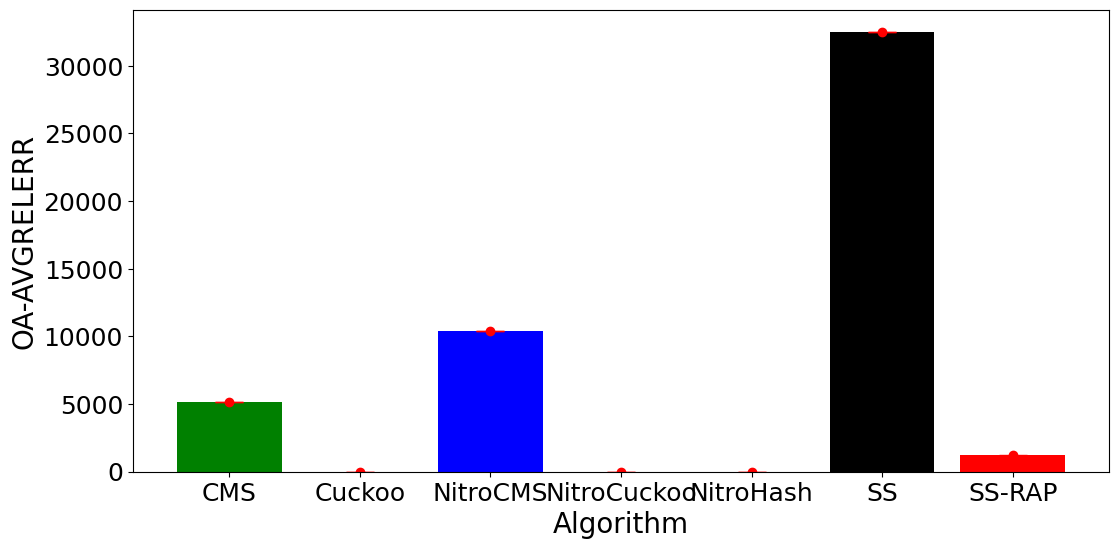}\label{fig:oa-avgrelerr:SJ14}}
	\subfloat[][Chicago15]{\includegraphics[width=0.32\textwidth]{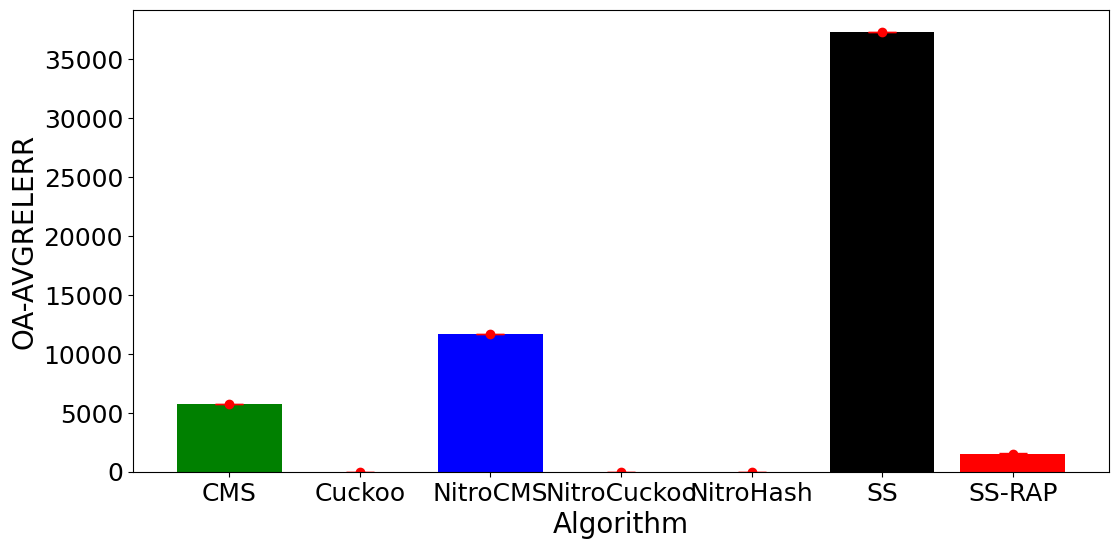}\label{fig:oa-avgrelerr:Chicago15}}\\
	\subfloat[][Chicago16Small]{\includegraphics[width=0.32\textwidth]{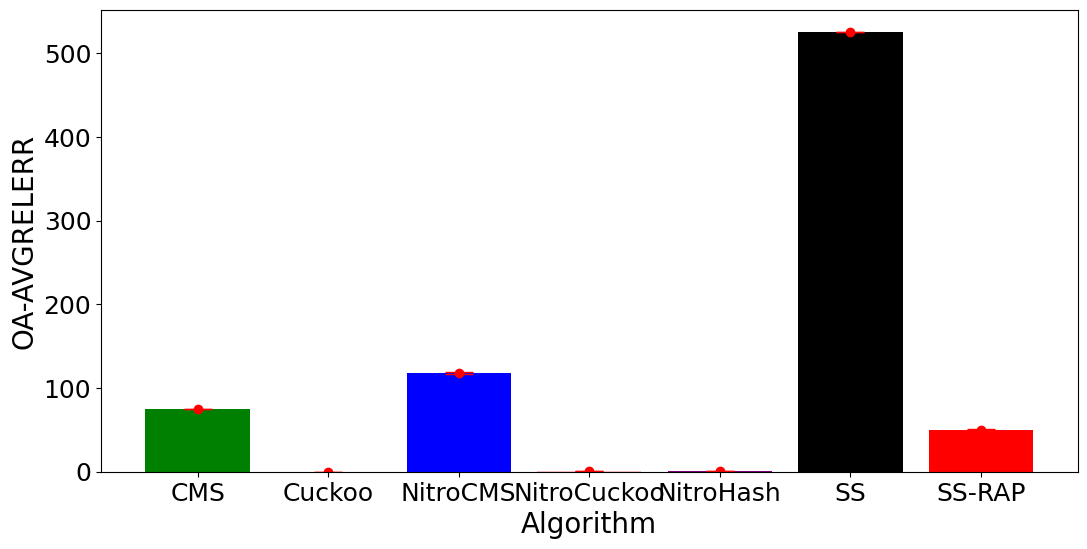}\label{fig:oa-avgrelerr:Chicago16Small}}
	\subfloat[][Chicago1610Mil]{\includegraphics[width=0.32\textwidth]{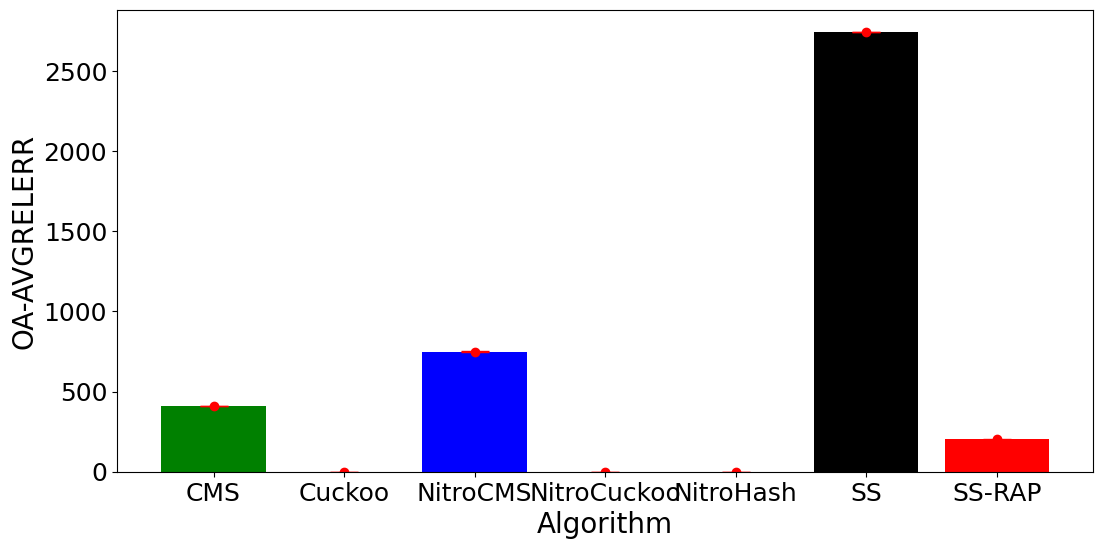}\label{fig:oa-avgrelerr:Chicago1610Mil}}
	\subfloat[][Chicago16]{\includegraphics[width=0.32\textwidth]{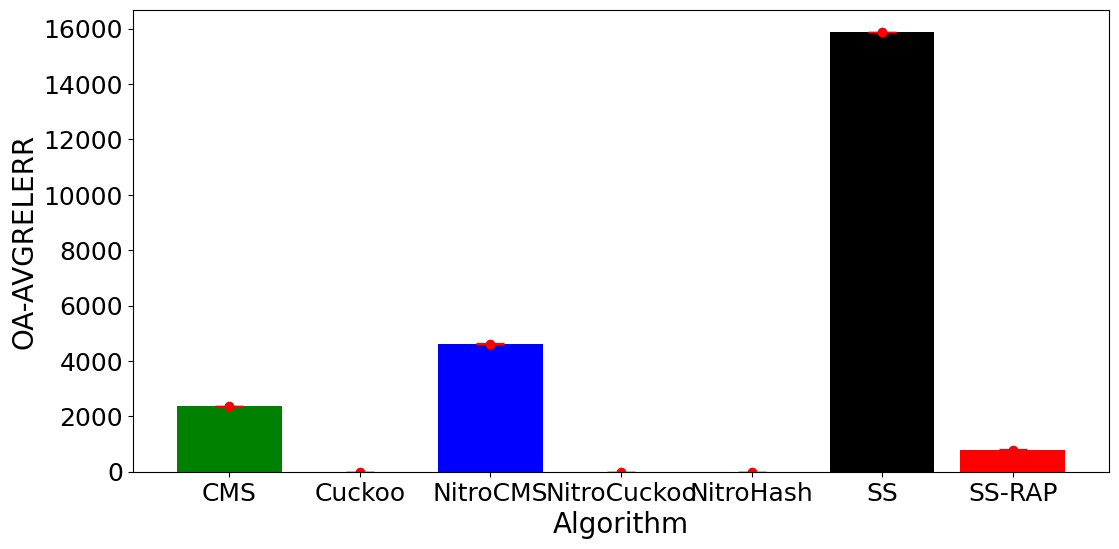}\label{fig:oa-avgrelerr:Chicago16}}
	\caption{OA Arrival AVGRELERR}
	\label{fig:oa-avgrelerr}
\end{figure*}

The results for the OA MSRE, OA AVGERR, and OA AVGRELERR metrics appear in Figures~\ref{fig:oa-msre},~\ref{fig:oa-avgerr}, and~\ref{fig:oa-avgrelerr}, respectively.
As can be seen, Cuckoo has a negligible error, regardless of the exact metric, compared to the other methods.
We notice small but non-negligible error for these traces with the NitroHash and NitroCuckoo metrics.
This indicates that a few heavy-hitters has non-negligible errors, as they were no properly represented by the sampling method.
Yet, the errors become very small as we are moving from Chicago16Small to Chicago1610Mil and then to the full Chicago16.
This can be expected since as the trace becomes longer, the probabilistic process becomes more representative.
Also, echoing the results of~\cite{RAP}, the error of SpaceSaving-RAP is much lower than the basic SpaceSaving, regardless of the metric, due to its more effective usage of the counters as explained in Section~\ref{sec:background}.
Interestingly, it also presents a significantly lower error than CMS and NitroCMS, whereas NitroCMS almost doubles the error compared to CMS.
We explore the impact of the minimal increment optimization in Section~\ref{sec:cms-deepdive} below.

We notice small but non-negligible errors for the Chicago16, Chicago1610Mil, and Chicago16Small traces with the NitroHash and NitroCuckoo algorithms and the MSRE and AVGERR metrics.
This indicates that a few heavy-hitters have non-negligible errors, as they were not properly represented by the sampling method.
Even here, the errors become very small as we are moving from Chicago16Small to Chicago1610Mil and then to the full Chicago16 trace.
This can be expected since as the trace becomes longer, the probabilistic process becomes more representative.


\begin{figure*}[t]
	\centering
	\subfloat[][ny19A]{\includegraphics[width=0.32\textwidth]{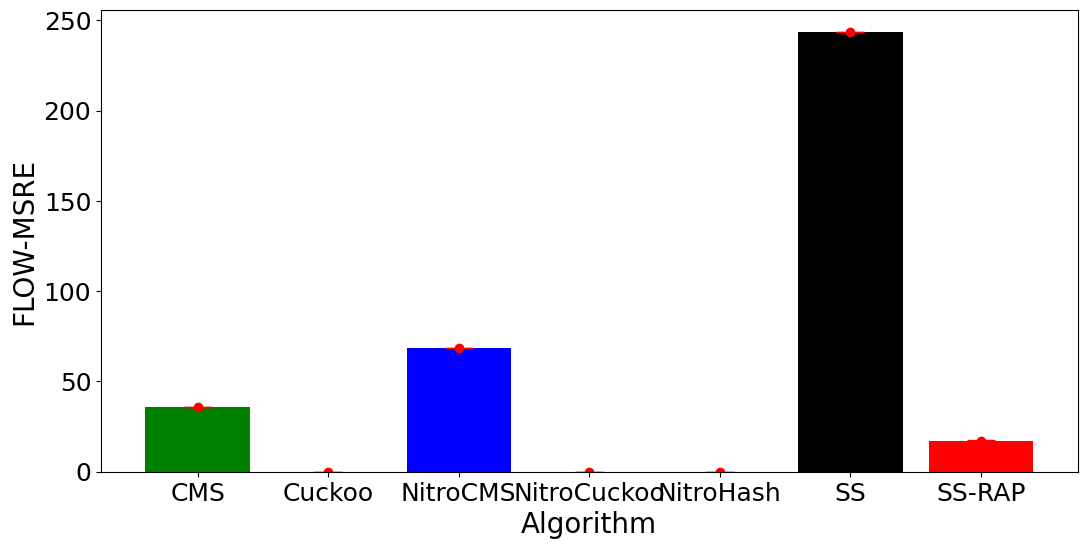}\label{fig:flow-msre:ny19A}}
	\subfloat[][SJ14]{\includegraphics[width=0.32\textwidth]{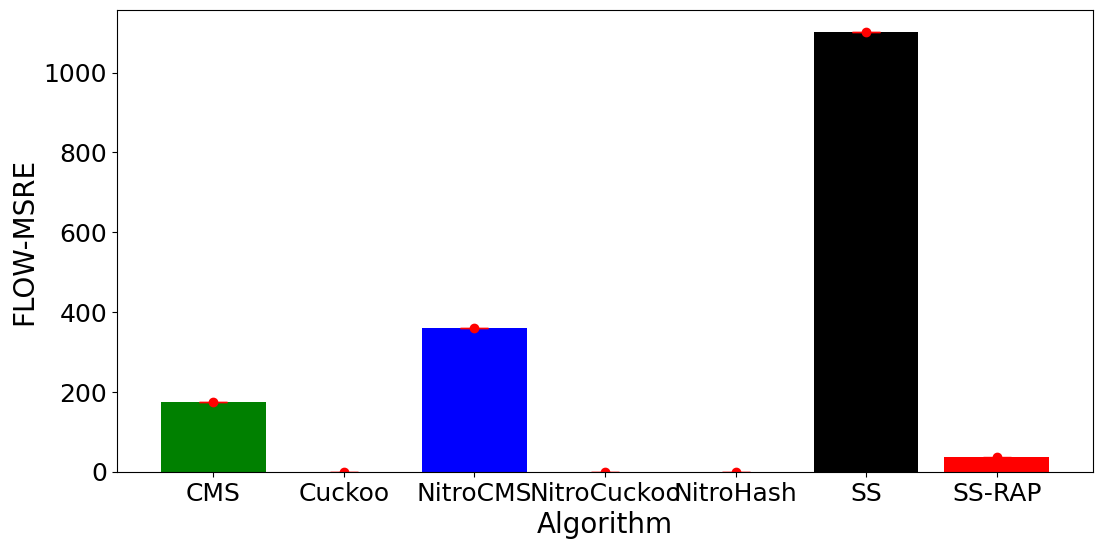}\label{fig:flow-msre:SJ14}}
	\subfloat[][Chicago15]{\includegraphics[width=0.32\textwidth]{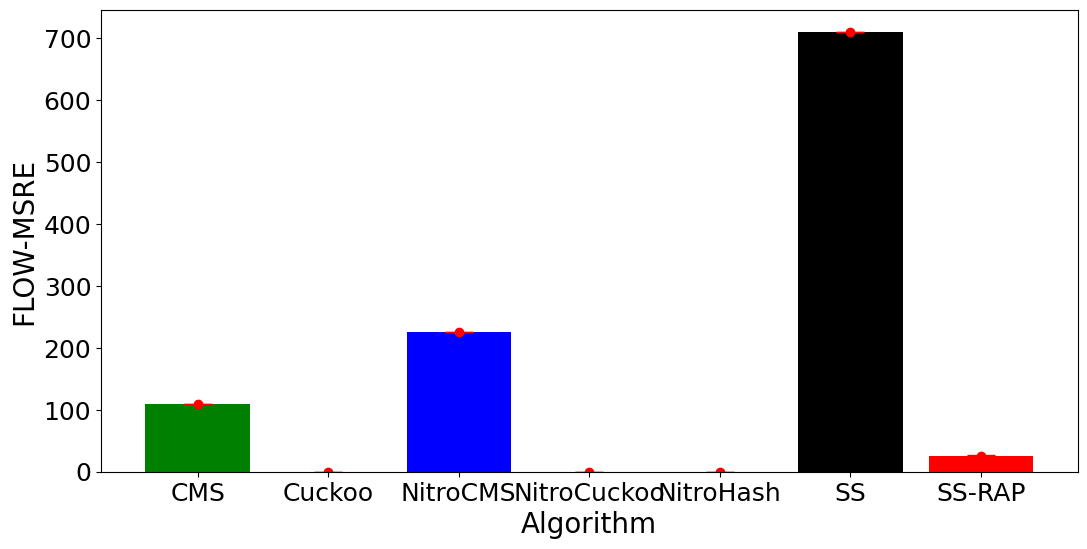}\label{fig:flow-msre:Chicago15}}\\
	\subfloat[][Chicago16Small]{\includegraphics[width=0.32\textwidth]{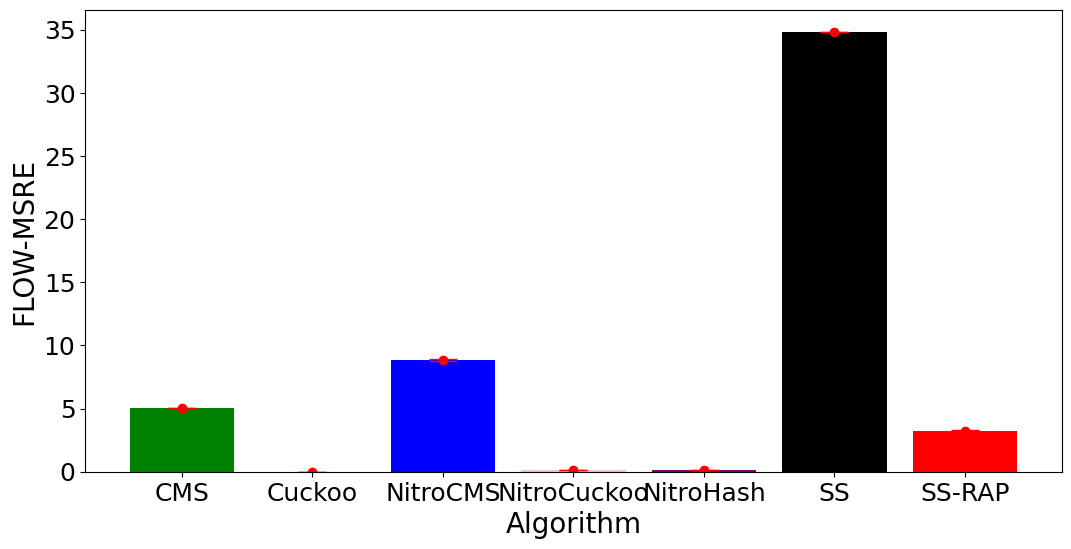}\label{fig:flow-msre:Chicago16Small}}
	\subfloat[][Chicago1610Mil]{\includegraphics[width=0.32\textwidth]{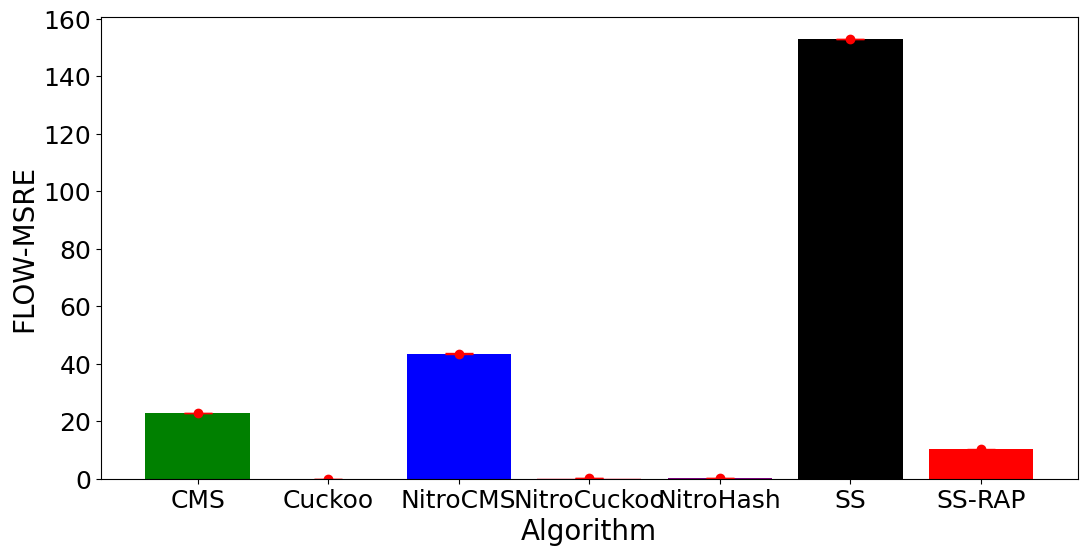}\label{fig:flow-msre:Chicago1610Mil}}
	\subfloat[][Chicago16]{\includegraphics[width=0.32\textwidth]{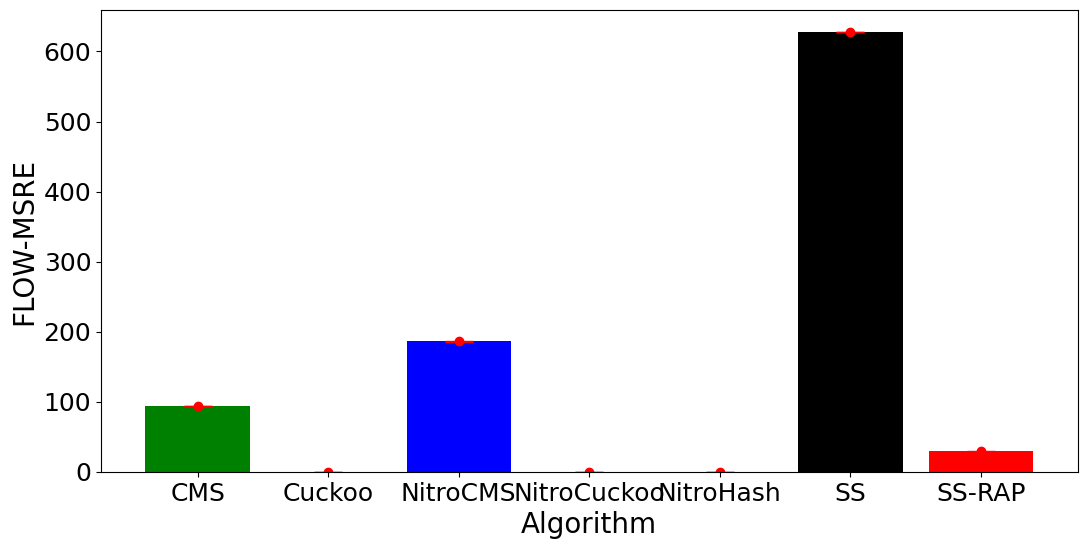}\label{fig:flow-msre:Chicago16}}
	\caption{Per Flow MSRE}
	\label{fig:flow-msre}
	%
	\subfloat[][ny19A]{\includegraphics[width=0.32\textwidth]{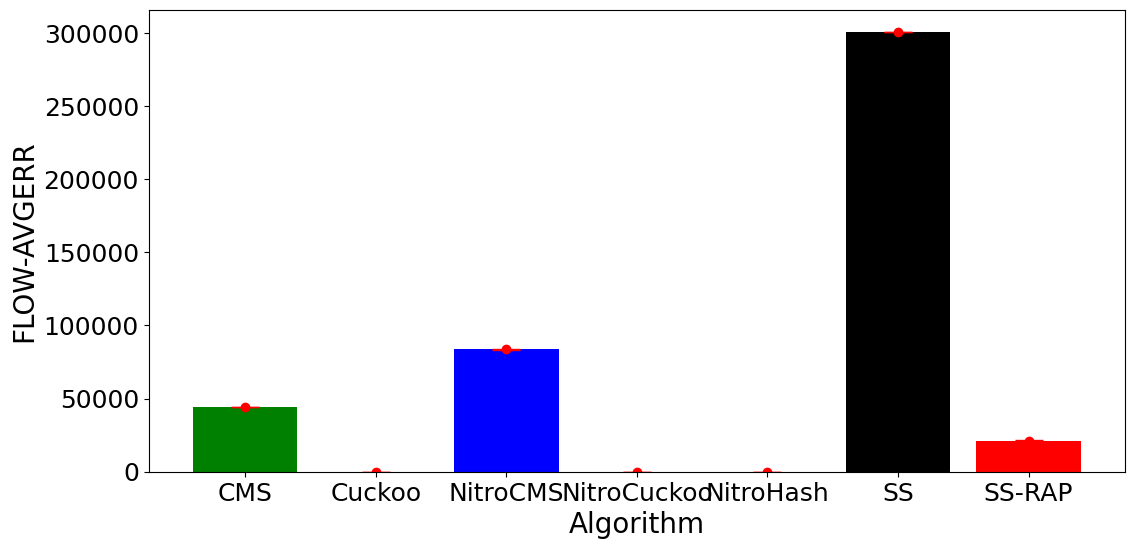}\label{fig:flow-avgerr:ny19A}}
	\subfloat[][SJ14]{\includegraphics[width=0.32\textwidth]{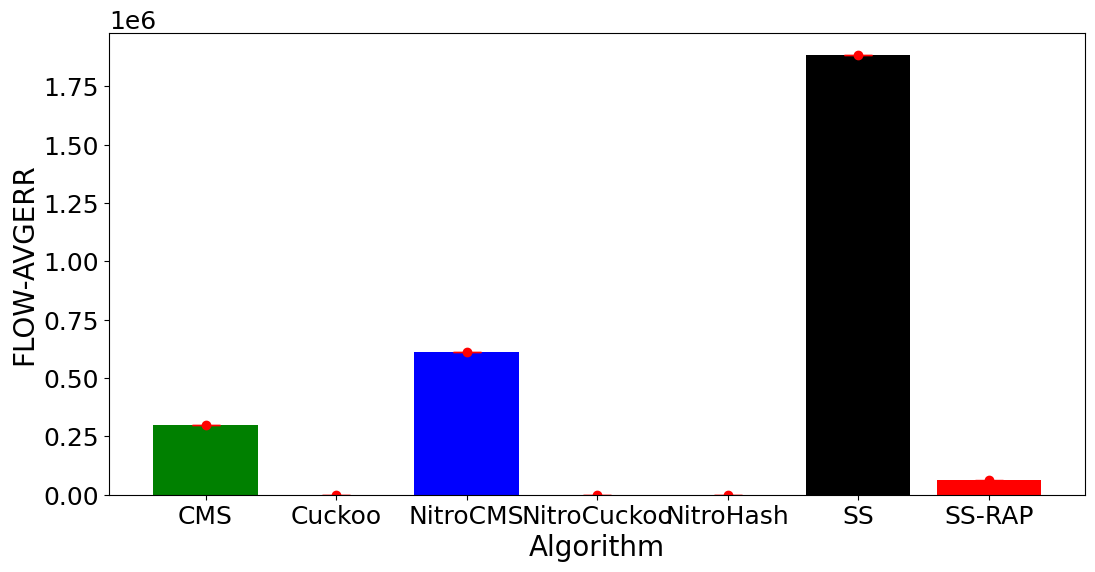}\label{fig:flow-avgerr:SJ14}}
	\subfloat[][Chicago15]{\includegraphics[width=0.32\textwidth]{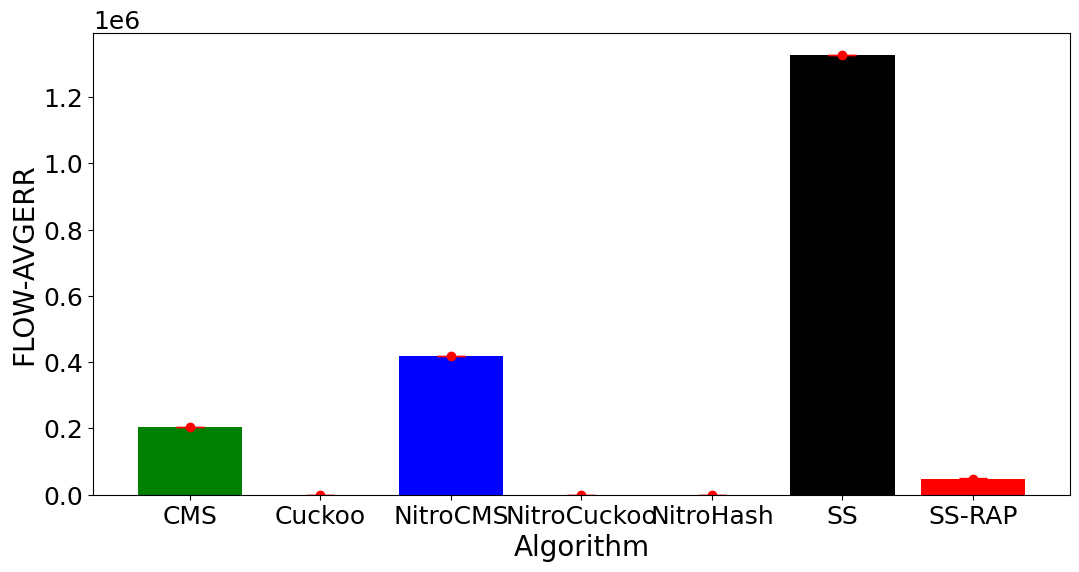}\label{fig:flow-avgerr:Chicago15}}\\
	\subfloat[][Chicago16Small]{\includegraphics[width=0.32\textwidth]{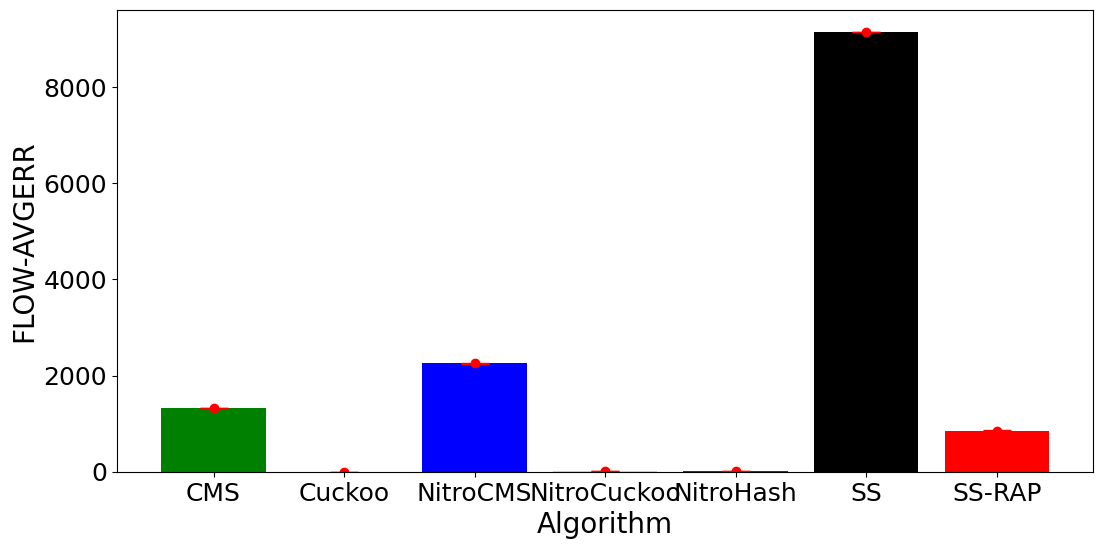}\label{fig:flow-avgerr:Chicago16Small}}
	\subfloat[][Chicago1610Mil]{\includegraphics[width=0.32\textwidth]{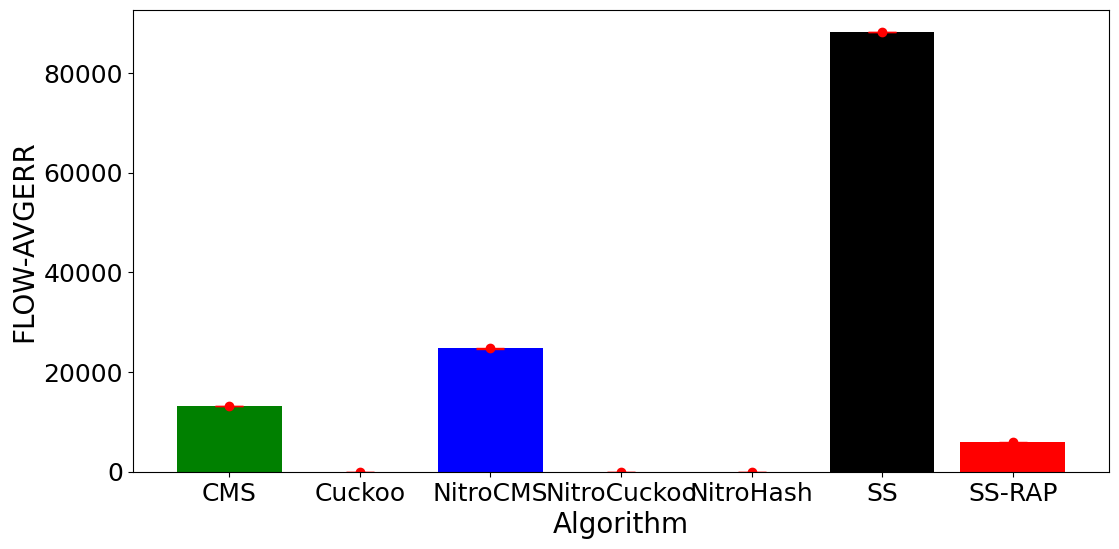}\label{fig:flow-avgerr:Chicago1610Mil}}
	\subfloat[][Chicago16]{\includegraphics[width=0.32\textwidth]{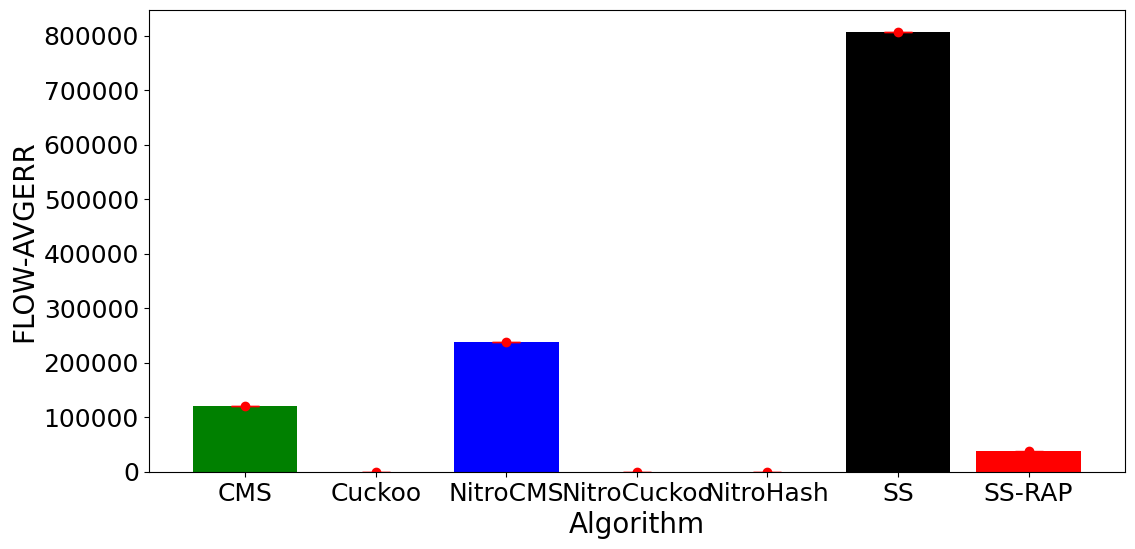}\label{fig:flow-avgerr:Chicago16}}
	\caption{Per FLOW AVGERR}
	\label{fig:flow-avgerr}
	%
	\subfloat[][ny19A]{\includegraphics[width=0.32\textwidth]{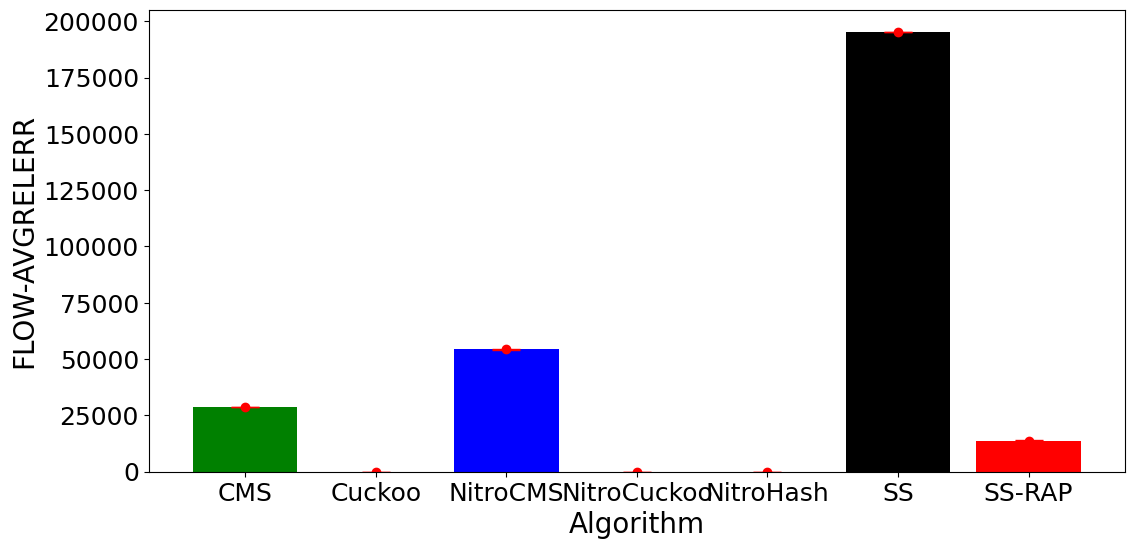}\label{fig:flow-avgrelerr:ny19A}}
	\subfloat[][SJ14]{\includegraphics[width=0.32\textwidth]{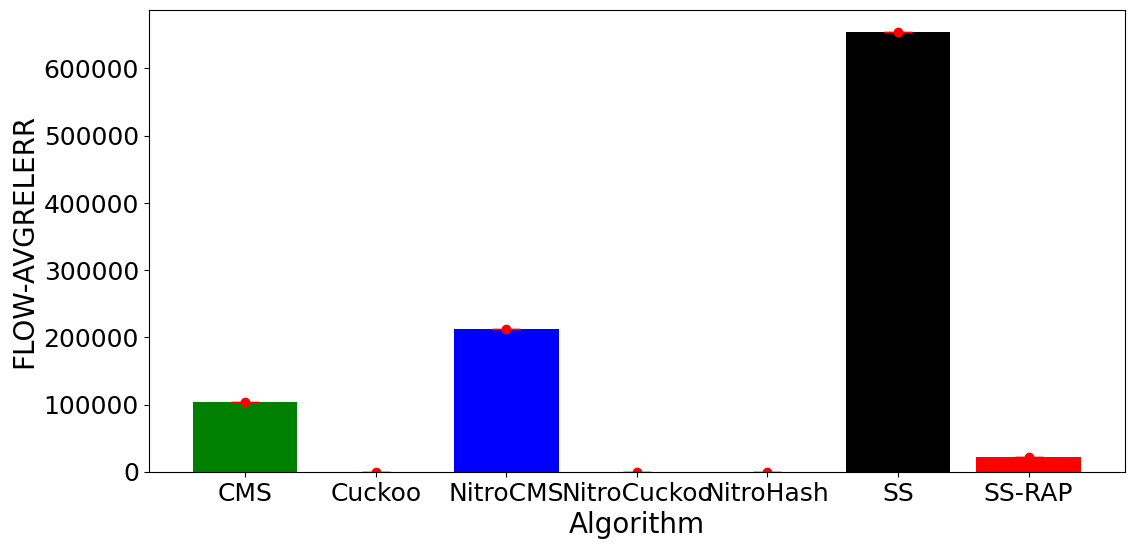}\label{fig:flow-avgrelerr:SJ14}}
	\subfloat[][Chicago15]{\includegraphics[width=0.32\textwidth]{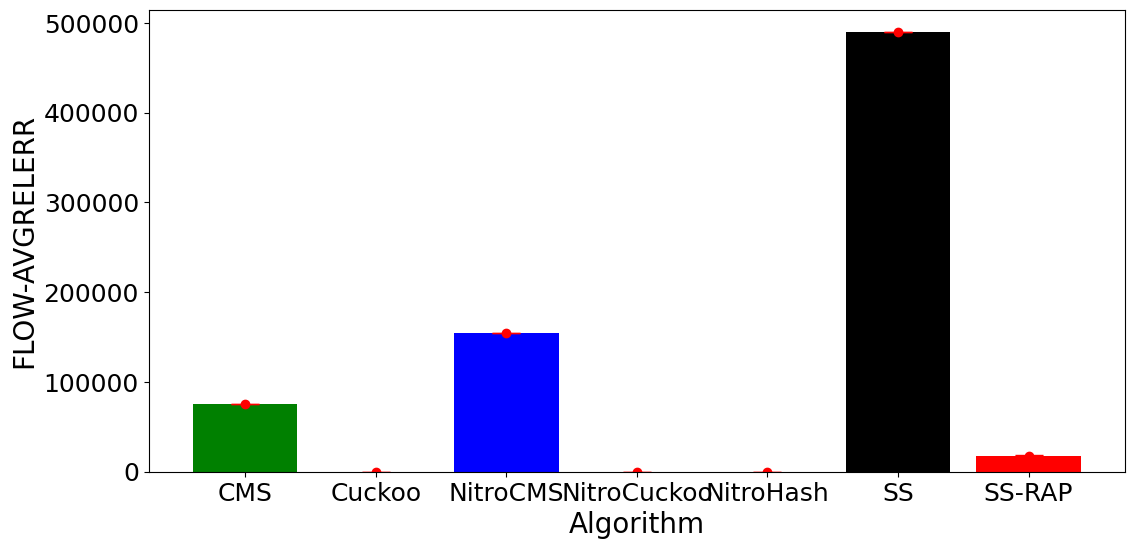}\label{fig:flow-avgrelerr:Chicago15}}\\
	\subfloat[][Chicago16Small]{\includegraphics[width=0.32\textwidth]{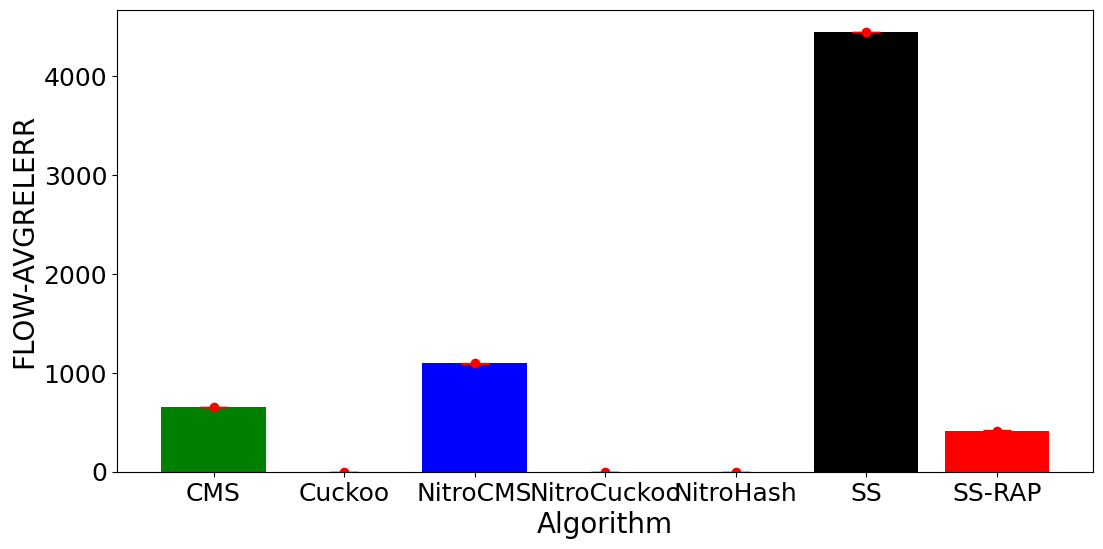}\label{fig:flow-avgrelerr:Chicago16Small}}
	\subfloat[][Chicago1610Mil]{\includegraphics[width=0.32\textwidth]{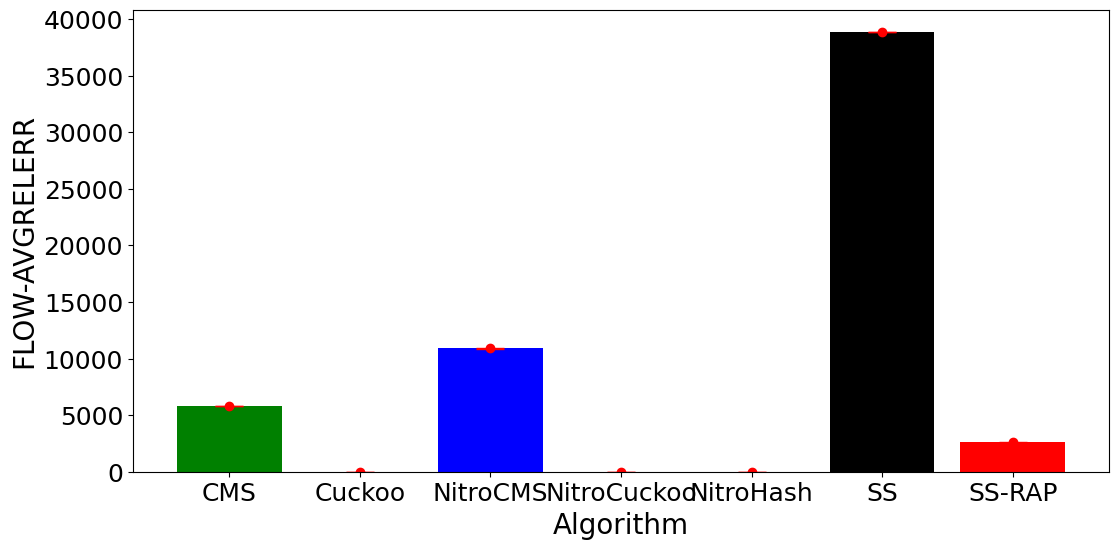}\label{fig:flow-avgrelerr:Chicago1610Mil}}
	\subfloat[][Chicago16]{\includegraphics[width=0.32\textwidth]{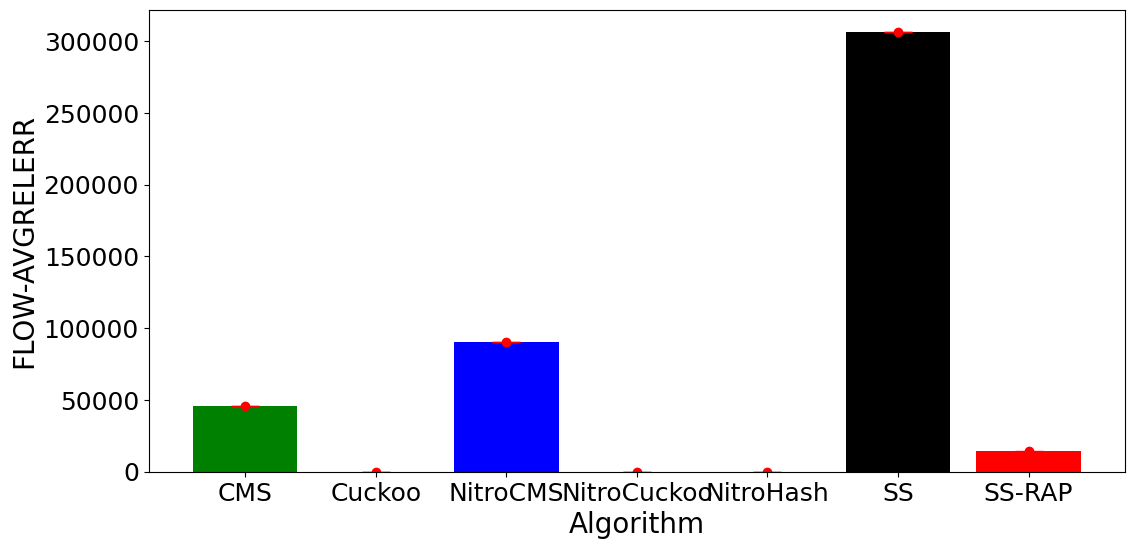}\label{fig:flow-avgrelerr:Chicago16}}
	\caption{Per FLOW AVGRELERR}
	\label{fig:flow-avgrelerr}
\end{figure*}

The Per Flow MSRE, Per Flow AVGERR, and Per Flow AVGRELERR results appear in Figures~\ref{fig:flow-msre},~\ref{fig:flow-avgerr}, and~\ref{fig:flow-avgrelerr}, respectively.
The results are qualitatively similar to the on-arrival error metrics.

\begin{figure*}[t]
	\centering
	\subfloat[][ny19A]{\includegraphics[width=0.32\textwidth]{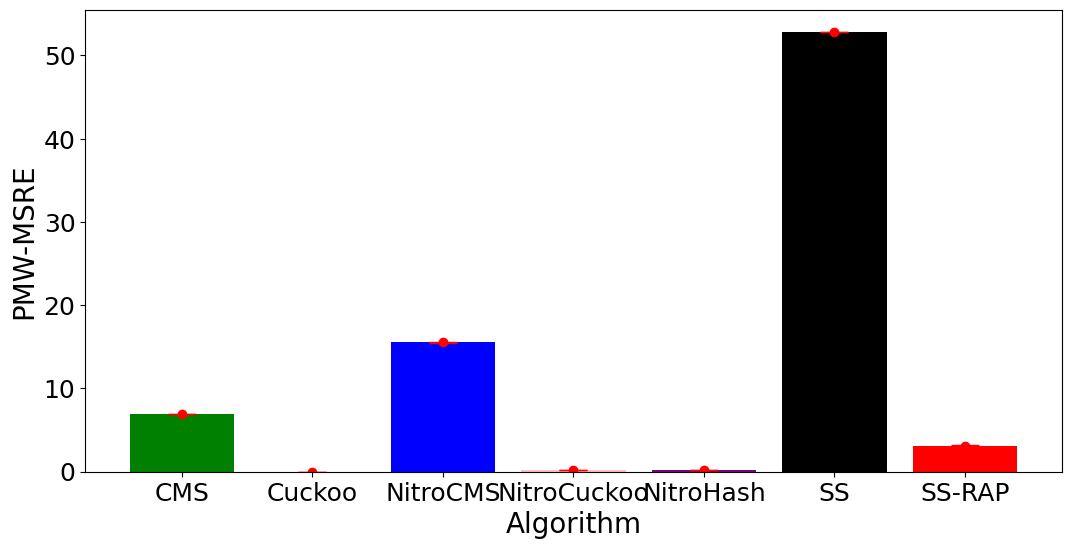}\label{fig:pm-msre:ny19A}}
	\subfloat[][SJ14]{\includegraphics[width=0.32\textwidth]{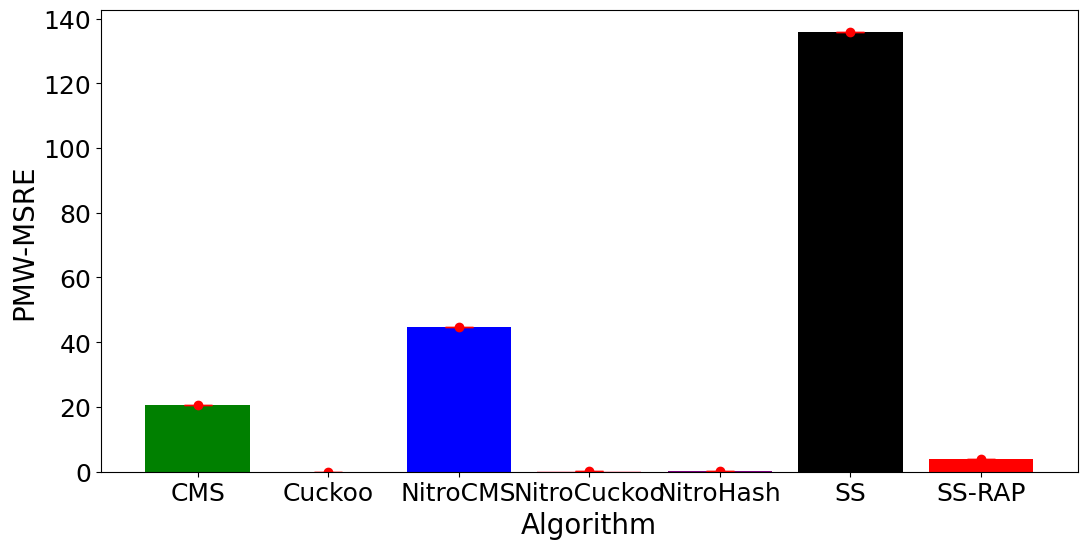}\label{fig:pm-msre:SJ14}}
	\subfloat[][Chicago15]{\includegraphics[width=0.32\textwidth]{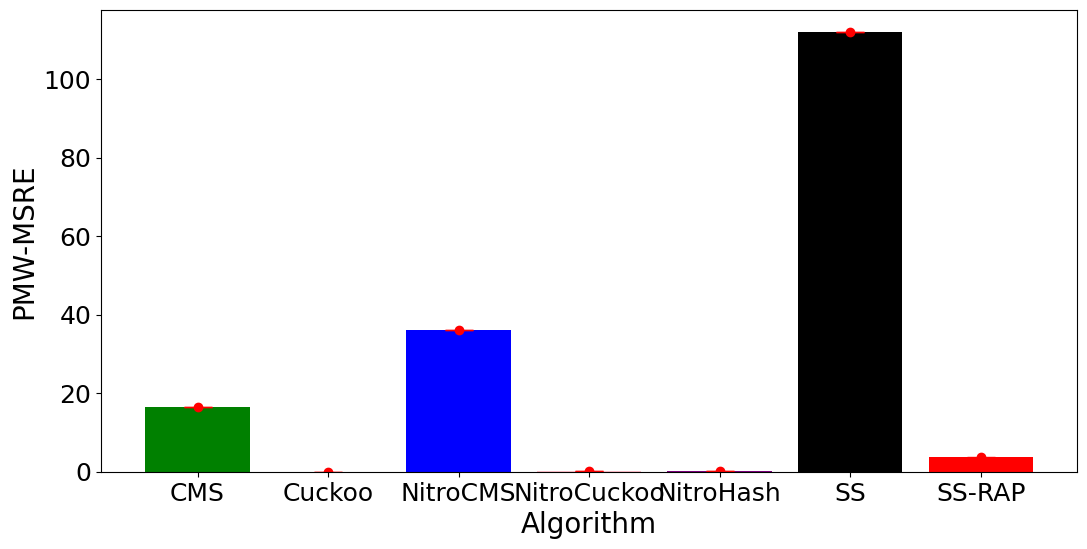}\label{fig:pm-msre:Chicago15}}\\
	\subfloat[][Chicago16Small]{\includegraphics[width=0.32\textwidth]{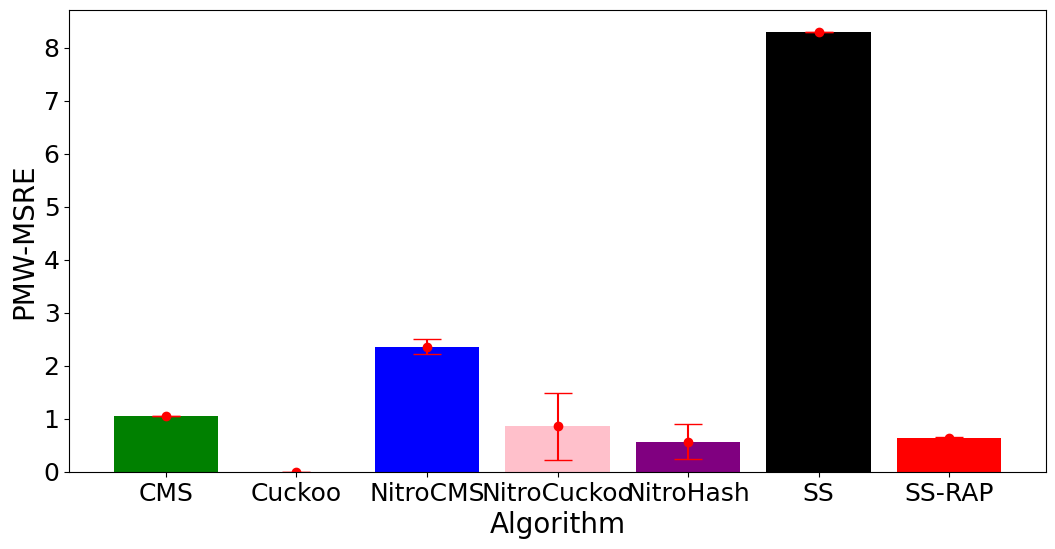}\label{fig:pm-msre:Chicago16Small}}
	\subfloat[][Chicago1610Mil]{\includegraphics[width=0.32\textwidth]{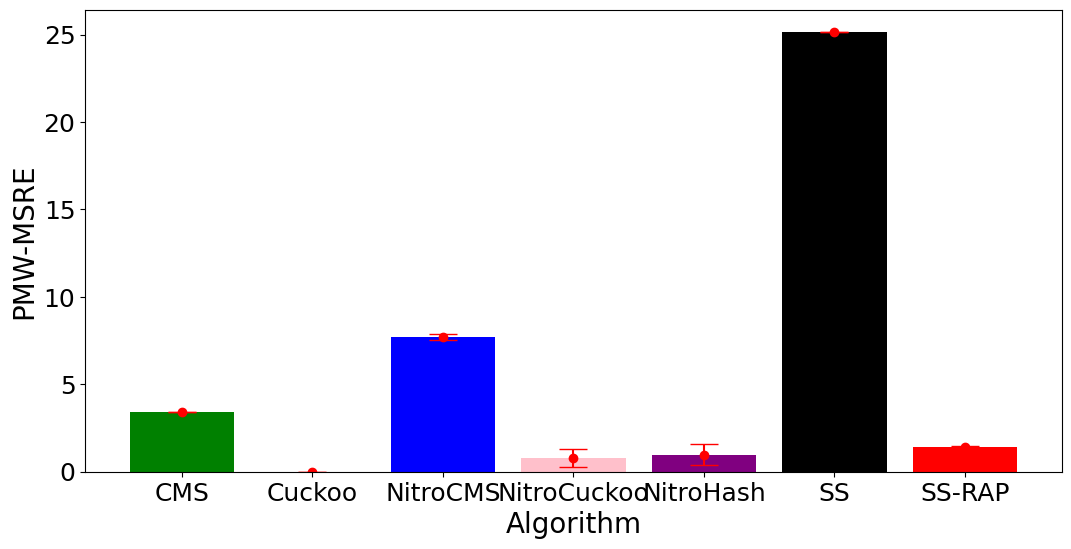}\label{fig:pm-msre:Chicago1610Mil}}
	\subfloat[][Chicago16]{\includegraphics[width=0.32\textwidth]{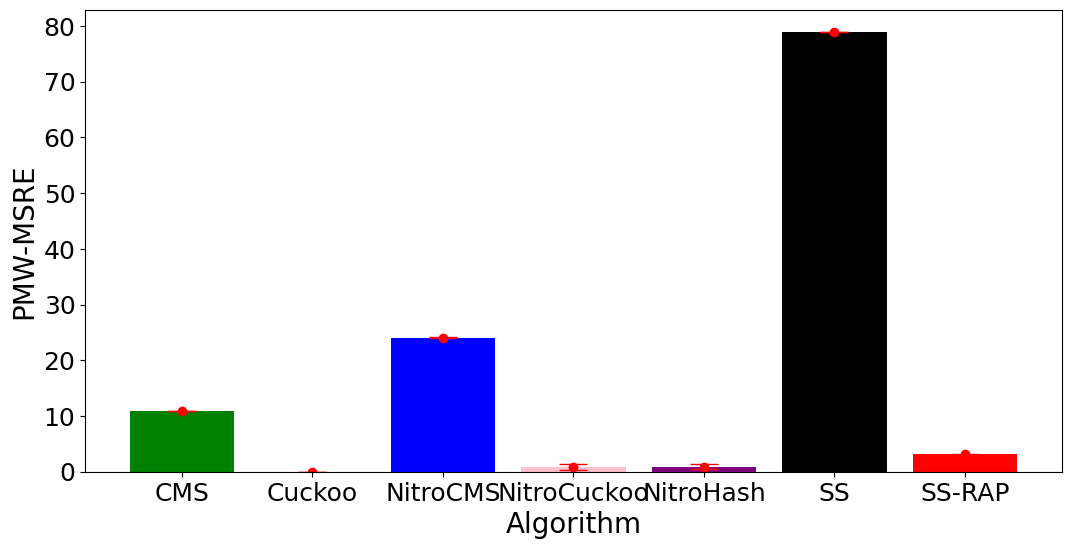}\label{fig:pm-msre:Chicago16}}
	\caption{Postmortem MSRE}
	\label{fig:pm-msre}
	%
	\subfloat[][ny19A]{\includegraphics[width=0.32\textwidth]{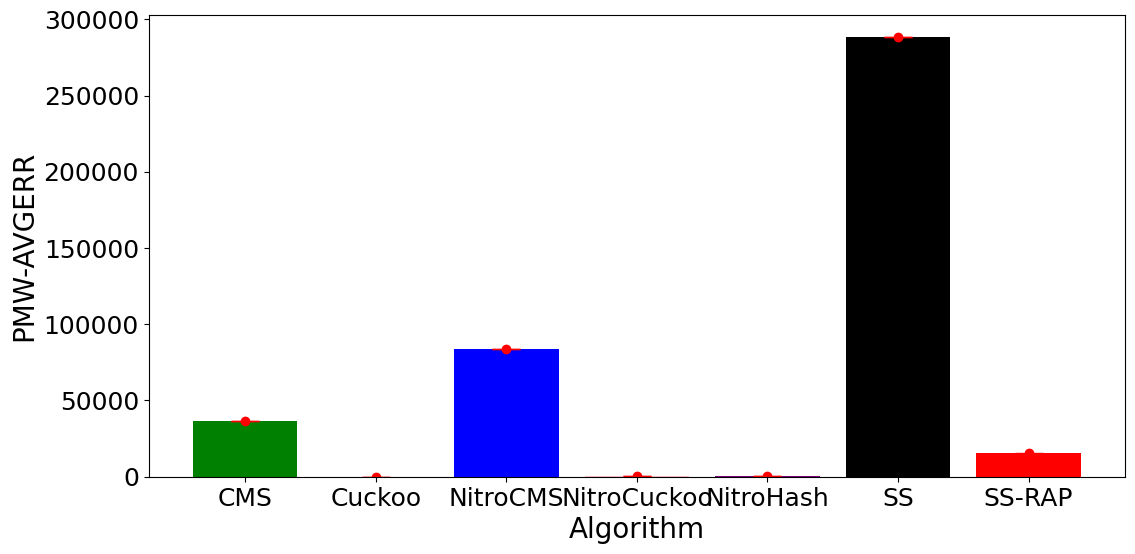}\label{fig:pm-avgerr:ny19A}}
	\subfloat[][SJ14]{\includegraphics[width=0.32\textwidth]{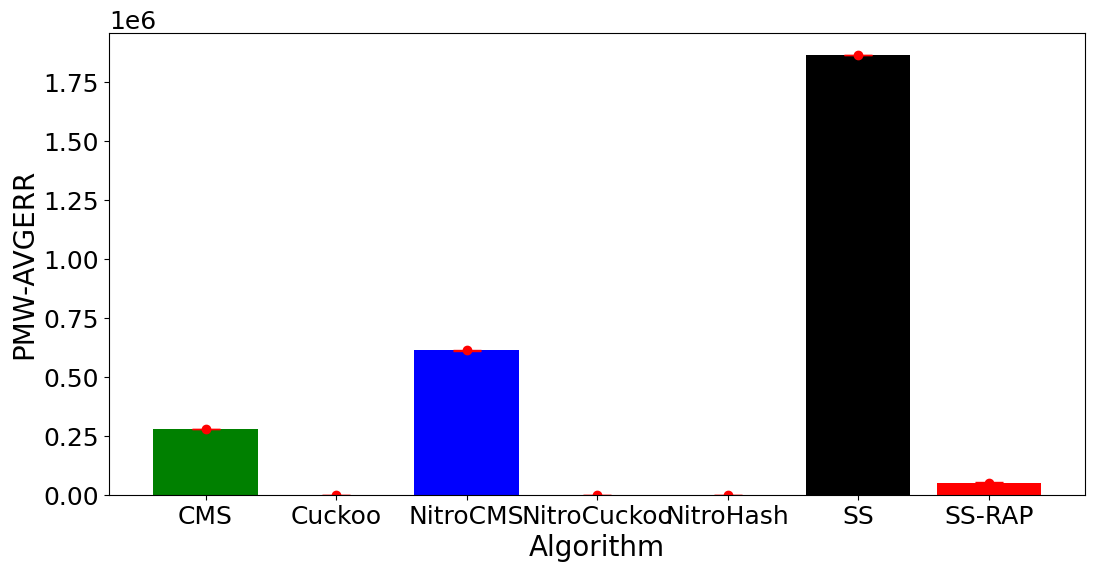}\label{fig:pm-avgerr:SJ14}}
	\subfloat[][Chicago15]{\includegraphics[width=0.32\textwidth]{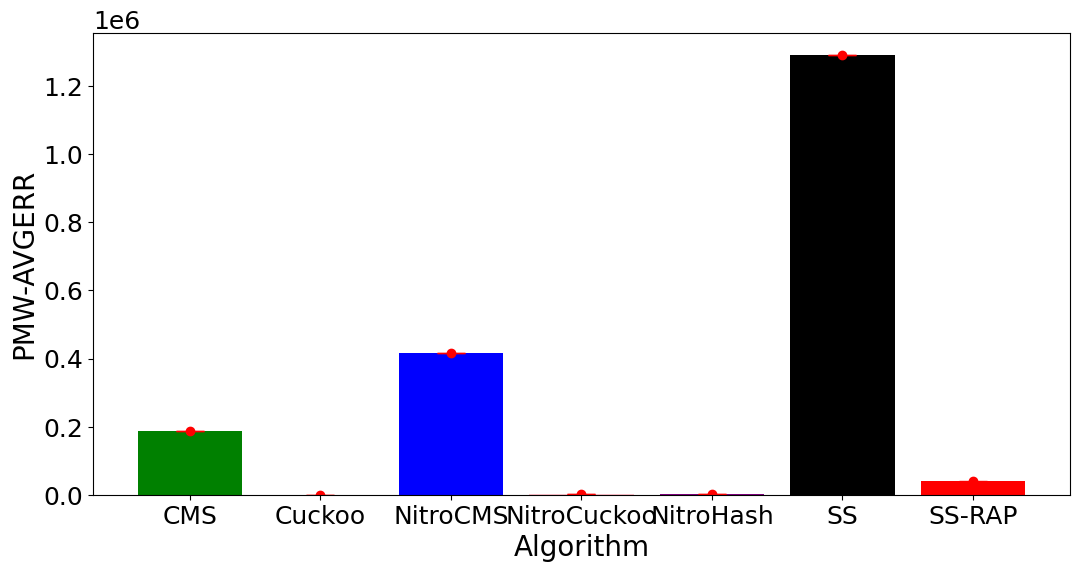}\label{fig:pm-avgerr:Chicago15}}\\
	\subfloat[][Chicago16Small]{\includegraphics[width=0.32\textwidth]{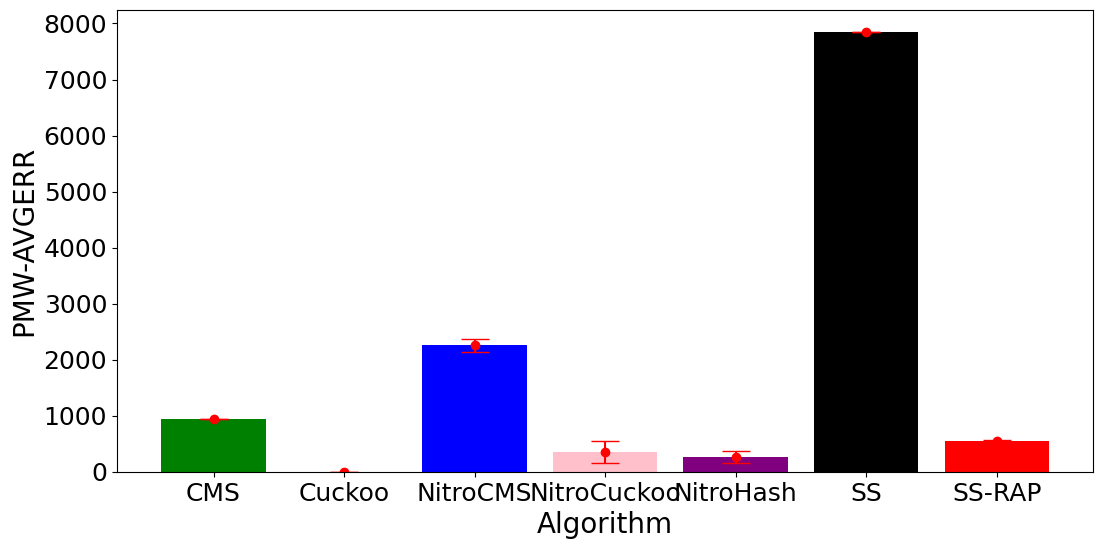}\label{fig:pm-avgerr:Chicago16Small}}
	\subfloat[][Chicago1610Mil]{\includegraphics[width=0.32\textwidth]{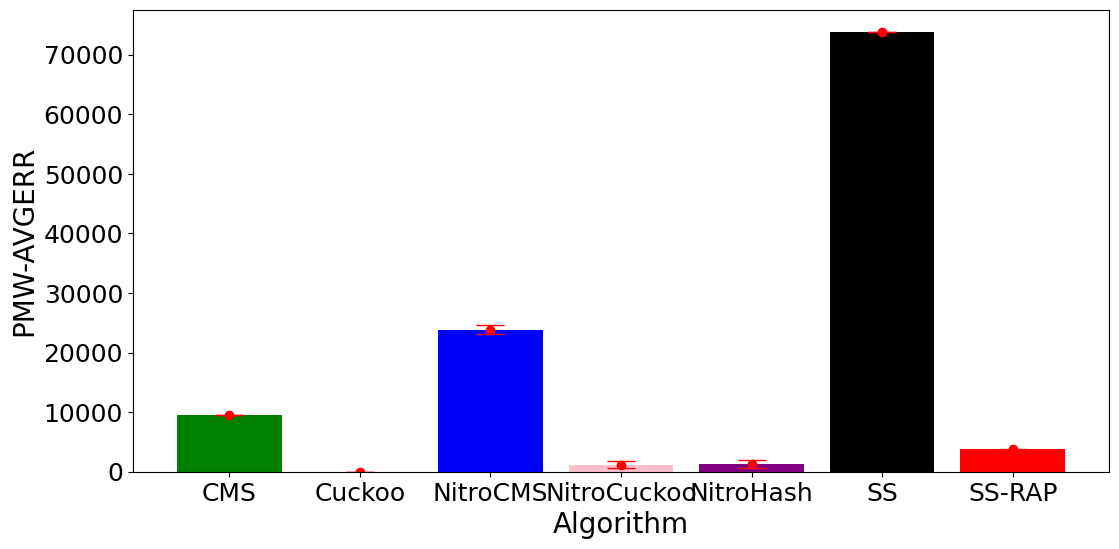}\label{fig:pm-avgerr:Chicago1610Mil}}
	\subfloat[][Chicago16]{\includegraphics[width=0.32\textwidth]{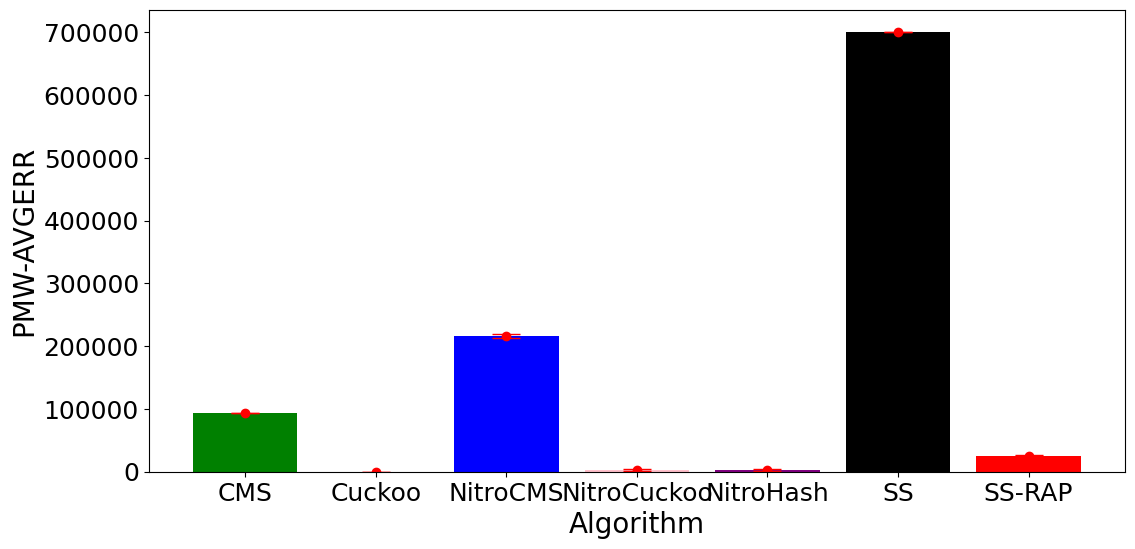}\label{fig:pm-avgerr:Chicago16}}
	\caption{Postmortem AVGERR}
	\label{fig:pm-avgerr}
	%
	\subfloat[][ny19A]{\includegraphics[width=0.32\textwidth]{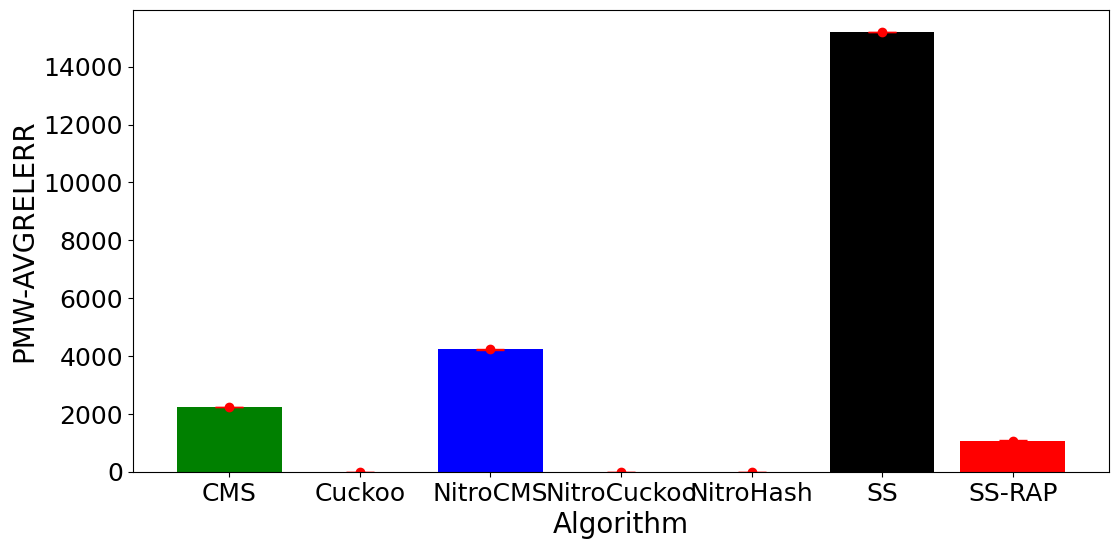}\label{fig:pm-avgrelerr:ny19A}}
	\subfloat[][SJ14]{\includegraphics[width=0.32\textwidth]{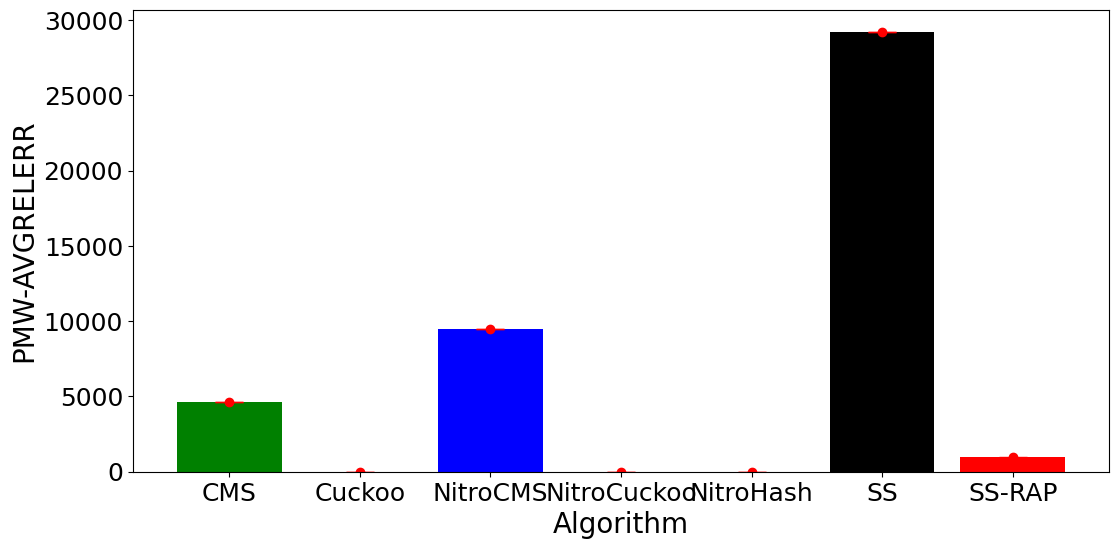}\label{fig:pm-avgrelerr:SJ14}}
	\subfloat[][Chicago15]{\includegraphics[width=0.32\textwidth]{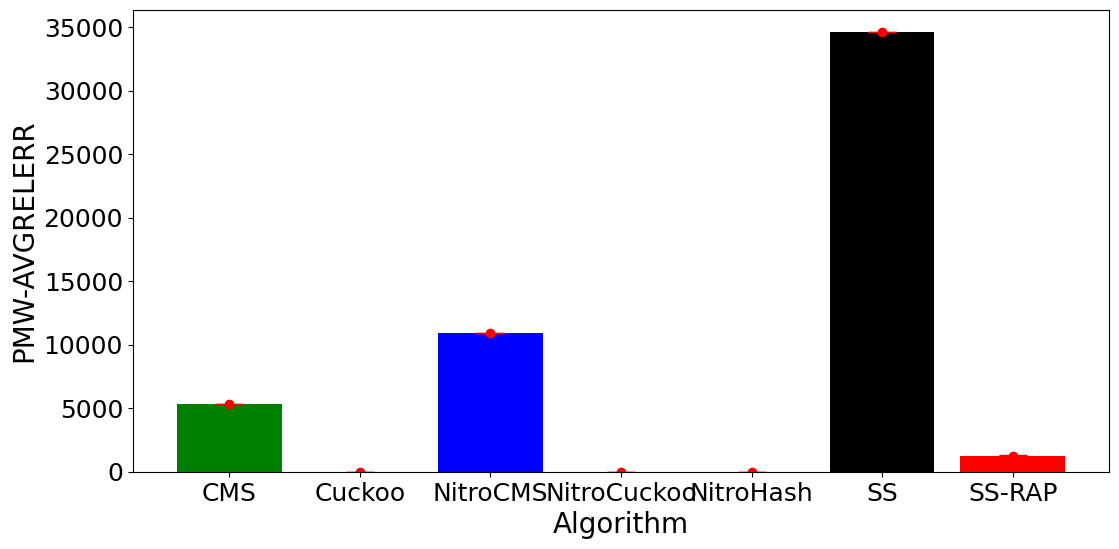}\label{fig:pm-avgrelerr:Chicago15}}\\
	\subfloat[][Chicago16Small]{\includegraphics[width=0.32\textwidth]{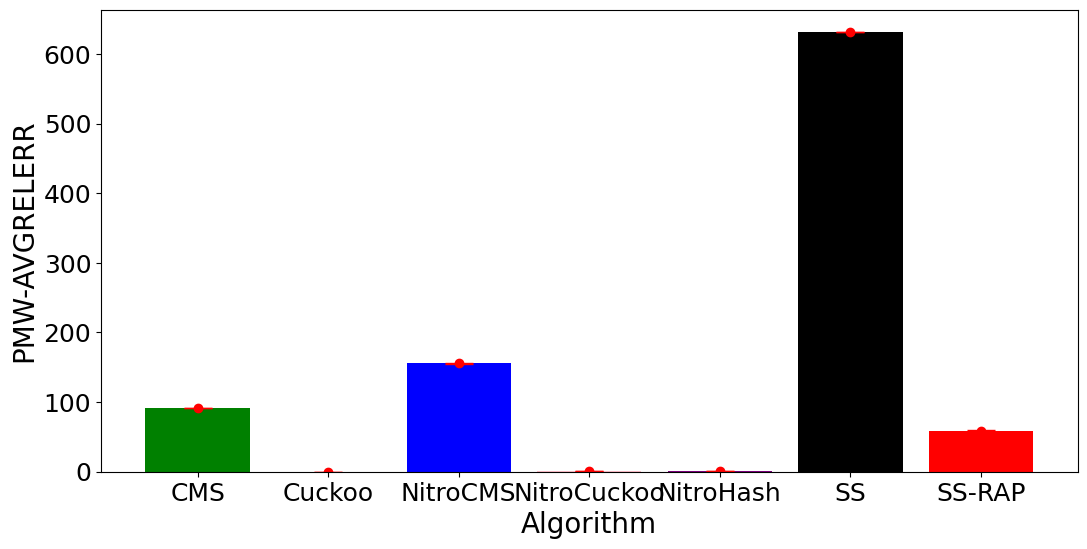}\label{fig:pm-avgrelerr:Chicago16Small}}
	\subfloat[][Chicago1610Mil]{\includegraphics[width=0.32\textwidth]{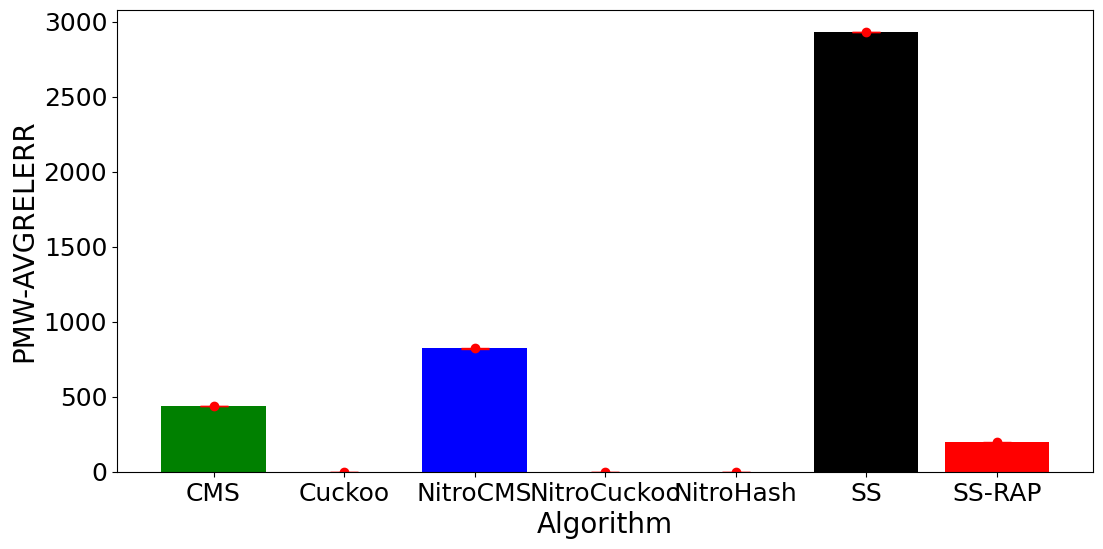}\label{fig:pm-avgrelerr:Chicago1610Mil}}
	\subfloat[][Chicago16]{\includegraphics[width=0.32\textwidth]{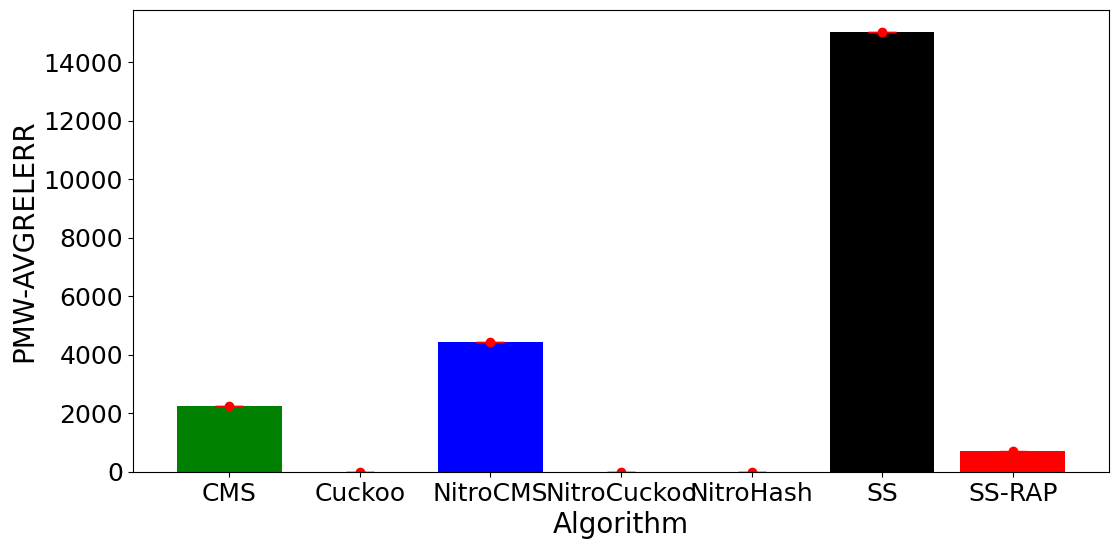}\label{fig:pm-avgrelerr:Chicago16}}
	\caption{Postmortem AVGRELERR}
	\label{fig:pm-avgrelerr}
\end{figure*}

The Postmortem MSRE, Postmortem AVGERR, and Postmortem AVGRELERR results appear in Figures~\ref{fig:pm-msre},~\ref{fig:pm-avgerr}, and~\ref{fig:pm-avgrelerr}, respectively.
Here, too, we notice small but non-negligible errors for the Chicago16, Chicago1610Mil, and Chicago16Small traces with the NitroHash and NitroCuckoo algorithms for the MSRE metric, which diminishes for the longer trace (the full Chicago16), similar to the on arrival error metric discussed above.

\subsection{Memory Consumption Results}
\label{sec:space}

\begin{figure*}[t]
	\centering
	\subfloat[][ny19A]{\includegraphics[width=0.32\textwidth]{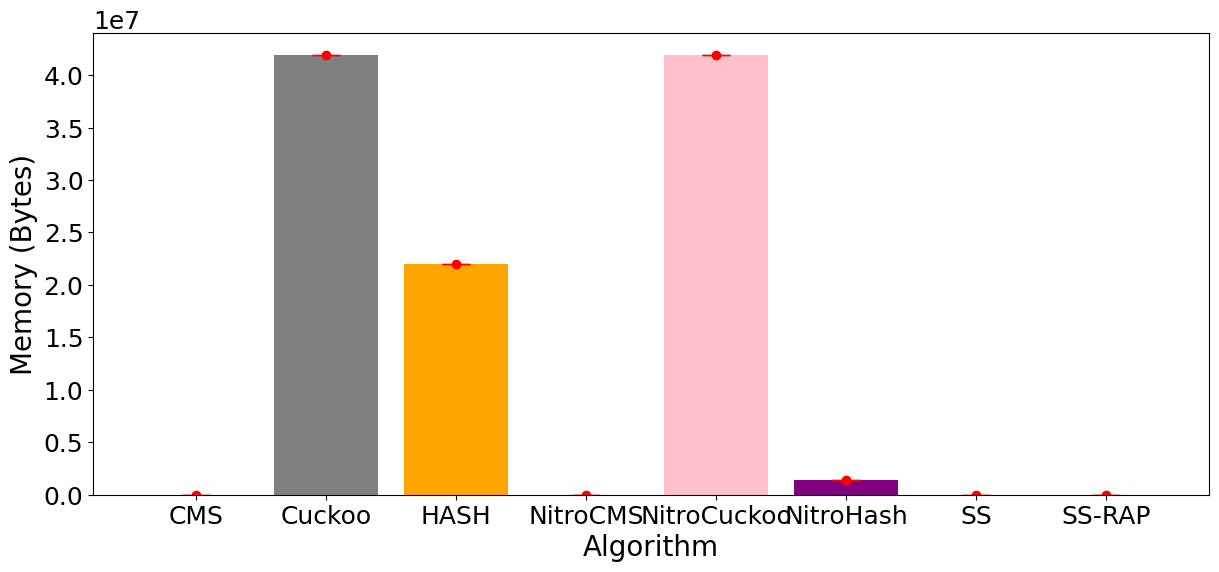}\label{fig:memory:ny19A}}
	\subfloat[][SJ14]{\includegraphics[width=0.32\textwidth]{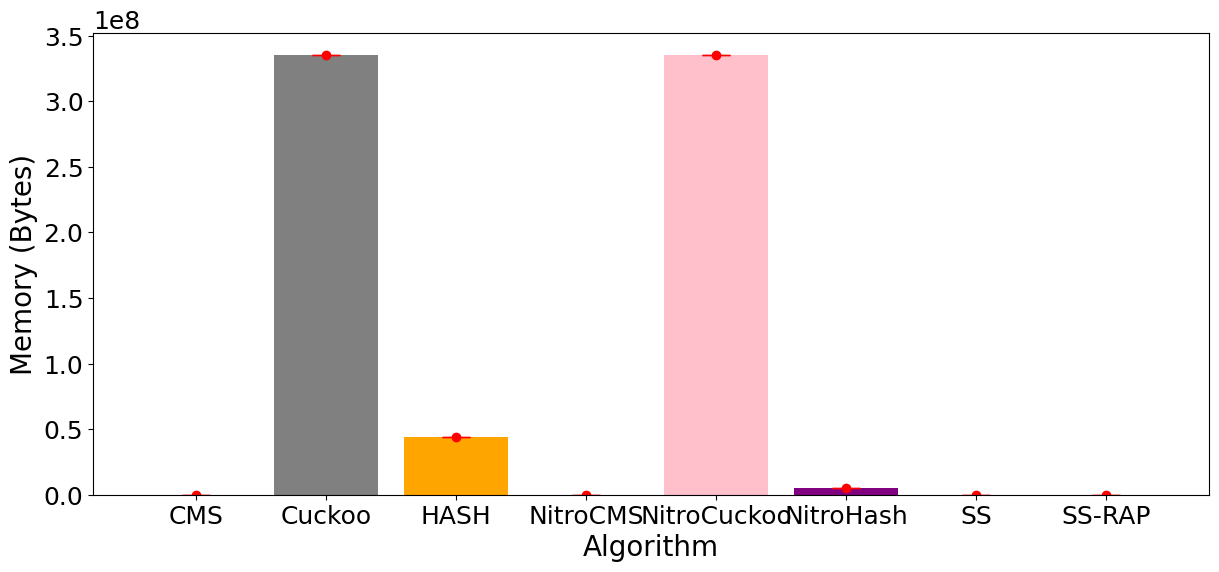}\label{fig:memory:SJ14}}
	\subfloat[][Chicago15]{\includegraphics[width=0.32\textwidth]{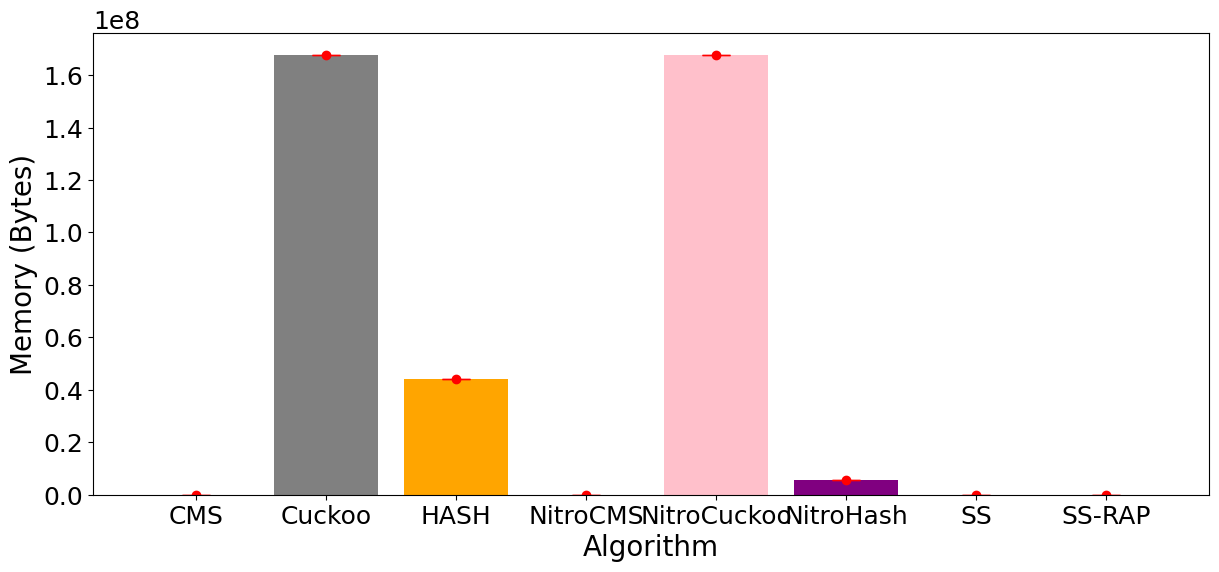}\label{fig:memory:Chicago15}}\\
	\subfloat[][Chicago16Small]{\includegraphics[width=0.32\textwidth]{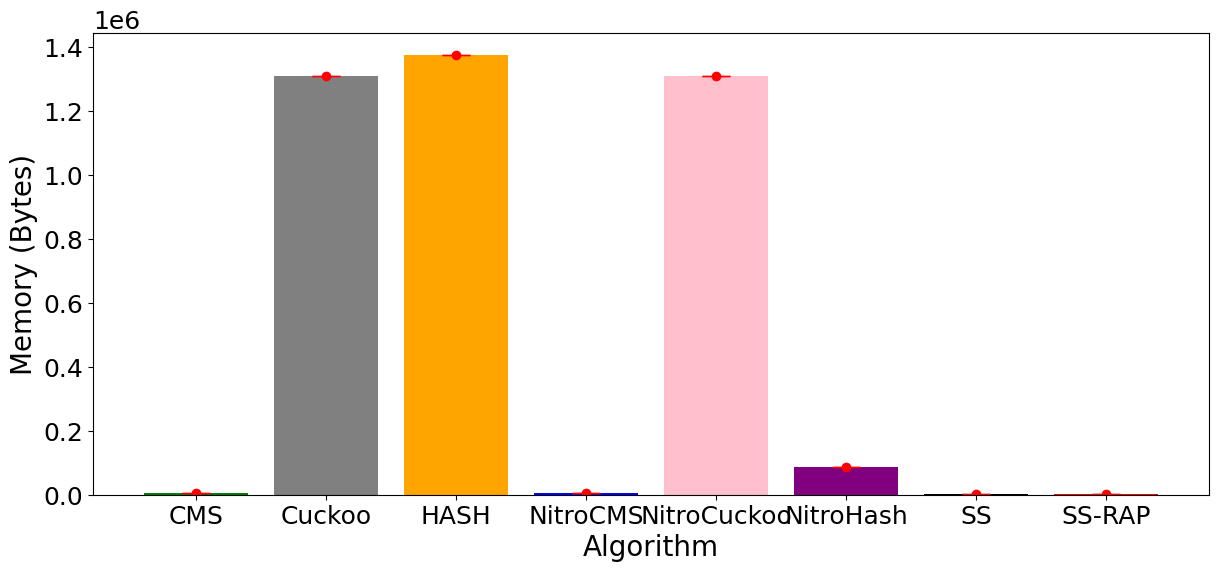}\label{fig:memory:Chicago16Small}}
	\subfloat[][Chicago1610Mil]{\includegraphics[width=0.32\textwidth]{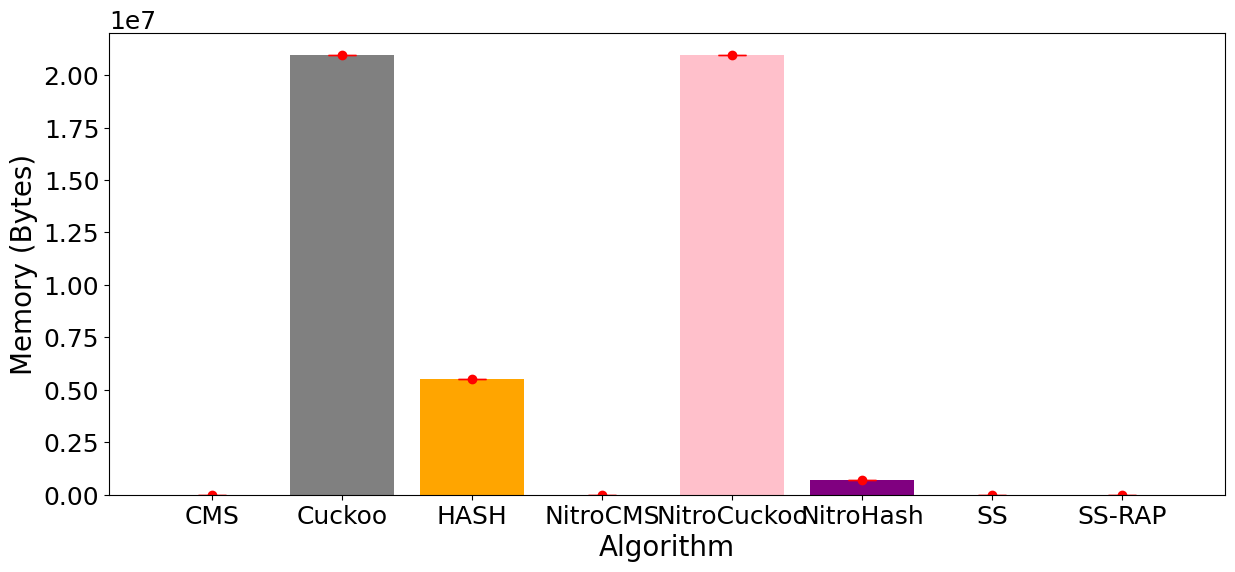}\label{fig:memory:Chicago1610Mil}}
	\subfloat[][Chicago16]{\includegraphics[width=0.32\textwidth]{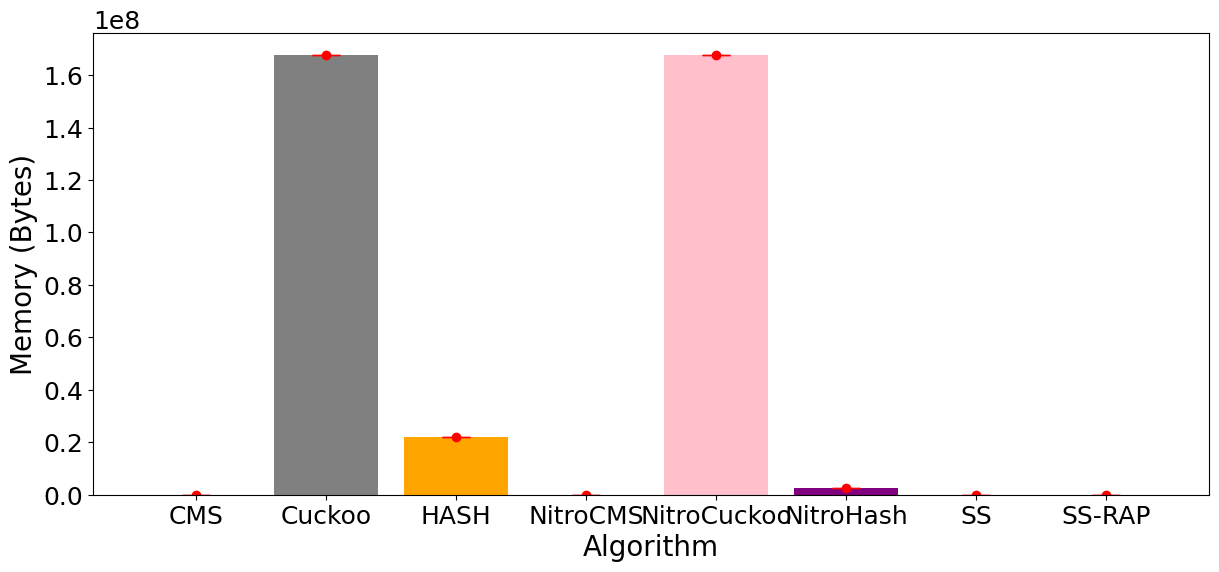}\label{fig:memory:Chicago16}}
	\caption{Memory consumption in bytes.}
	\label{fig:memory}
\end{figure*}

The memory consumed by each of the algorithms appears in Figure~\ref{fig:memory}.
For all algorithms, we used 32 bits counters.
The complete identifier size (IP 4-tuple) is 32 bits.
In the case of HASH and NitroHash, the size is the final hash table capacity multiplied by the size of each item (identifier + counter).
In these implementations, that table is initiated with a very small capacity, and is enlarged (doubled) whenever Rust's default implementation decides it is not large enough.
Cuckoo and NitroCuckoo are configured to hold all items in the trace, since the Cuckoo filter implementation is not elastic.
In this case, we multiply the capacity of the table by the size of each item, this time made of an 8-Byte fingerprint + counter.
CMS and Nitro CMS are configured according to their specification for $\epsilon = 0.01$ and $\delta = 0.01$; the space is computed as the number of resulting counters (1024) times counter size.
Finally, Space Saving (with and without RAP) is also configured according to its specification for $\epsilon = 0.01$; the space is computed as the number of resulting entries (100) times the counter size + identifier size.

\begin{figure*}[t]
	\centering
	\subfloat[][ny19A]{\includegraphics[width=0.3\textwidth]{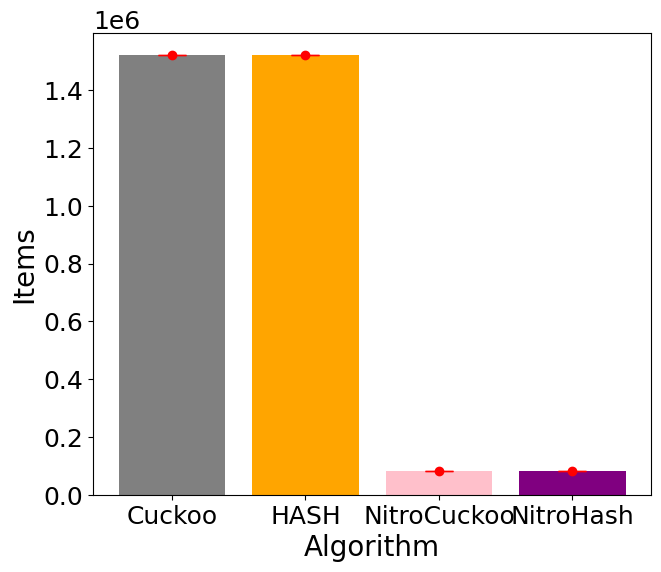}\label{fig:items:ny19A}}
	\subfloat[][SJ14]{\includegraphics[width=0.3\textwidth]{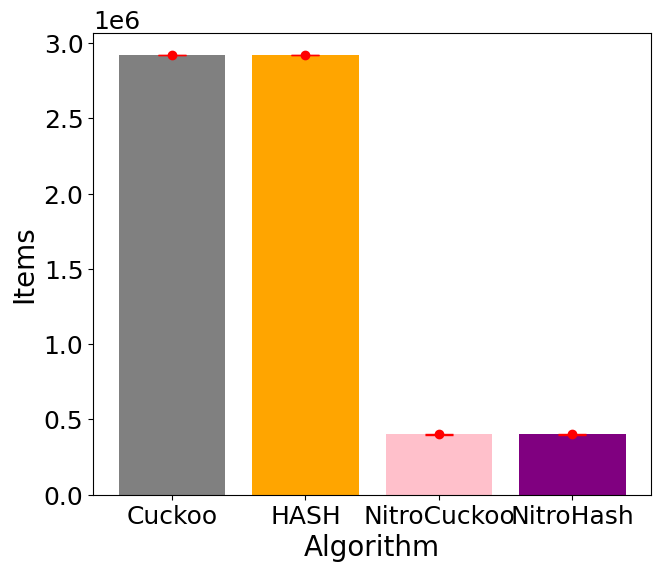}\label{fig:items:SJ14}}
	\subfloat[][Chicago15]{\includegraphics[width=0.3\textwidth]{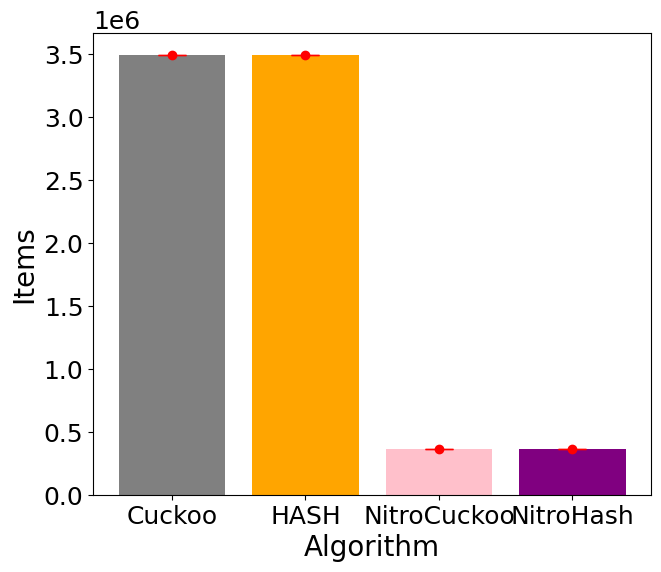}\label{fig:items:Chicago15}}\\
	\subfloat[][Chicago16Small]{\includegraphics[width=0.3\textwidth]{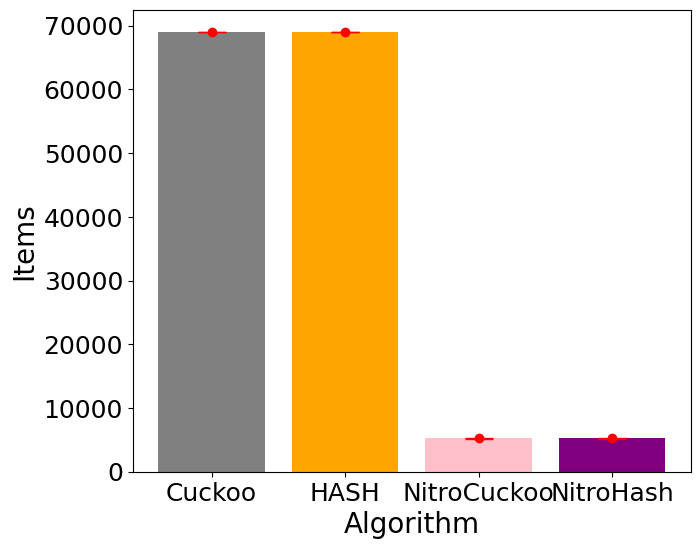}\label{fig:items:Chicago16Small}}
	\subfloat[][Chicago1610Mil]{\includegraphics[width=0.3\textwidth]{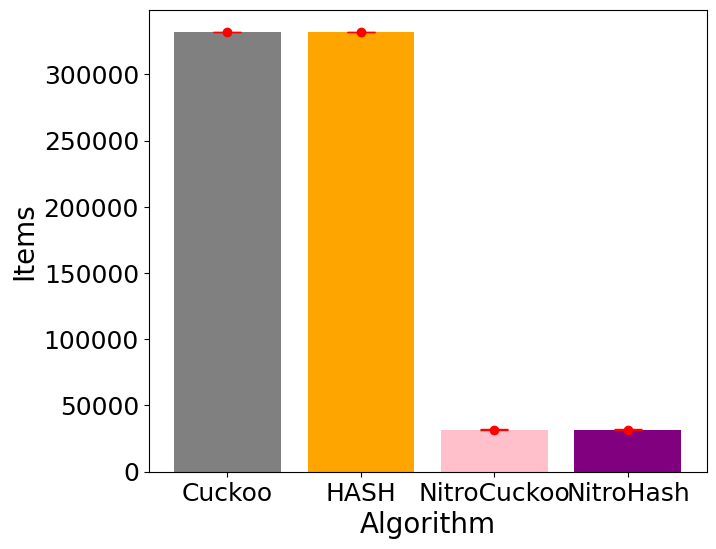}\label{fig:items:Chicago1610Mil}}
	\subfloat[][Chicago16]{\includegraphics[width=0.3\textwidth]{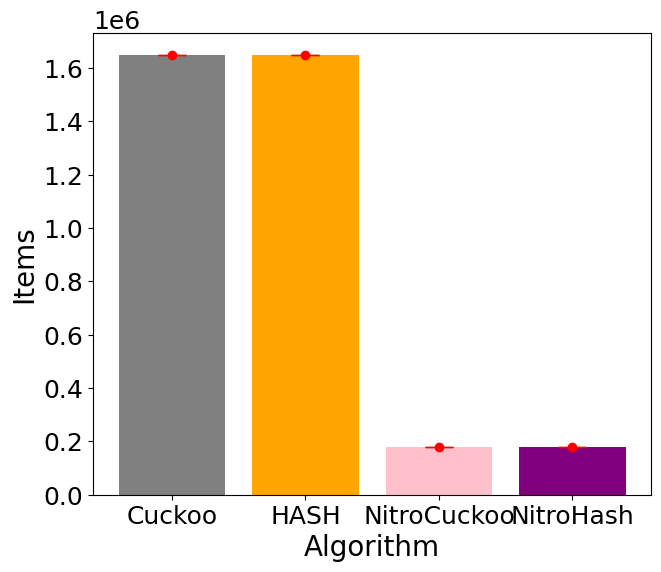}\label{fig:items:Chicago16}}
	\caption{Number of (unique) items inserted by each algorithm into its main data structure.}
	\label{fig:items}
	%
	\subfloat[][ny19A]{\includegraphics[width=0.3\textwidth]{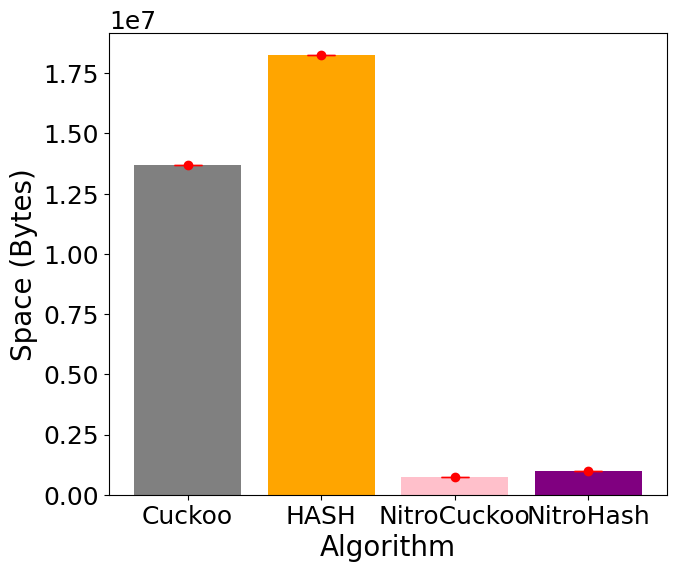}\label{fig:item:ny19A}}
	\subfloat[][SJ14]{\includegraphics[width=0.3\textwidth]{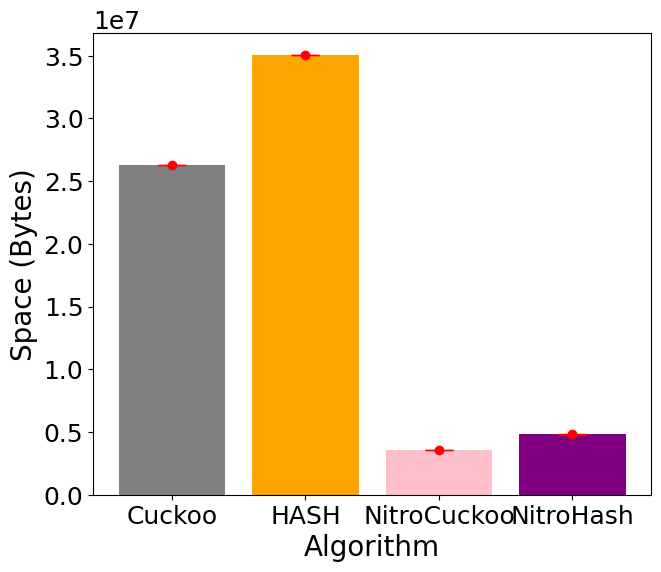}\label{fig:item:SJ14}}
	\subfloat[][Chicago15]{\includegraphics[width=0.3\textwidth]{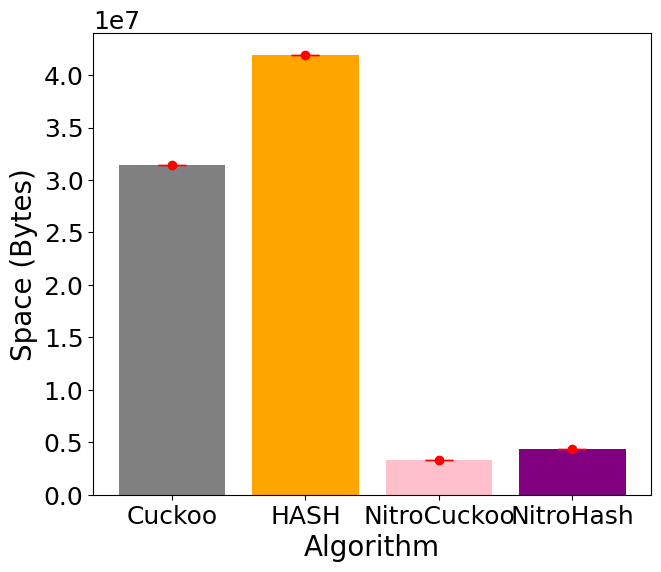}\label{fig:item:Chicago15}}\\
	\subfloat[][Chicago16Small]{\includegraphics[width=0.3\textwidth]{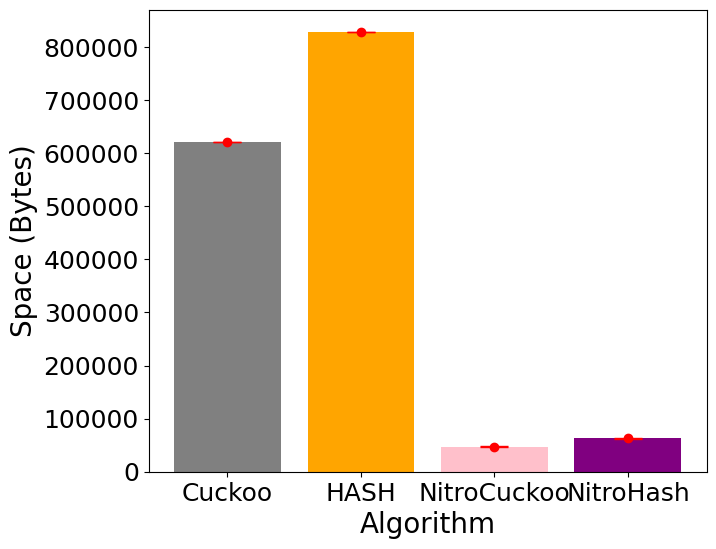}\label{fig:item:Chicago16Small}}
	\subfloat[][Chicago1610Mil]{\includegraphics[width=0.3\textwidth]{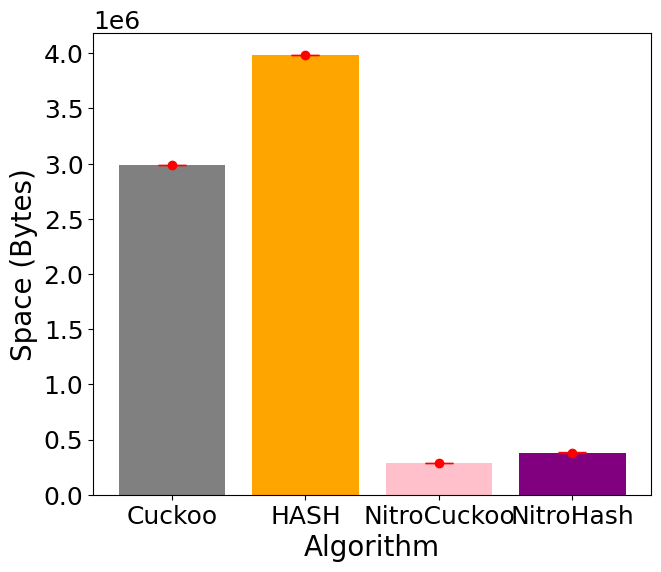}\label{fig:item:Chicago1610Mil}}
	\subfloat[][Chicago16]{\includegraphics[width=0.3\textwidth]{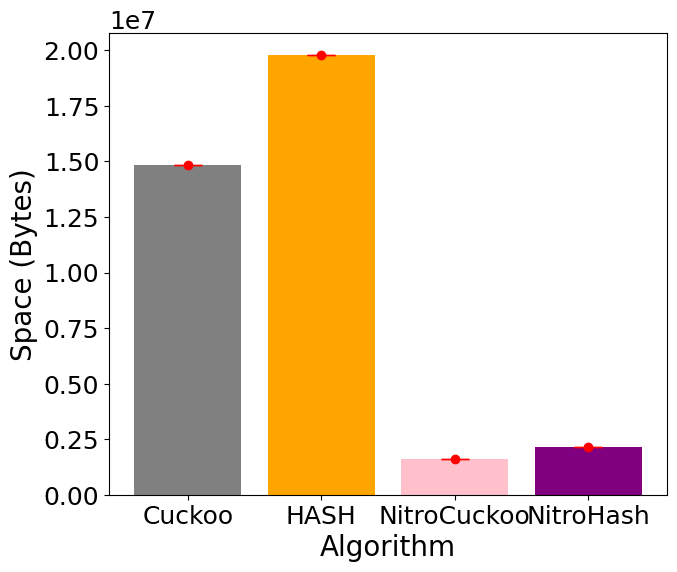}\label{fig:item:Chicago16}}
	\caption{Space (bytes) consumed by the items inserted by each algorithm into its main data structure.}
	\label{fig:space}
\end{figure*}

For the HASH, NitroHash, Cuckoo, and NitroCuckoo, we also report the number of (unique) items stored in the respective data structure in Figure~\ref{fig:items} as well as the amount of space these items consume in Figure~\ref{fig:space}.
The rational here is that if we have a good estimate for how many unique items would be stored, we could configure the data structure accordingly and save space.
In particular, when applying the Nitro optimization, the use of sampling translates into inserting much fewer items into the data structure, and so in addition to the improved throughput, it can be used to reduce the memory consumption, as explored in Section~\ref{sec:cuckoo-deepdive} below.
The reason HASH requires more space than Cuckoo when they hold the same number of items is related to the size of full identifiers vs. fingerprints.
Notice that with CMS, NitroCMS, and Space Saving (w/o RAP) the amount of counters or entries is fixed by the error guarantees and do not depend on the workload, and therefore these measures are meaningless for them.

\subsection{CMS W/O Minimal Increment}
\label{sec:cms-deepdive}

\begin{figure*}[t]
	\centering
	\subfloat[][ny19A]{\includegraphics[width=0.15\textwidth]{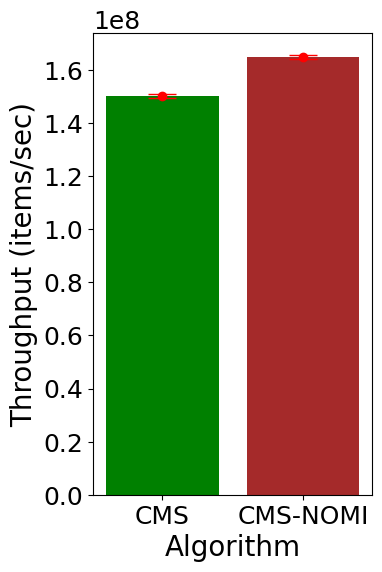}\label{fig:thpt-write:NOMI-ny19A}}
	\subfloat[][SJ14]{\includegraphics[width=0.15\textwidth]{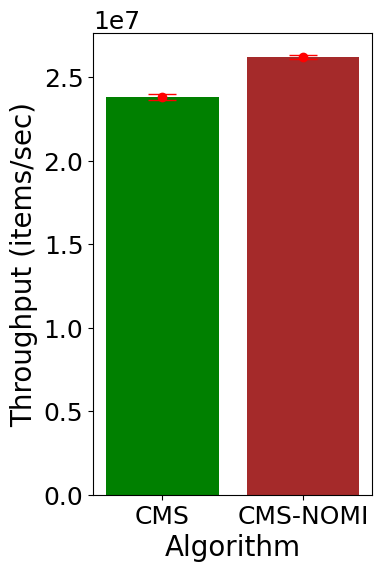}\label{fig:thpt-write:NOMI-SJ14}}
	\subfloat[][Chicago15]{\includegraphics[width=0.15\textwidth]{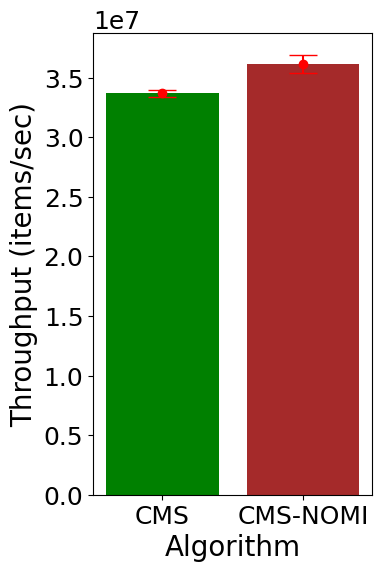}\label{fig:thpt-write:NOMI-Chicago15}}
	\subfloat[][Chic16Small]{\includegraphics[width=0.15\textwidth]{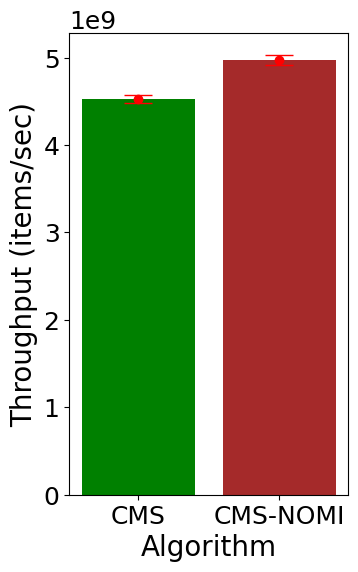}\label{fig:thpt-write:NOMI-Chicago16Small}}
	\subfloat[][Chic1610Mil]{\includegraphics[width=0.15\textwidth]{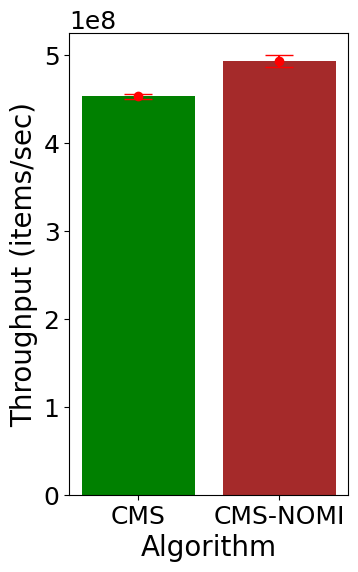}\label{fig:thpt-write:NOMI-Chicago1610Mil}}
	\subfloat[][Chicago16]{\includegraphics[width=0.15\textwidth]{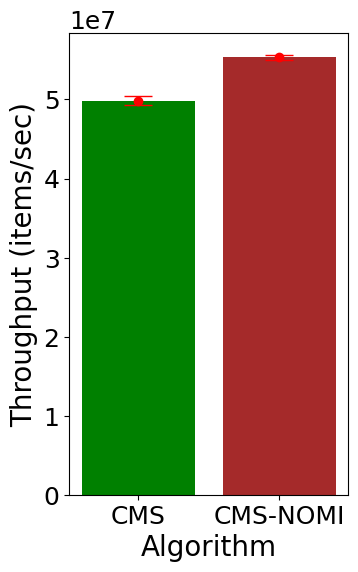}\label{fig:thpt-write:NOMI-Chicago16}}
	\caption{Throughput results for the write only test w/o minimal increment}
	\label{fig:NOMI-thpt-write}
%
	\centering
	\subfloat[][ny19A]{\includegraphics[width=0.15\textwidth]{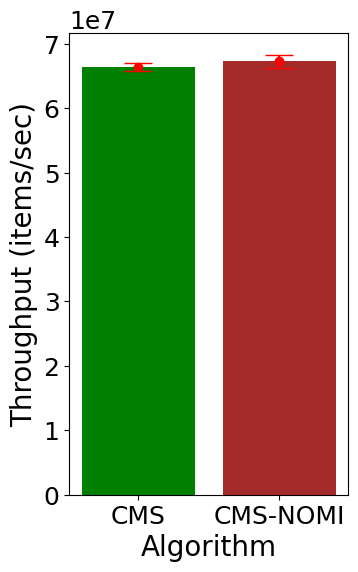}\label{fig:thpt-rw:NOMI-ny19A}}
	\subfloat[][SJ14]{\includegraphics[width=0.15\textwidth]{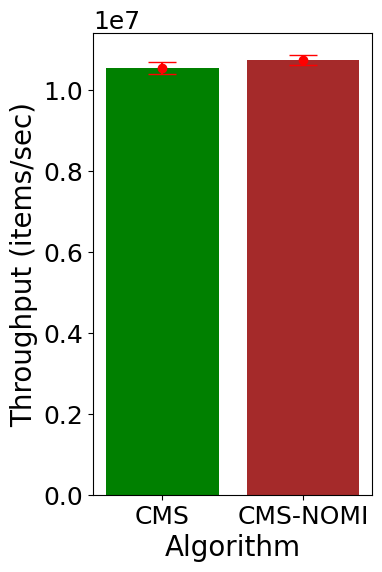}\label{fig:thpt-rw:NOMI-SJ14}}
	\subfloat[][Chicago15]{\includegraphics[width=0.15\textwidth]{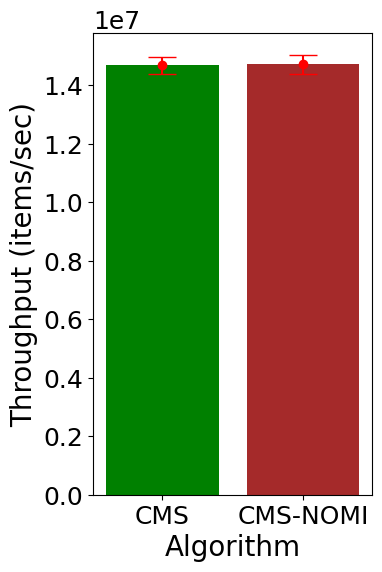}\label{fig:thpt-rw:CNOMI-hicago15}}
	\subfloat[][Chic16Small]{\includegraphics[width=0.15\textwidth]{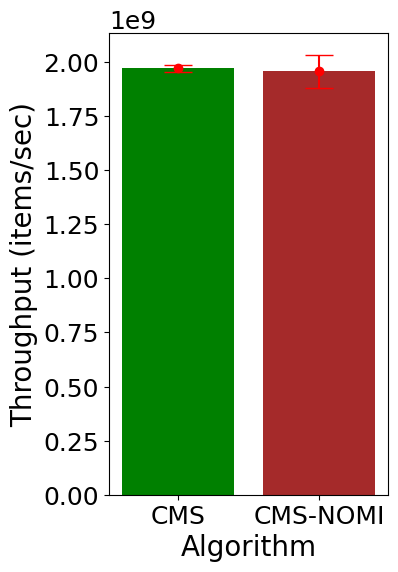}\label{fig:thpt-rw:NOMI-Chicago16Small}}
	\subfloat[][Chic1610Mil]{\includegraphics[width=0.15\textwidth]{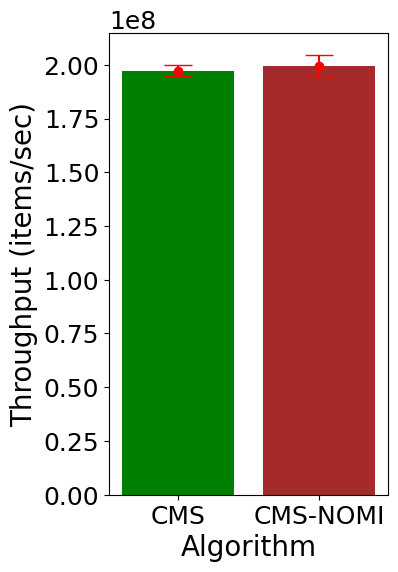}\label{fig:thpt-rw:NOMI-Chicago1610Mil}}
	\subfloat[][Chicago16]{\includegraphics[width=0.15\textwidth]{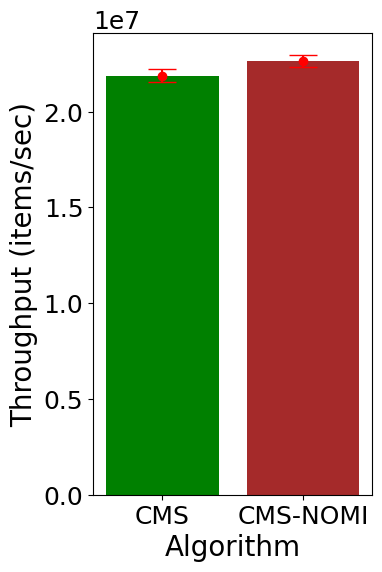}\label{fig:thpt-rw:NOMI-Chicago16}}
	\caption{Throughput results for the write-read test w/o minimal increment}
	\label{fig:NOMI-thpt-rw}
%
	\centering
	\subfloat[][ny19A]{\includegraphics[width=0.15\textwidth]{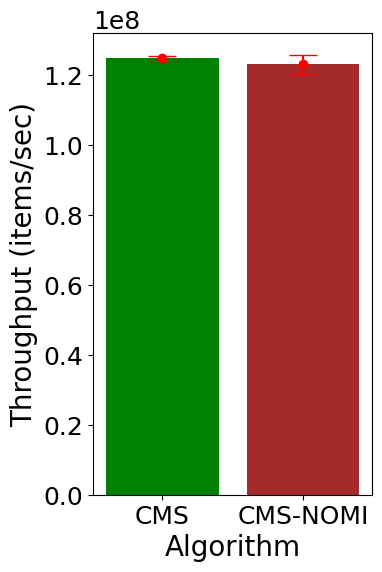}\label{fig:thpt-read:NOMI-ny19A}}
	\subfloat[][SJ14]{\includegraphics[width=0.15\textwidth]{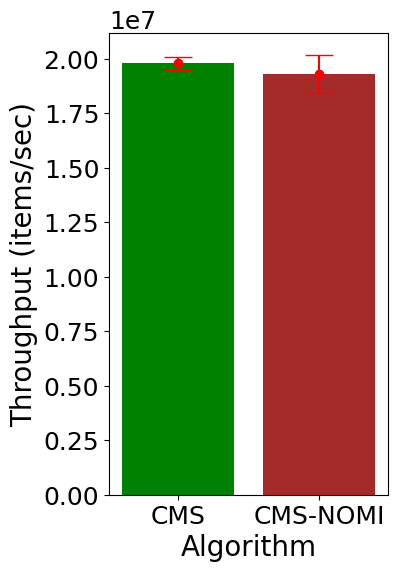}\label{fig:thpt-read:NOMI-SJ14}}
	\subfloat[][Chicago15]{\includegraphics[width=0.15\textwidth]{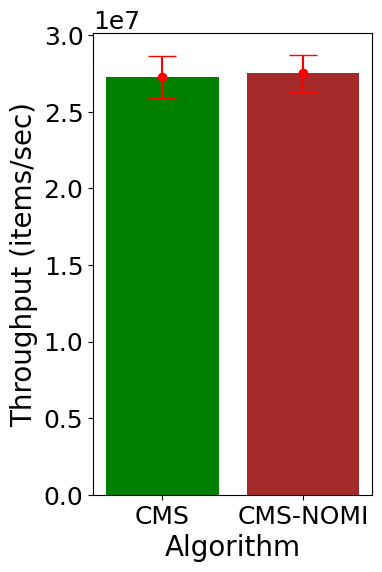}\label{fig:thpt-read:NOMI-Chicago15}}
	\subfloat[][Chic16Small]{\includegraphics[width=0.15\textwidth]{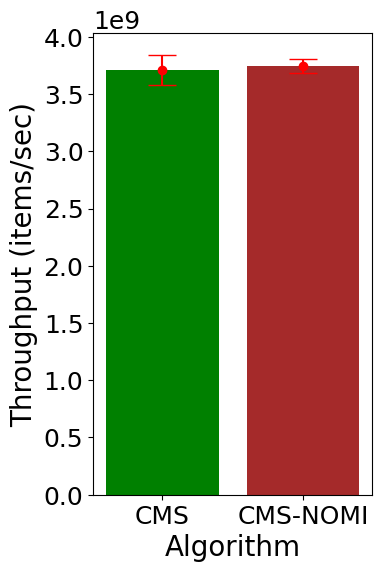}\label{fig:thpt-read:NOMI-Chicago16Small}}
	\subfloat[][Chic1610Mil]{\includegraphics[width=0.15\textwidth]{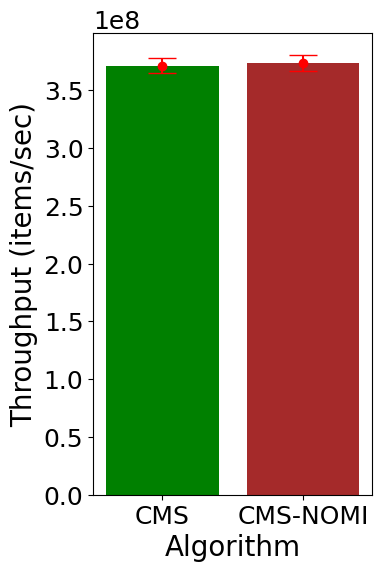}\label{fig:thpt-read:NOMI-Chicago1610Mil}}
	\subfloat[][Chicago16]{\includegraphics[width=0.15\textwidth]{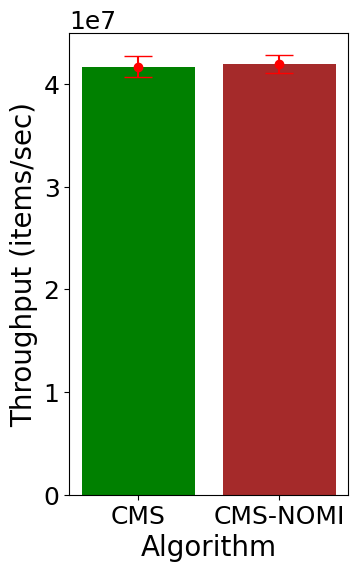}\label{fig:thpt-read:NOMI-Chicago16}}
	\caption{Throughput results for the read only test w/o minimal increment}
	\label{fig:NOMI-thpt-read}
\end{figure*}

We now discuss the impact of the minimal increment (conservative update) optimization of CMS.
We compare an implementation without minimal increment, nicknamed CMS-NOMI, to the default implementation.
As can be seen in Figure~\ref{fig:NOMI-thpt-write}, the minimal increment optimization increases performance in the write only test by roughly $10$\%.
This is naturally reduced in the read-write test, shown in Figure~\ref{fig:NOMI-thpt-rw}, and completely diminishes in the read only test, reported in Figure~\ref{fig:NOMI-thpt-read}.

\begin{figure*}[t]
	\centering
	\subfloat[][ny19A]{\includegraphics[width=0.15\textwidth]{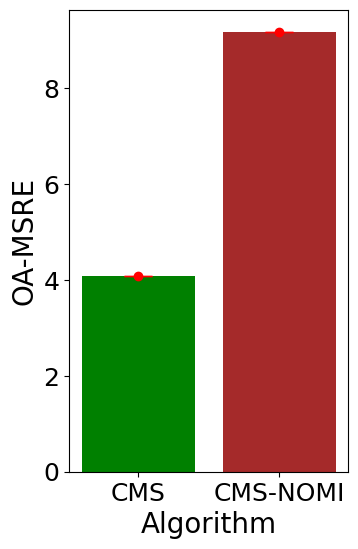}\label{fig:oa-msre:NOMI-ny19A}}
	\subfloat[][SJ14]{\includegraphics[width=0.15\textwidth]{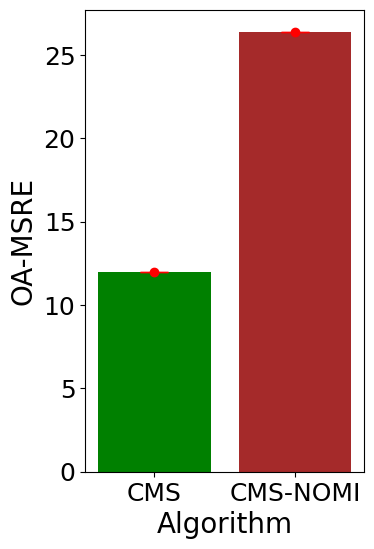}\label{fig:oa-msre:NOMI-SJ14}}
	\subfloat[][Chicago15]{\includegraphics[width=0.15\textwidth]{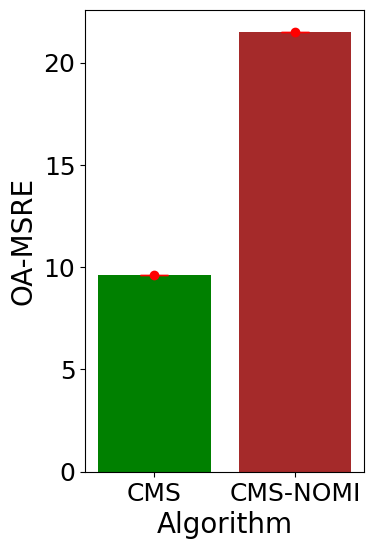}\label{fig:oa-msre:NOMI-Chicago15}}
	\subfloat[][Chic16Small]{\includegraphics[width=0.15\textwidth]{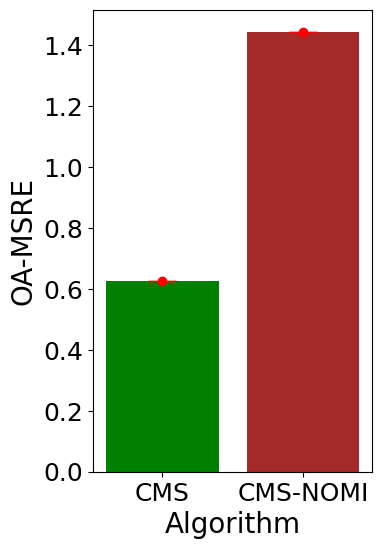}\label{fig:oa-msre:NOMI-Chicago16Small}}
	\subfloat[][Chic1610Mil]{\includegraphics[width=0.15\textwidth]{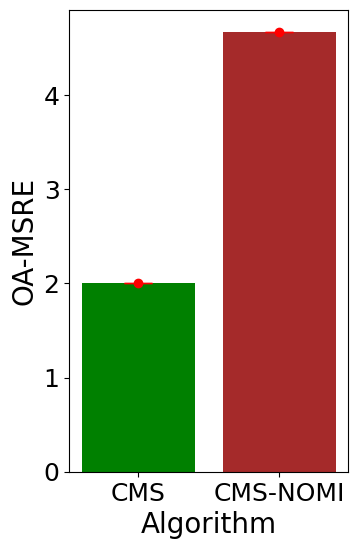}\label{fig:oa-msre:NOMI-Chicago1610Mil}}
	\subfloat[][Chicago16]{\includegraphics[width=0.15\textwidth]{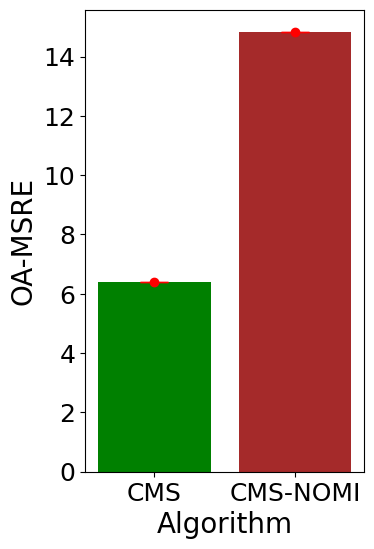}\label{fig:oa-msre:NOMI-Chicago16}}
	\caption{OA Arrival MSRE w/o minimal increment}
	\label{fig:NOMI-oa-msre}
	%
	\subfloat[][ny19A]{\includegraphics[width=0.15\textwidth]{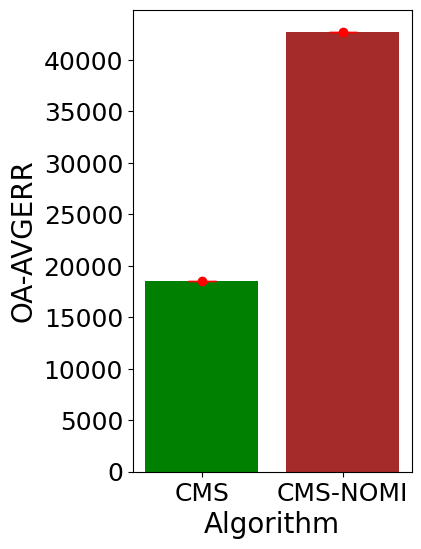}\label{fig:oa-avgerr:NOMI-ny19A}}
	\subfloat[][SJ14]{\includegraphics[width=0.15\textwidth]{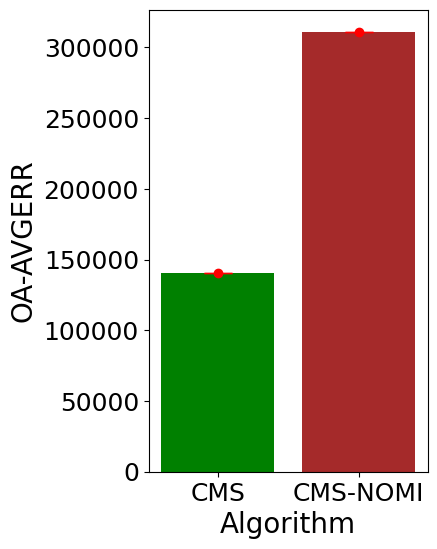}\label{fig:oa-avgerr:NOMI-SJ14}}
	\subfloat[][Chicago15]{\includegraphics[width=0.15\textwidth]{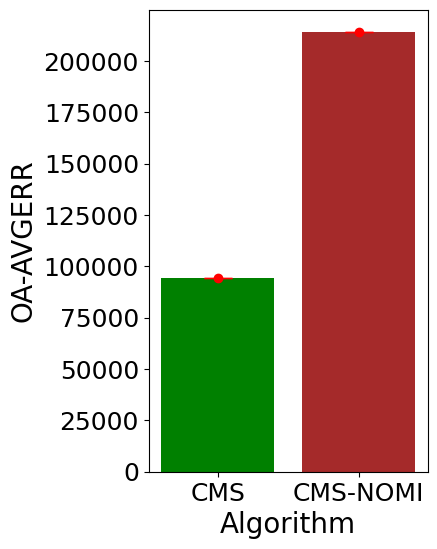}\label{fig:oa-avgerr:NOMI-Chicago15}}
	\subfloat[][Chic16Small]{\includegraphics[width=0.15\textwidth]{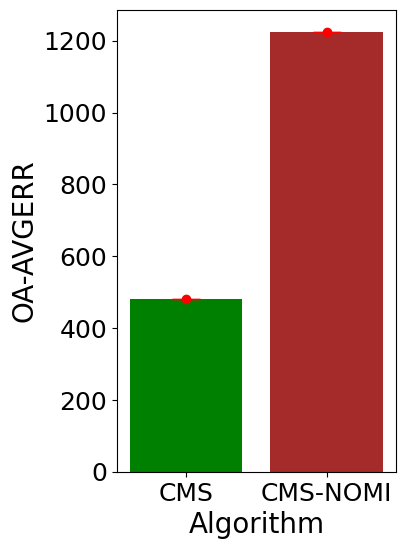}\label{fig:oa-avgerr:NOMI-Chicago16Small}}
	\subfloat[][Chic1610Mil]{\includegraphics[width=0.15\textwidth]{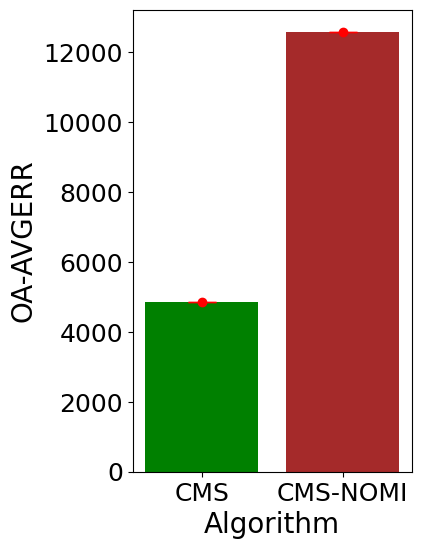}\label{fig:oa-avgerr:NOMI-Chicago1610Mil}}
	\subfloat[][Chicago16]{\includegraphics[width=0.15\textwidth]{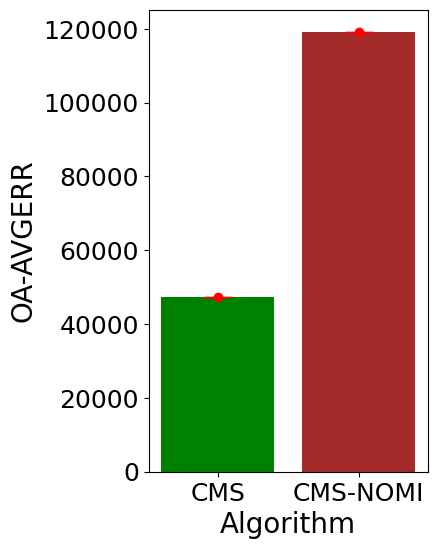}\label{fig:oa-avgerr:NOMI-Chicago16}}
	\caption{OA Arrival AVGERR w/o minimal increment}
	\label{fig:NOMI-oa-avgerr}
	%
	\subfloat[][ny19A]{\includegraphics[width=0.15\textwidth]{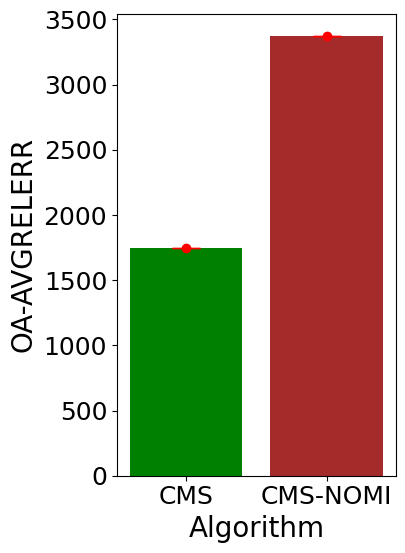}\label{fig:oa-avgrelerr:NOMI-ny19A}}
	\subfloat[][SJ14]{\includegraphics[width=0.15\textwidth]{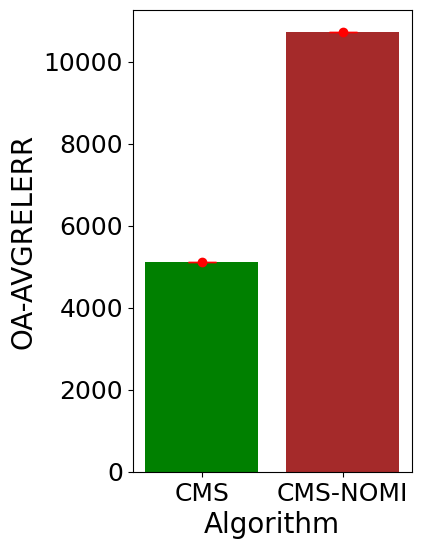}\label{fig:oa-avgrelerr:NOMI-SJ14}}
	\subfloat[][Chicago15]{\includegraphics[width=0.15\textwidth]{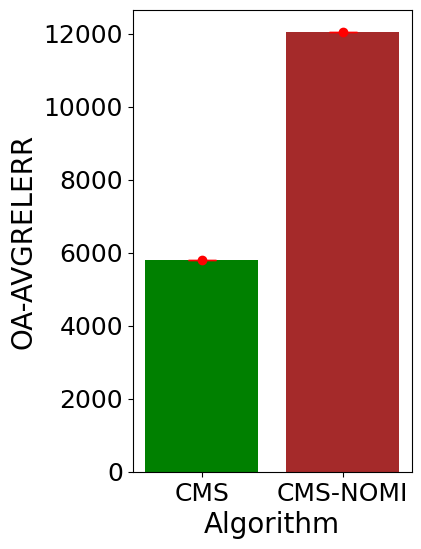}\label{fig:oa-avgrelerr:NOMI-Chicago15}}
	\subfloat[][Chic16Small]{\includegraphics[width=0.15\textwidth]{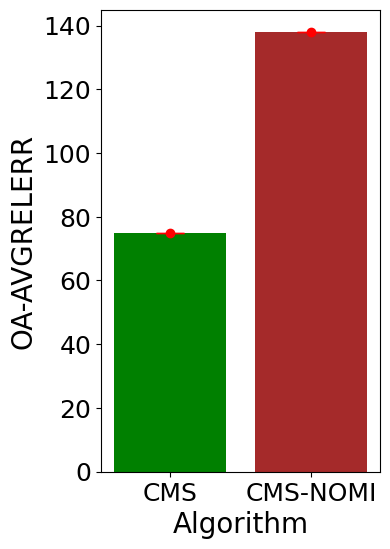}\label{fig:oa-avgrelerr:NOMI-Chicago16Small}}
	\subfloat[][Chic1610Mil]{\includegraphics[width=0.15\textwidth]{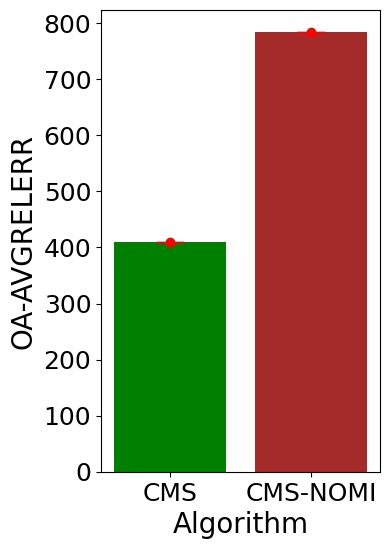}\label{fig:oa-avgrelerr:NOMI-Chicago1610Mil}}
	\subfloat[][Chicago16]{\includegraphics[width=0.15\textwidth]{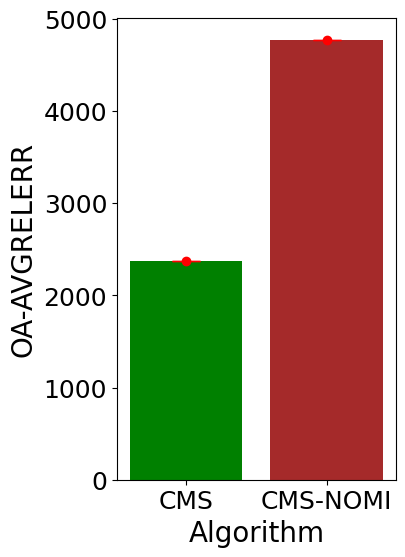}\label{fig:oa-avgrelerr:NOMI-Chicago16}}
	\caption{OA Arrival AVGRELERR w/o minimal increment}
	\label{fig:NOMI-oa-avgrelerr}
\end{figure*}

\begin{figure*}[t]
	\centering
	\subfloat[][ny19A]{\includegraphics[width=0.15\textwidth]{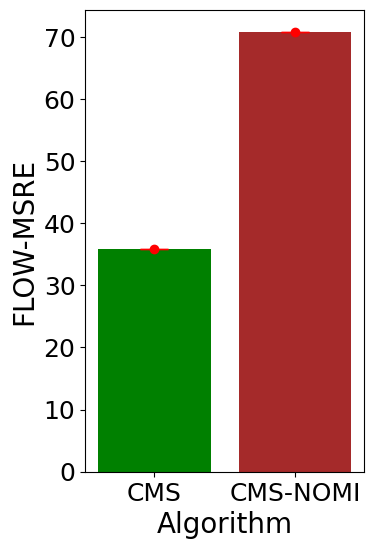}\label{fig:flow-msre:NOMI-ny19A}}
	\subfloat[][SJ14]{\includegraphics[width=0.15\textwidth]{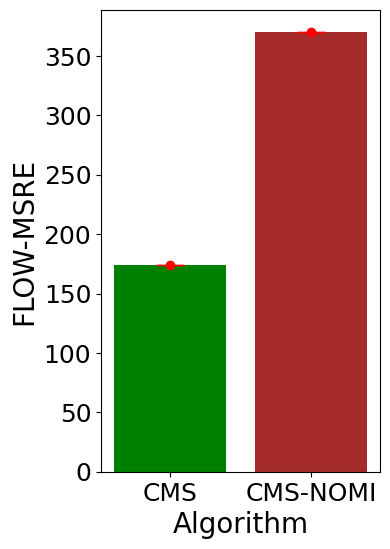}\label{fig:flow-msre:NOMI-SJ14}}
	\subfloat[][Chicago15]{\includegraphics[width=0.15\textwidth]{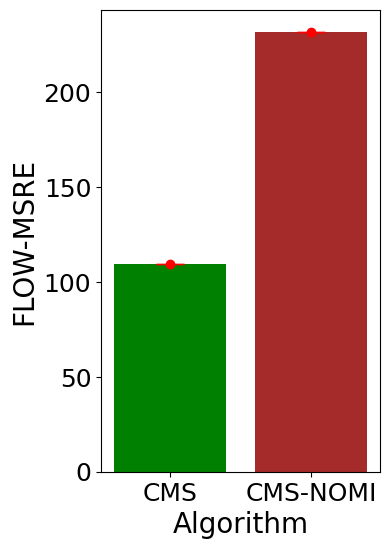}\label{fig:flow-msre:NOMI-Chicago15}}
	\subfloat[][Chic16Small]{\includegraphics[width=0.15\textwidth]{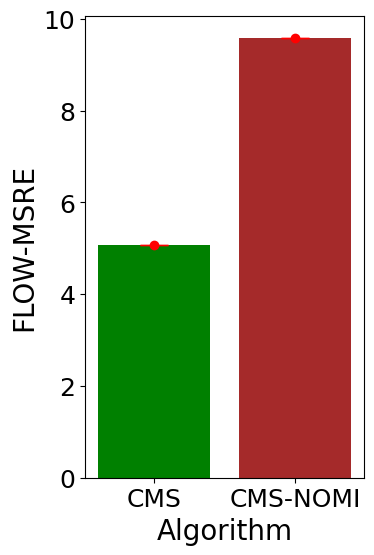}\label{fig:flow-msre:NOMI-Chicago16Small}}
	\subfloat[][Chic1610Mil]{\includegraphics[width=0.15\textwidth]{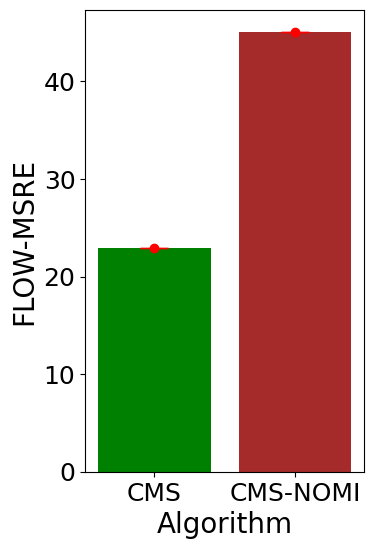}\label{fig:flow-msre:NOMI-Chicago1610Mil}}
	\subfloat[][Chicago16]{\includegraphics[width=0.15\textwidth]{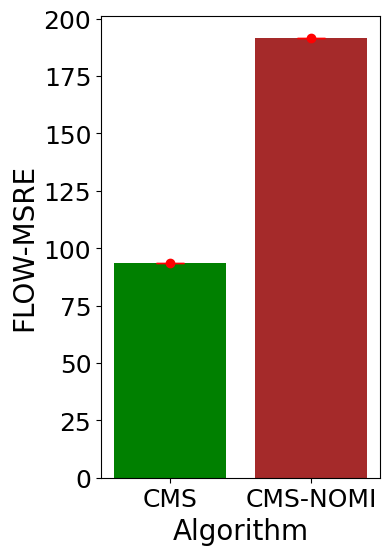}\label{fig:flow-msre:NOMI-Chicago16}}
	\caption{Per Flow MSRE w/o minimal increment}
	\label{fig:NOMI-flow-msre}
	%
	\subfloat[][ny19A]{\includegraphics[width=0.15\textwidth]{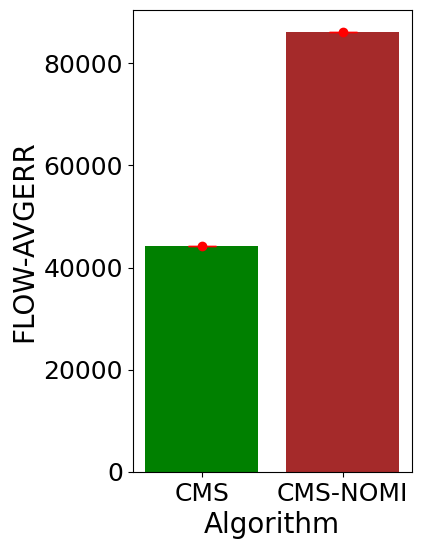}\label{fig:flow-avgerr:NOMI-ny19A}}
	\subfloat[][SJ14]{\includegraphics[width=0.15\textwidth]{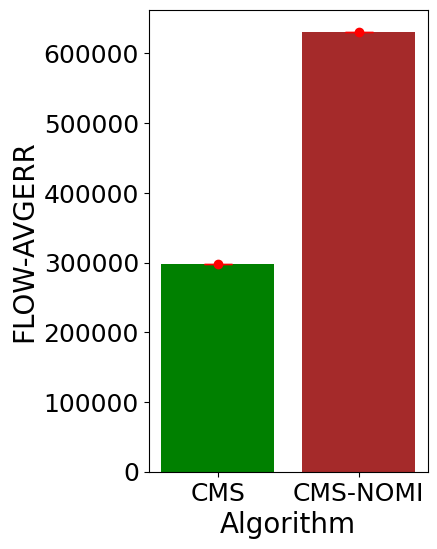}\label{fig:flow-avgerr:NOMI-SJ14}}
	\subfloat[][Chicago15]{\includegraphics[width=0.15\textwidth]{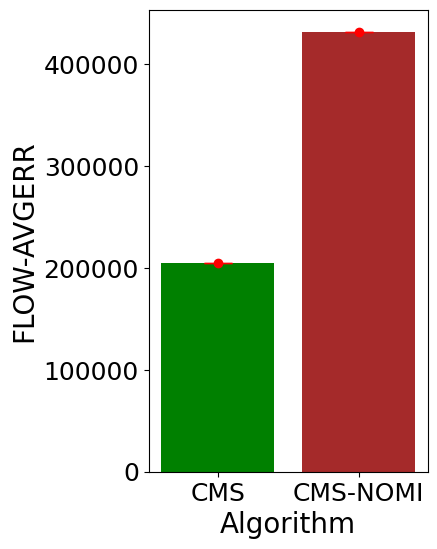}\label{fig:flow-avgerr:NOMI-Chicago15}}
	\subfloat[][Chic16Small]{\includegraphics[width=0.15\textwidth]{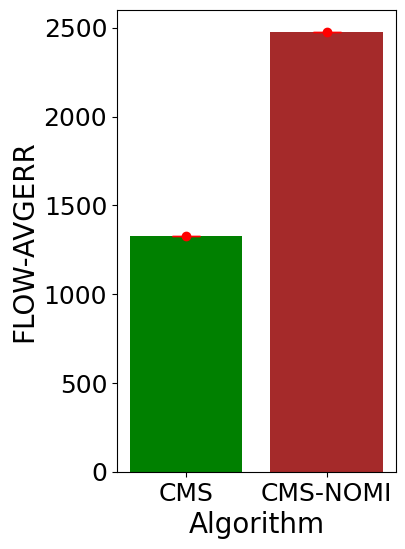}\label{fig:flow-avgerr:NOMI-Chicago16Small}}
	\subfloat[][Chic1610Mil]{\includegraphics[width=0.15\textwidth]{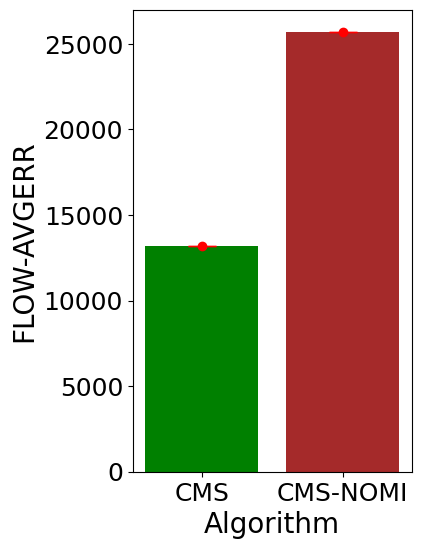}\label{fig:flow-avgerr:NOMI-Chicago1610Mil}}
	\subfloat[][Chicago16]{\includegraphics[width=0.15\textwidth]{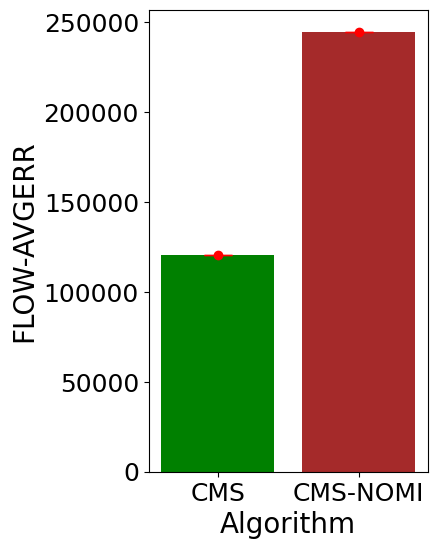}\label{fig:flow-avgerr:NOMI-Chicago16}}
	\caption{Per FLOW AVGERR w/o minimal increment}
	\label{fig:NOMI-flow-avgerr}
	%
	\subfloat[][ny19A]{\includegraphics[width=0.15\textwidth]{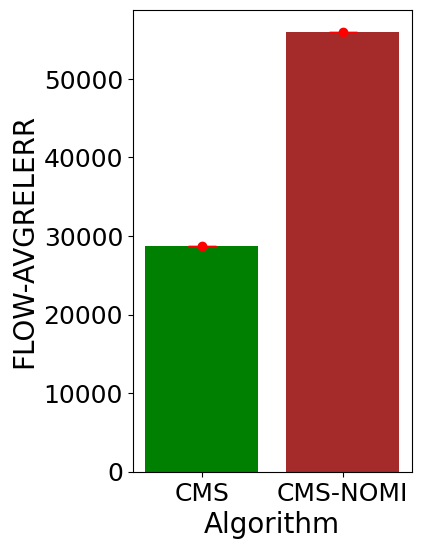}\label{fig:flow-avgrelerr:NOMI-ny19A}}
	\subfloat[][SJ14]{\includegraphics[width=0.15\textwidth]{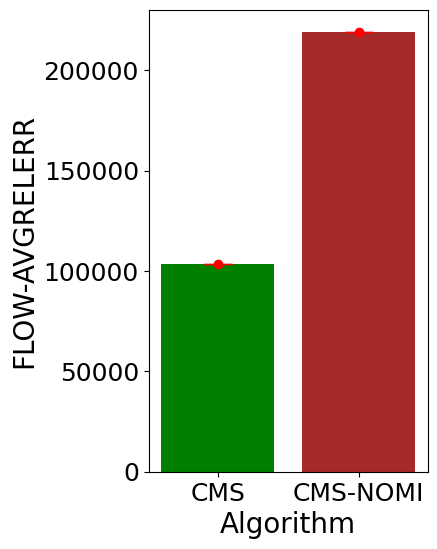}\label{fig:flow-avgrelerr:NOMI-SJ14}}
	\subfloat[][Chicago15]{\includegraphics[width=0.15\textwidth]{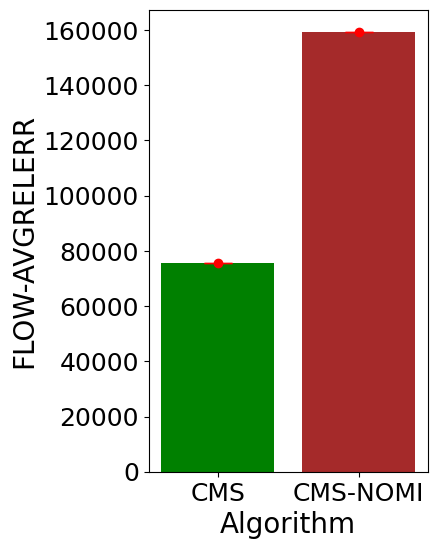}\label{fig:flow-avgrelerr:NOMI-Chicago15}}
	\subfloat[][Chic16Small]{\includegraphics[width=0.15\textwidth]{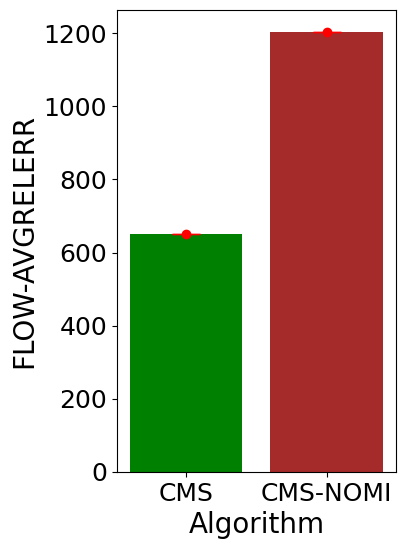}\label{fig:flow-avgrelerr:NOMI-Chicago16Small}}
	\subfloat[][Chic1610Mil]{\includegraphics[width=0.15\textwidth]{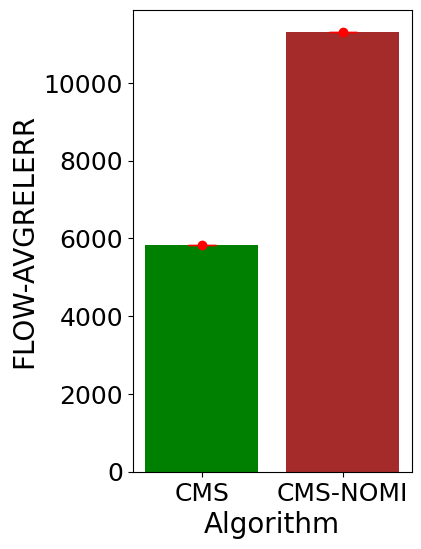}\label{fig:flow-avgrelerr:NOMI-Chicago1610Mil}}
	\subfloat[][Chicago16]{\includegraphics[width=0.15\textwidth]{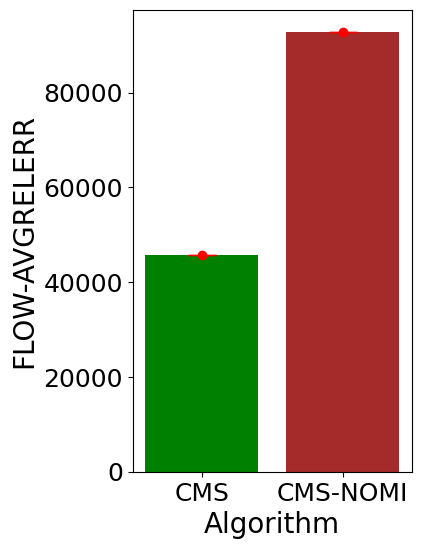}\label{fig:flow-avgrelerr:NOMI-Chicago16}}
	\caption{Per FLOW AVGRELERR w/o minimal increment}
	\label{fig:NOMI-flow-avgrelerr}
\end{figure*}

\begin{figure*}[t]
	\centering
	\subfloat[][ny19A]{\includegraphics[width=0.15\textwidth]{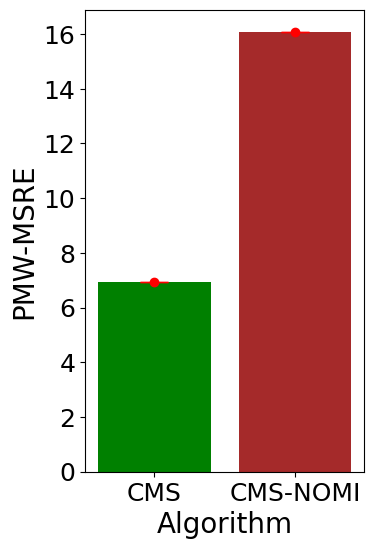}\label{fig:pm-msre:NOMI-ny19A}}
	\subfloat[][SJ14]{\includegraphics[width=0.15\textwidth]{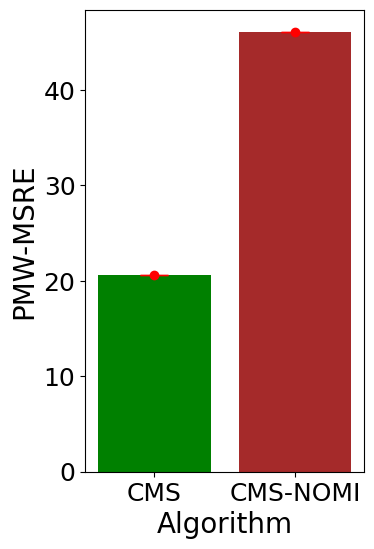}\label{fig:pm-msre:NOMI-SJ14}}
	\subfloat[][Chicago15]{\includegraphics[width=0.15\textwidth]{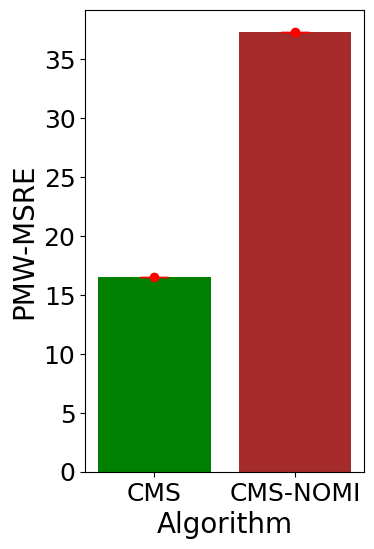}\label{fig:pm-msre:NOMI-Chicago15}}
	\subfloat[][Chic16Small]{\includegraphics[width=0.15\textwidth]{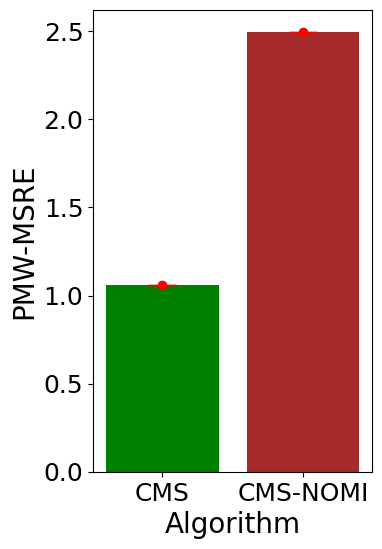}\label{fig:pm-msre:NOMI-Chicago16Small}}
	\subfloat[][Chic1610Mil]{\includegraphics[width=0.15\textwidth]{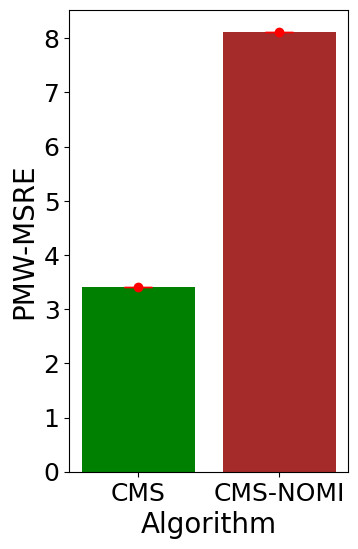}\label{fig:pm-msre:NOMI-Chicago1610Mil}}
	\subfloat[][Chicago16]{\includegraphics[width=0.15\textwidth]{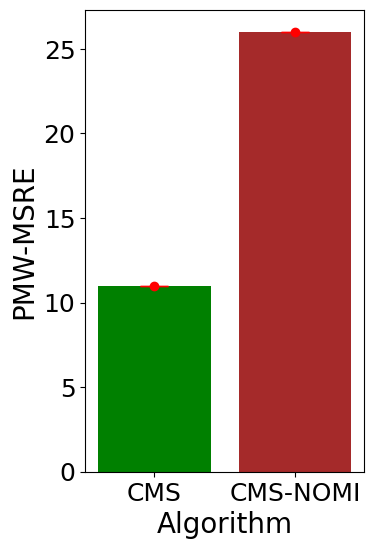}\label{fig:pm-msre:NOMI-Chicago16}}
	\caption{Postmortem MSRE w/o minimal increment}
	\label{fig:NOMI-pm-msre}
	%
	\subfloat[][ny19A]{\includegraphics[width=0.15\textwidth]{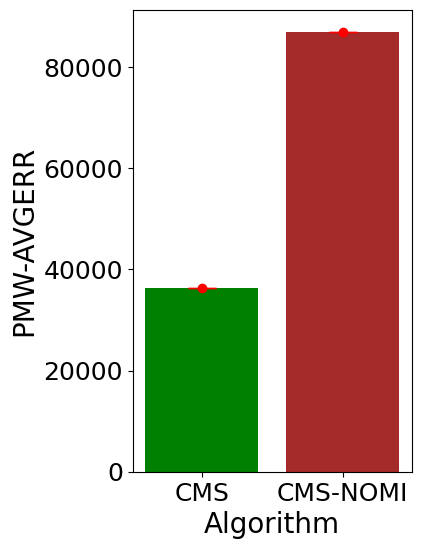}\label{fig:pm-avgerr:NOMI-ny19A}}
	\subfloat[][SJ14]{\includegraphics[width=0.15\textwidth]{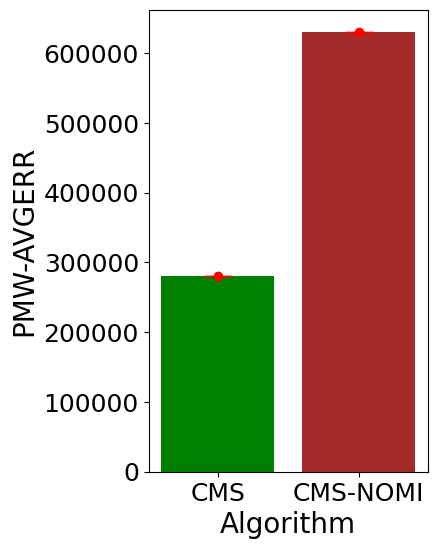}\label{fig:pm-avgerr:NOMI-SJ14}}
	\subfloat[][Chicago15]{\includegraphics[width=0.15\textwidth]{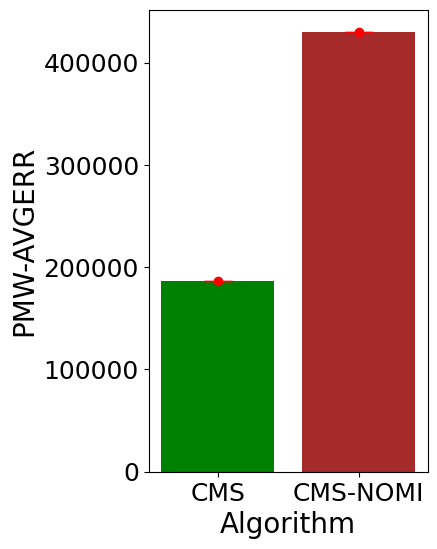}\label{fig:pm-avgerr:NOMI-Chicago15}}
	\subfloat[][Chic16Small]{\includegraphics[width=0.15\textwidth]{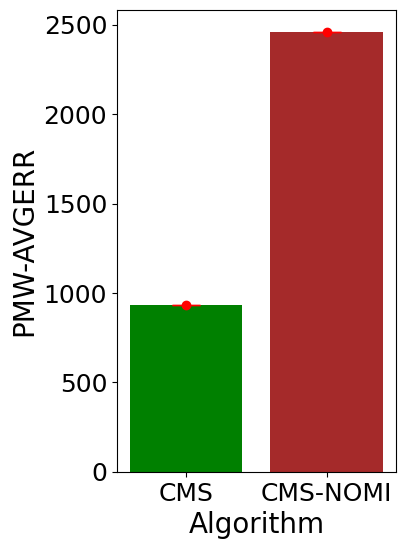}\label{fig:pm-avgerr:NOMI-Chicago16Small}}
	\subfloat[][Chic1610Mil]{\includegraphics[width=0.15\textwidth]{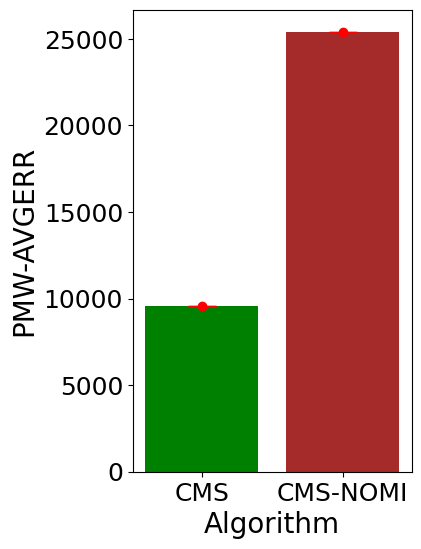}\label{fig:pm-avgerr:NOMI-Chicago1610Mil}}
	\subfloat[][Chicago16]{\includegraphics[width=0.15\textwidth]{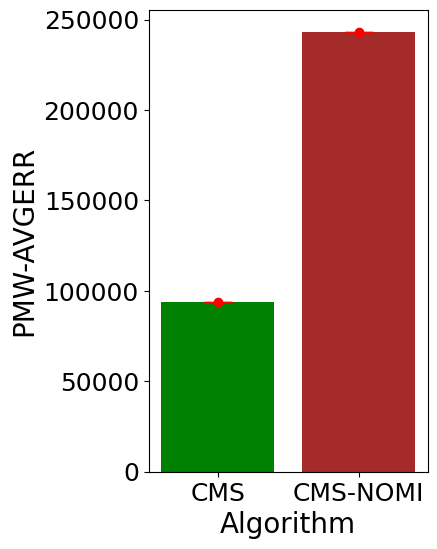}\label{fig:pm-avgerr:NOMI-Chicago16}}
	\caption{Postmortem AVGERR w/o minimal increment}
	\label{fig:NOMI-pm-avgerr}
	%
	\subfloat[][ny19A]{\includegraphics[width=0.15\textwidth]{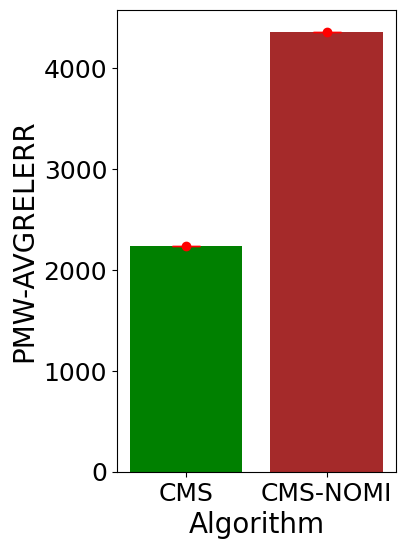}\label{fig:pm-avgrelerr:NOMI-ny19A}}
	\subfloat[][SJ14]{\includegraphics[width=0.15\textwidth]{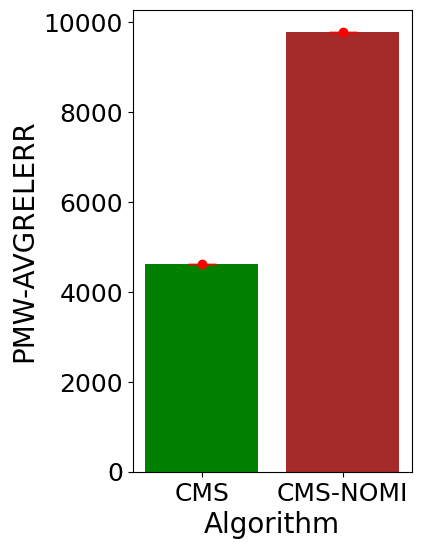}\label{fig:pm-avgrelerr:NOMI-SJ14}}
	\subfloat[][Chicago15]{\includegraphics[width=0.15\textwidth]{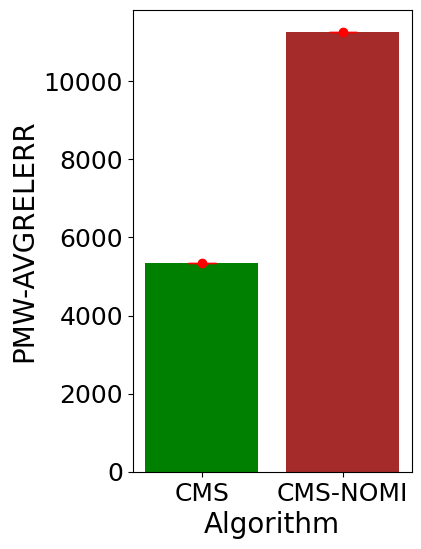}\label{fig:pm-avgrelerr:NOMI-Chicago15}}
	\subfloat[][Chic16Small]{\includegraphics[width=0.15\textwidth]{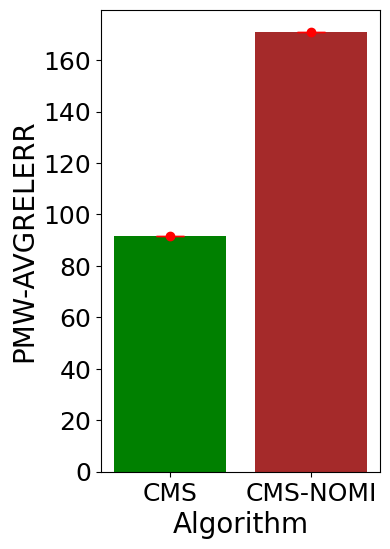}\label{fig:pm-avgrelerr:NOMI-Chicago16Small}}
	\subfloat[][Chic1610Mil]{\includegraphics[width=0.15\textwidth]{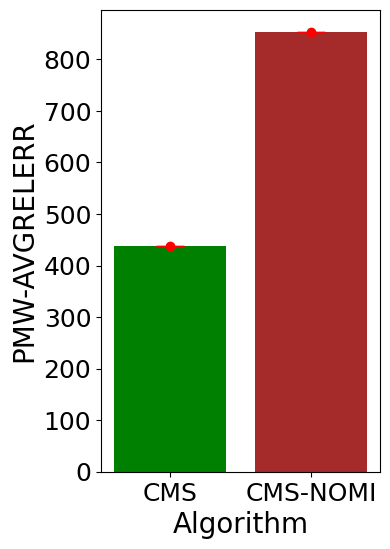}\label{fig:pm-avgrelerr:NOMI-Chicago1610Mil}}
	\subfloat[][Chicago16]{\includegraphics[width=0.15\textwidth]{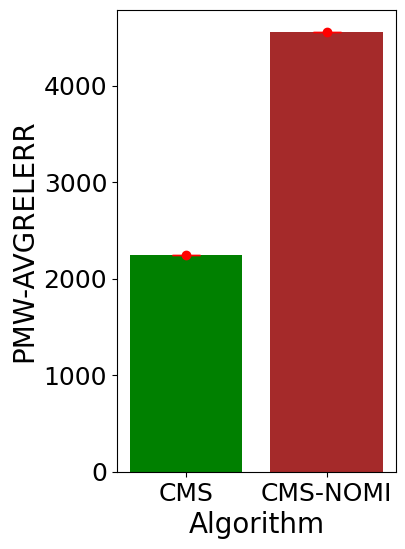}\label{fig:pm-avgrelerr:NOMI-Chicago16}}
	\caption{Postmortem AVGRELERR  w/o minimal increment}
	\label{fig:NOMI-pm-avgrelerr}
\end{figure*}

On the other hand, in terms of errors, the minimal increment increases the error by more than a factor of $2$ for the various metrics, as shown in Figures~\ref{fig:NOMI-oa-msre}--\ref{fig:NOMI-pm-avgrelerr}.
Given the very low memory requirement of CMS, it may make sense to avoid the minimal increment optimization for write only cases, and compensate for this by allocating more counters in each array.
However, when memory os very tight, or when the workload includes mostly reads, then the minimal increment is very effective.

\subsection{Space Tradoffs for NitroCuckoo}
\label{sec:cuckoo-deepdive}

In this section, we explore what happens when NitroCuckoo is allocated with a $p$ fraction of the memory, where $p$ is the sampling probability used by the Nitro optimization.
Specifically, we compare Cuckoo and NitroCuckoo when configured as before, with the NitroCuckoo algorithm whose Cuckoo filter table is allocated with $p$ times the number of entries, nicknamed NC-SMALL.
In our case, $p=0.01$, which means NC-SMALL's table is $100$ times smaller.

\begin{figure*}[t]
	\centering
	\subfloat[][ny19A]{\includegraphics[width=0.15\textwidth]{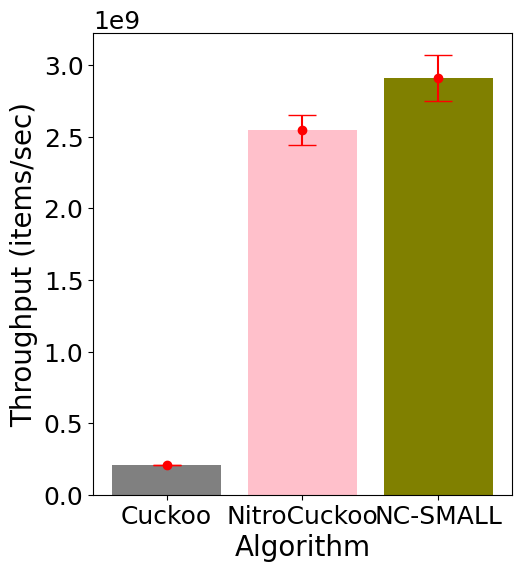}\label{fig:thpt-write:NITRO-ny19A}}
	\subfloat[][SJ14]{\includegraphics[width=0.15\textwidth]{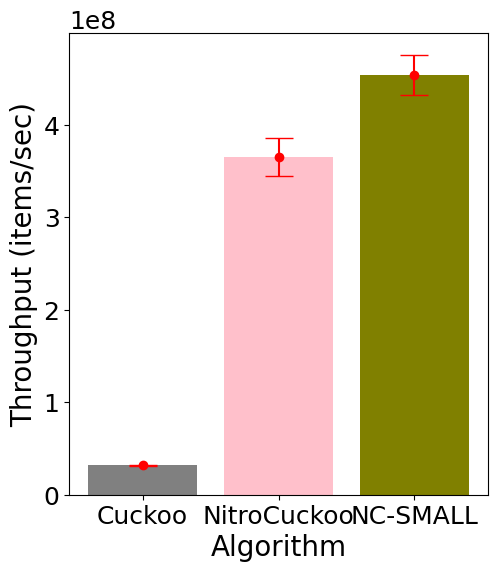}\label{fig:thpt-write:NITRO-SJ14}}
	\subfloat[][Chicago15]{\includegraphics[width=0.15\textwidth]{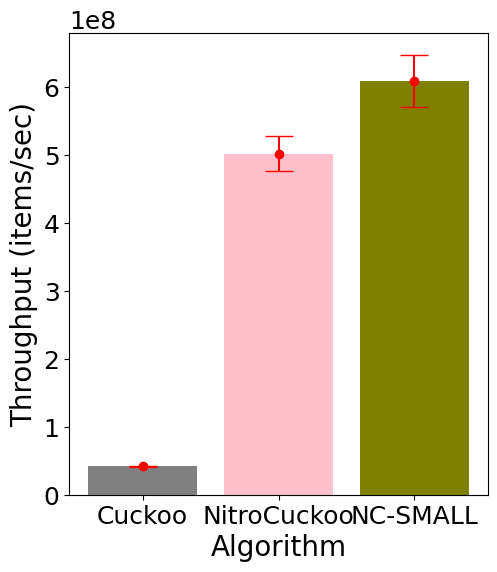}\label{fig:thpt-write:NITRO-Chicago15}}
	\subfloat[][Chic16Small]{\includegraphics[width=0.15\textwidth]{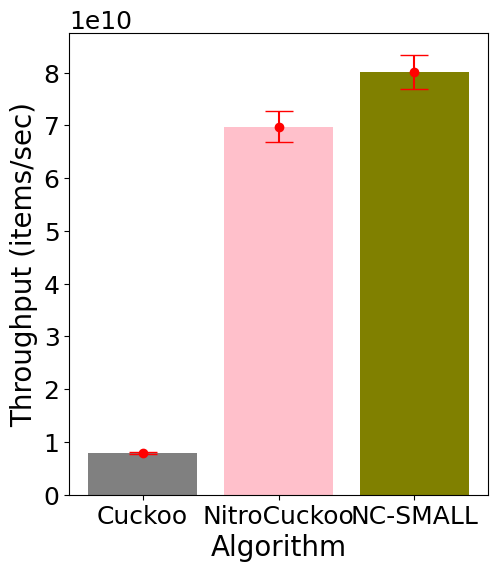}\label{fig:thpt-write:NITRO-Chicago16Small}}
	\subfloat[][Chic1610Mil]{\includegraphics[width=0.15\textwidth]{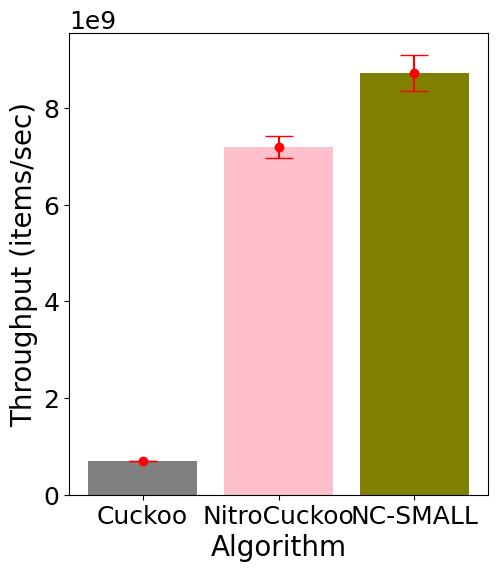}\label{fig:thpt-write:NITRO-Chicago1610Mil}}
	\subfloat[][Chicago16]{\includegraphics[width=0.15\textwidth]{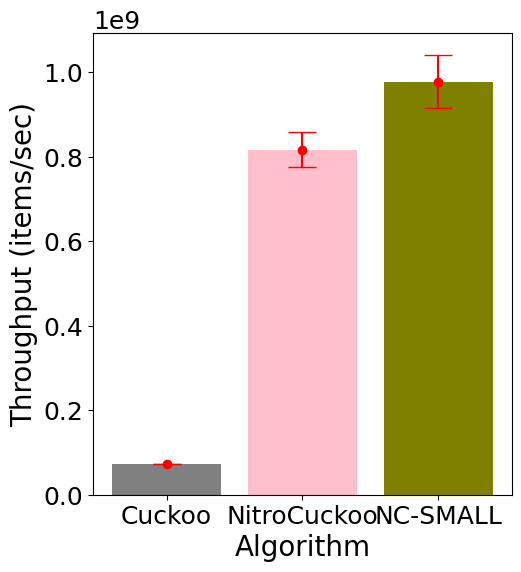}\label{fig:thpt-write:NITRO-Chicago16}}
	\caption{Throughput results for the write only test with different Cockoo configurations}
	\label{fig:NITRO-thpt-write}
	%
	\centering
	\subfloat[][ny19A]{\includegraphics[width=0.15\textwidth]{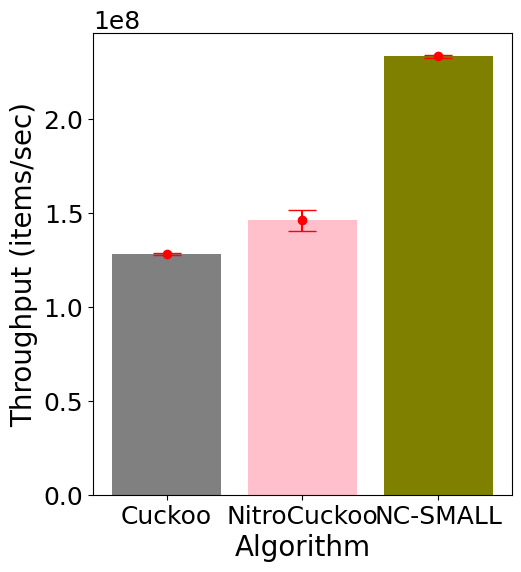}\label{fig:thpt-rw:NITRO-ny19A}}
	\subfloat[][SJ14]{\includegraphics[width=0.15\textwidth]{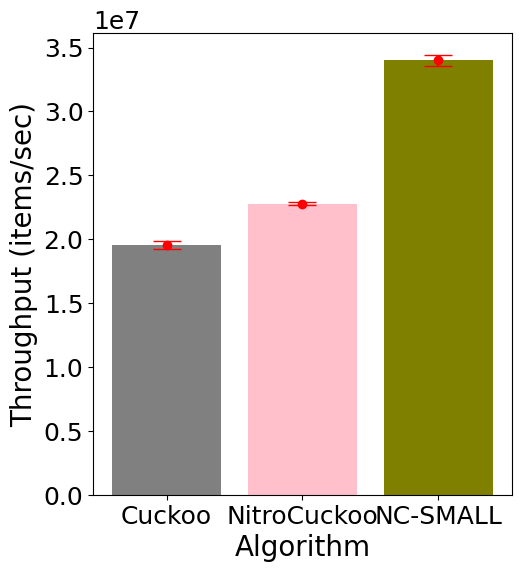}\label{fig:thpt-rw:NITRO-SJ14}}
	\subfloat[][Chicago15]{\includegraphics[width=0.15\textwidth]{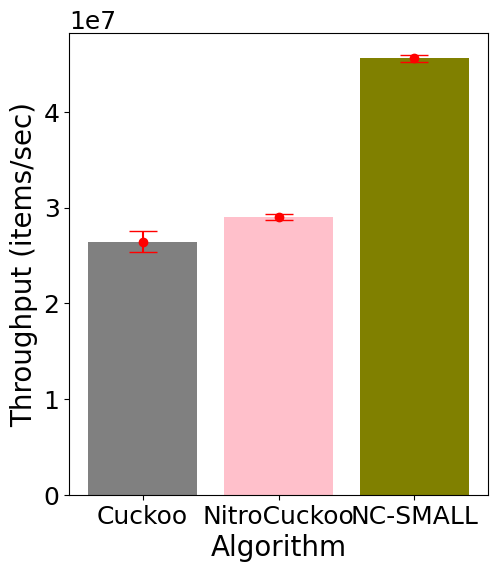}\label{fig:thpt-rw:CNITRO-hicago15}}
	\subfloat[][Chic16Small]{\includegraphics[width=0.15\textwidth]{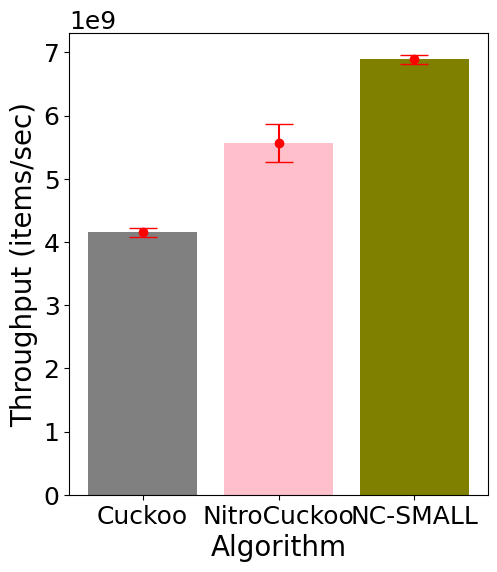}\label{fig:thpt-rw:NITRO-Chicago16Small}}
	\subfloat[][Chic1610Mil]{\includegraphics[width=0.15\textwidth]{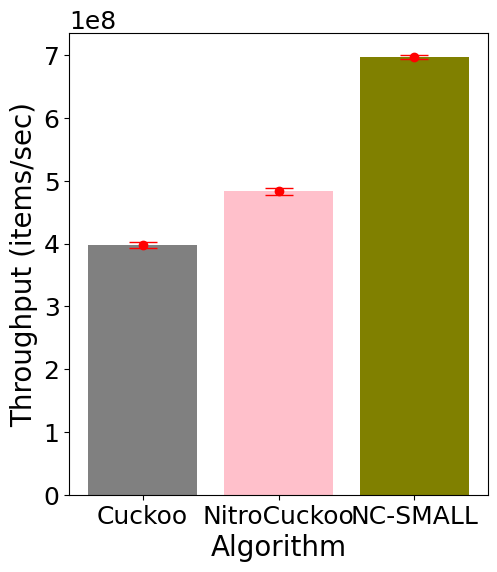}\label{fig:thpt-rw:NITRO-Chicago1610Mil}}
	\subfloat[][Chicago16]{\includegraphics[width=0.15\textwidth]{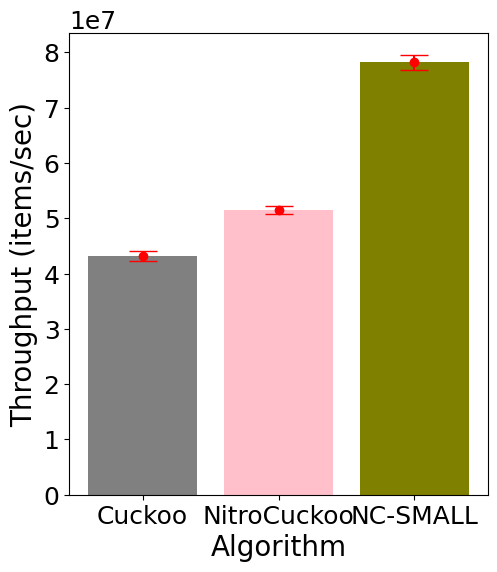}\label{fig:thpt-rw:NITRO-Chicago16}}
	\caption{Throughput results for the write-read test with different Cockoo configurations}
	\label{fig:NITRO-thpt-rw}
	%
	\centering
	\subfloat[][ny19A]{\includegraphics[width=0.15\textwidth]{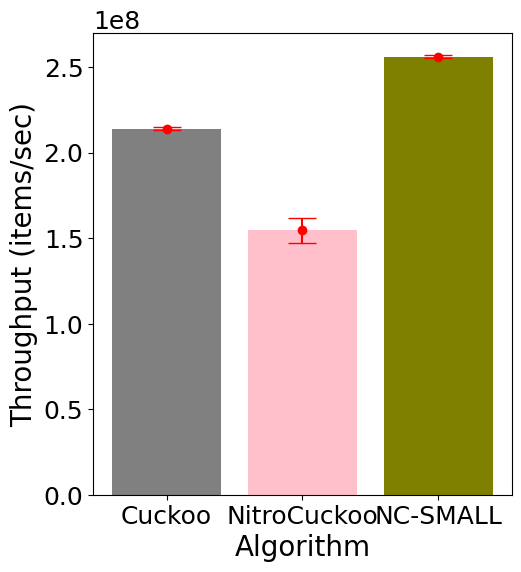}\label{fig:thpt-read:NITRO-ny19A}}
	\subfloat[][SJ14]{\includegraphics[width=0.15\textwidth]{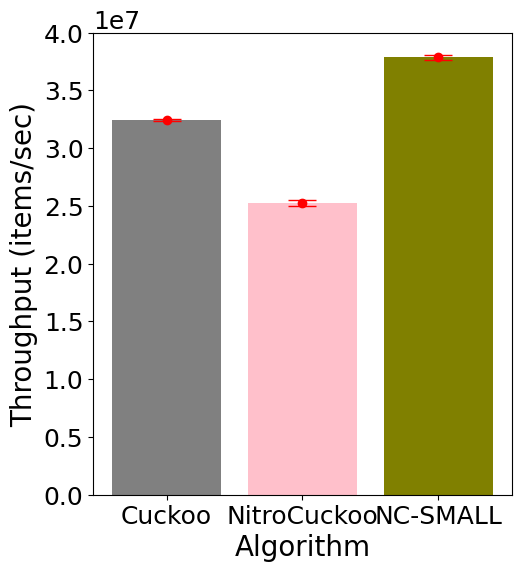}\label{fig:thpt-read:NITRO-SJ14}}
	\subfloat[][Chicago15]{\includegraphics[width=0.15\textwidth]{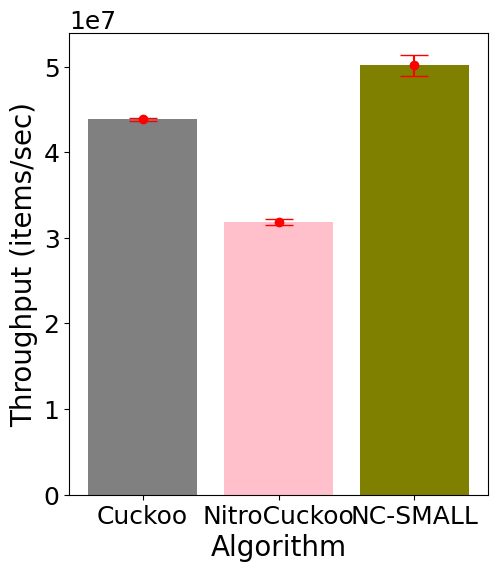}\label{fig:thpt-read:NITRO-Chicago15}}
	\subfloat[][Chic16Small]{\includegraphics[width=0.15\textwidth]{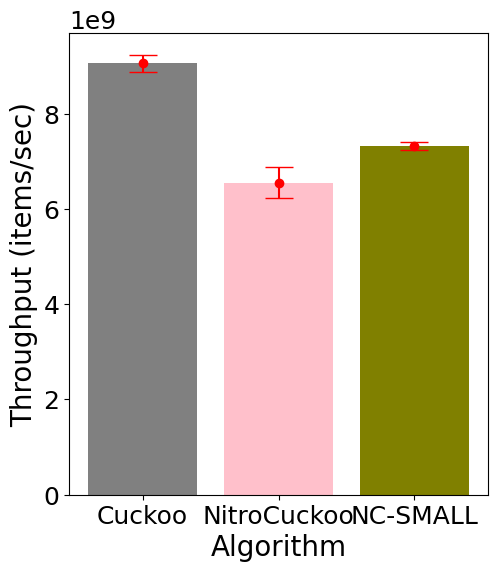}\label{fig:thpt-read:NITRO-Chicago16Small}}
	\subfloat[][Chic1610Mil]{\includegraphics[width=0.15\textwidth]{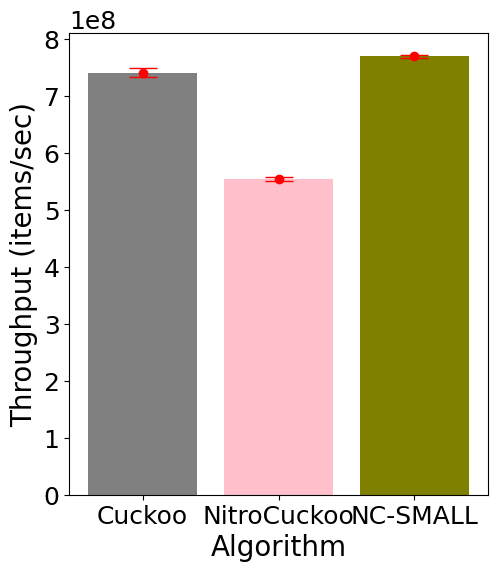}\label{fig:thpt-read:NITRO-Chicago1610Mil}}
	\subfloat[][Chicago16]{\includegraphics[width=0.15\textwidth]{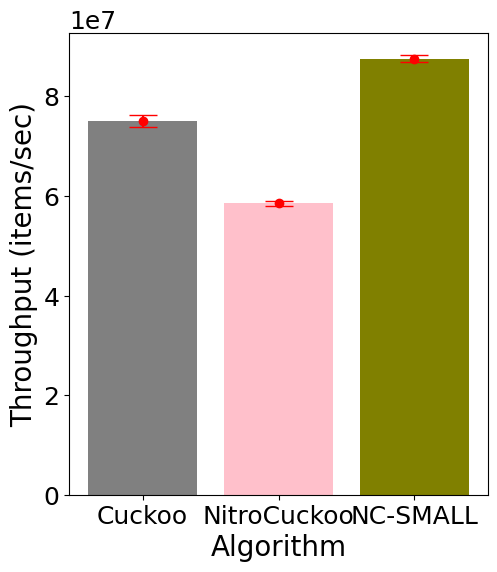}\label{fig:thpt-read:NITRO-Chicago16}}
	\caption{Throughput results for the read only test with different Cockoo configurations}
	\label{fig:NITRO-thpt-read}
\end{figure*}

As can be seen in Figure~\ref{fig:NITRO-thpt-write}, this gave a further $10$\% to $20$\% throughput improvement when compared to NitroCuckoo in the write only test, due to the smaller size of the table, which fits better in the hardware cache.
The improvement becomes even more significant in the case of the write-read test, as the impact of having a smaller size more than offsets the increased lookup time for non-existing items.
However, the difference between NC-SMALL and Cuckoo shrinks.

Finally, in the read only test, reported in Figure~\ref{fig:NITRO-thpt-read}, Cuckoo is significantly faster than NitroCuckoo as discussed above, but NC-SMALL is still the clear winner, meaning that the impact of fitting better in the hardware cache is stronger than the impact of the average longer query process when fewer items are found.

\begin{figure*}[t]
	\centering
	\subfloat[][ny19A]{\includegraphics[width=0.15\textwidth]{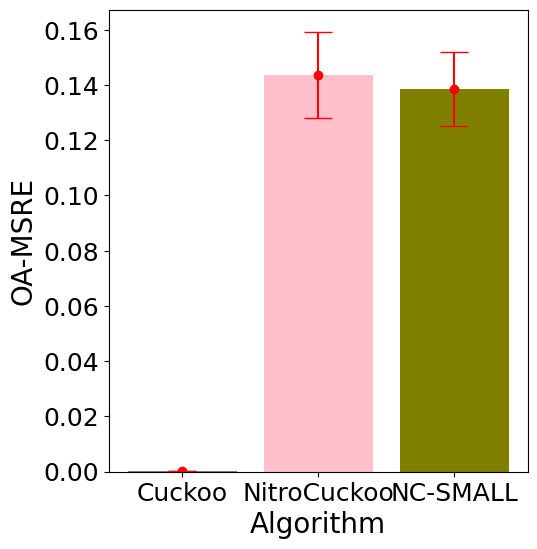}\label{fig:oa-msre:NITRO-ny19A}}
	\subfloat[][SJ14]{\includegraphics[width=0.15\textwidth]{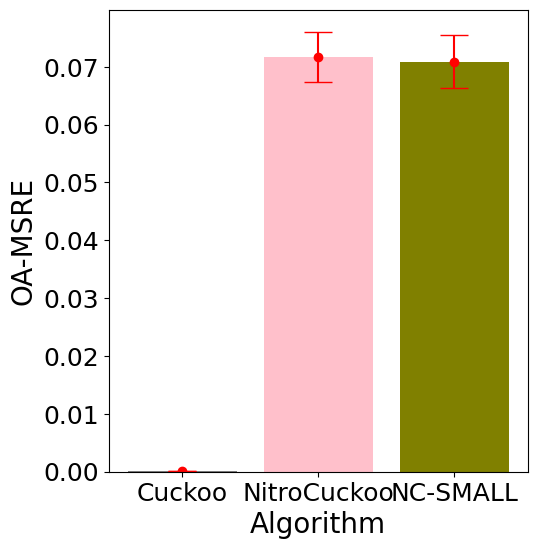}\label{fig:oa-msre:NITRO-SJ14}}
	\subfloat[][Chicago15]{\includegraphics[width=0.15\textwidth]{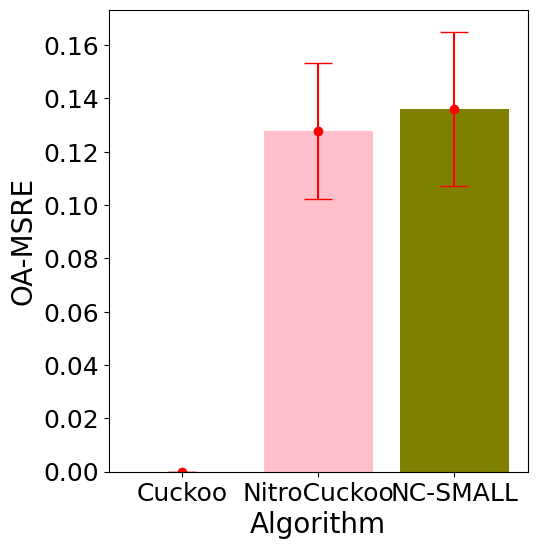}\label{fig:oa-msre:NITRO-Chicago15}}
	\subfloat[][Chic16Small]{\includegraphics[width=0.15\textwidth]{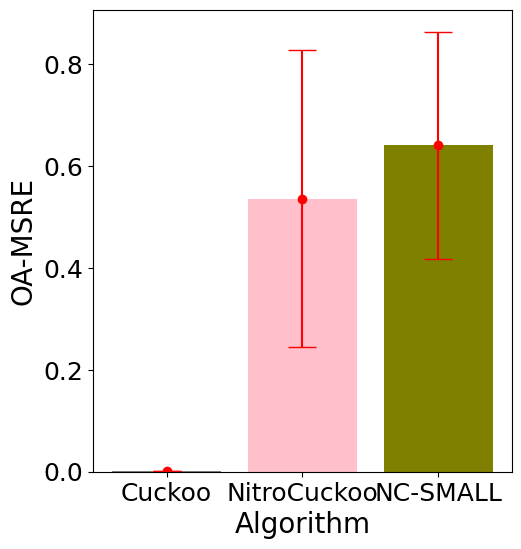}\label{fig:oa-msre:NITRO-Chicago16Small}}
	\subfloat[][Chic1610Mil]{\includegraphics[width=0.15\textwidth]{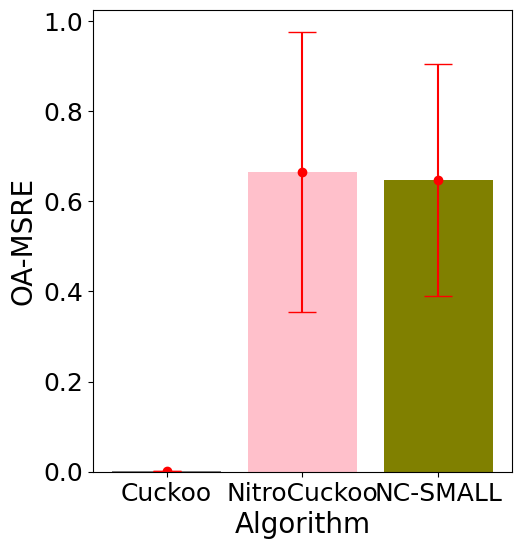}\label{fig:oa-msre:NITRO-Chicago1610Mil}}
	\subfloat[][Chicago16]{\includegraphics[width=0.15\textwidth]{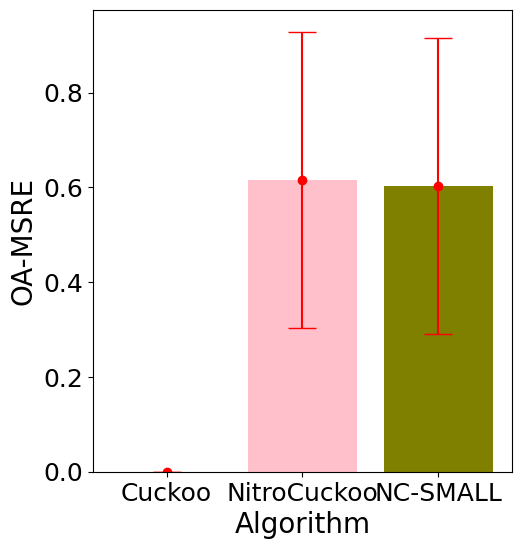}\label{fig:oa-msre:NITRO-Chicago16}}
	\caption{OA Arrival MSRE with different Cockoo configurations}
	\label{fig:NITRO-oa-msre}
	%
	\subfloat[][ny19A]{\includegraphics[width=0.15\textwidth]{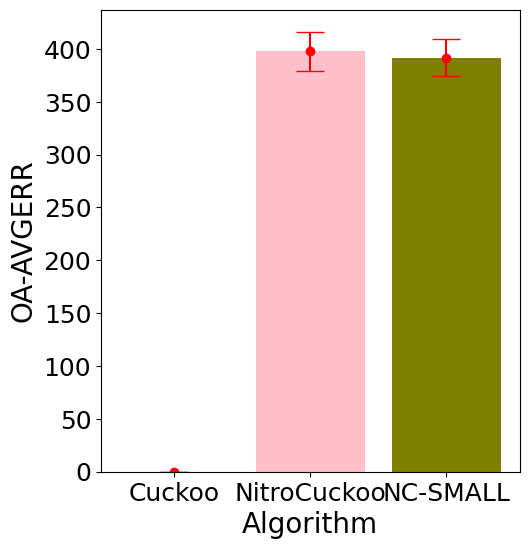}\label{fig:oa-avgerr:NITRO-ny19A}}
	\subfloat[][SJ14]{\includegraphics[width=0.15\textwidth]{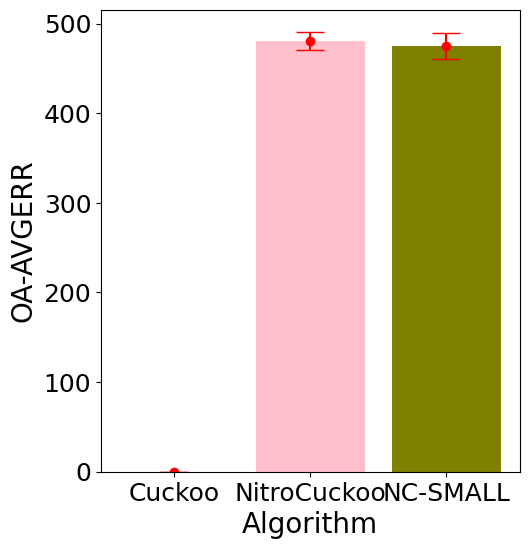}\label{fig:oa-avgerr:NITRO-SJ14}}
	\subfloat[][Chicago15]{\includegraphics[width=0.15\textwidth]{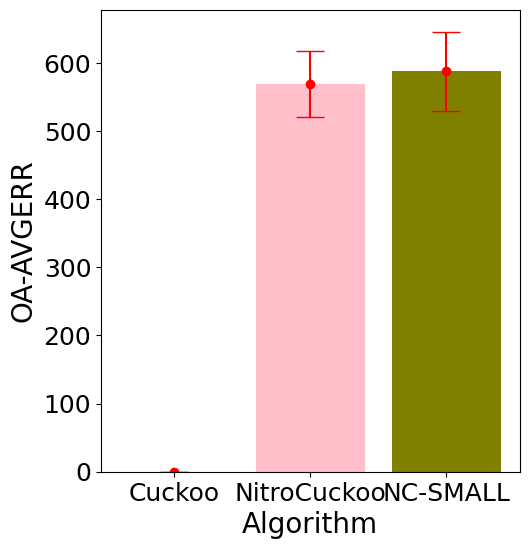}\label{fig:oa-avgerr:NITRO-Chicago15}}
	\subfloat[][Chic16Small]{\includegraphics[width=0.15\textwidth]{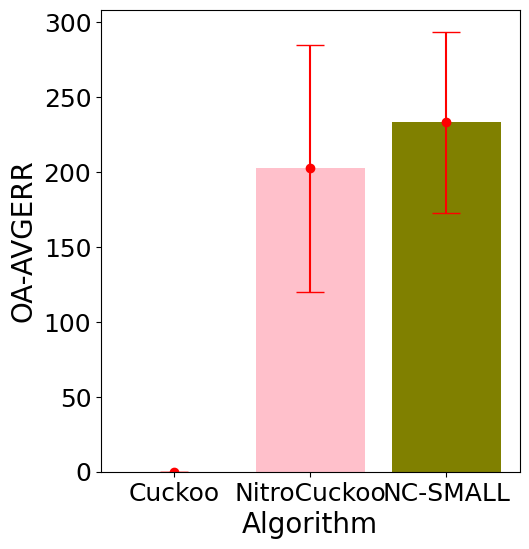}\label{fig:oa-avgerr:NITRO-Chicago16Small}}
	\subfloat[][Chic1610Mil]{\includegraphics[width=0.15\textwidth]{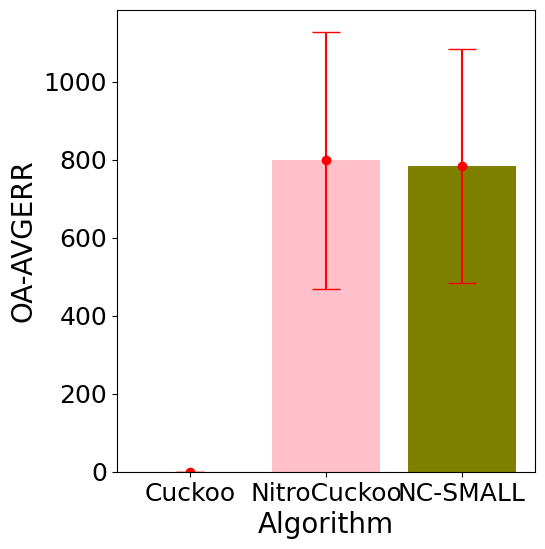}\label{fig:oa-avgerr:NITRO-Chicago1610Mil}}
	\subfloat[][Chicago16]{\includegraphics[width=0.15\textwidth]{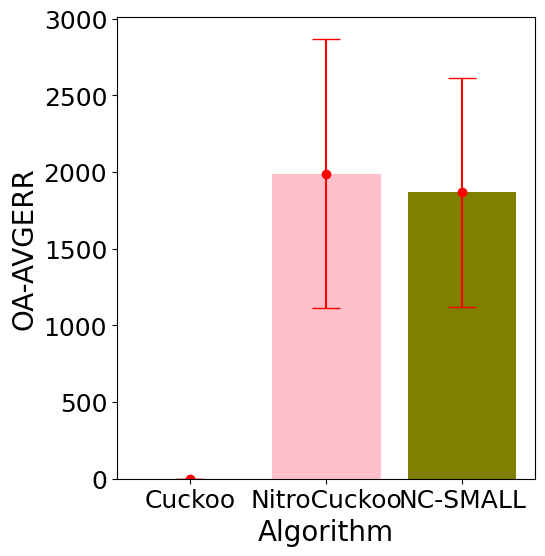}\label{fig:oa-avgerr:NITRO-Chicago16}}
	\caption{OA Arrival AVGERR with different Cockoo configurations}
	\label{fig:NITRO-oa-avgerr}
	%
	\subfloat[][ny19A]{\includegraphics[width=0.15\textwidth]{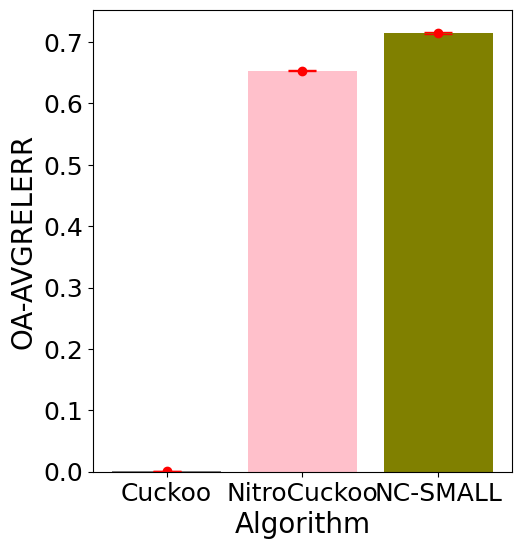}\label{fig:oa-avgrelerr:NITRO-ny19A}}
	\subfloat[][SJ14]{\includegraphics[width=0.15\textwidth]{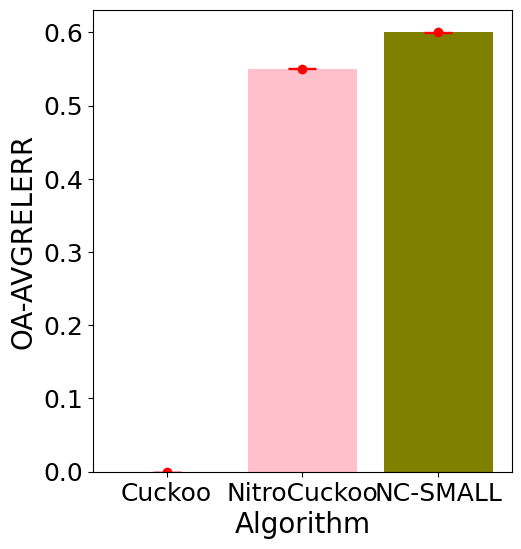}\label{fig:oa-avgrelerr:NITRO-SJ14}}
	\subfloat[][Chicago15]{\includegraphics[width=0.15\textwidth]{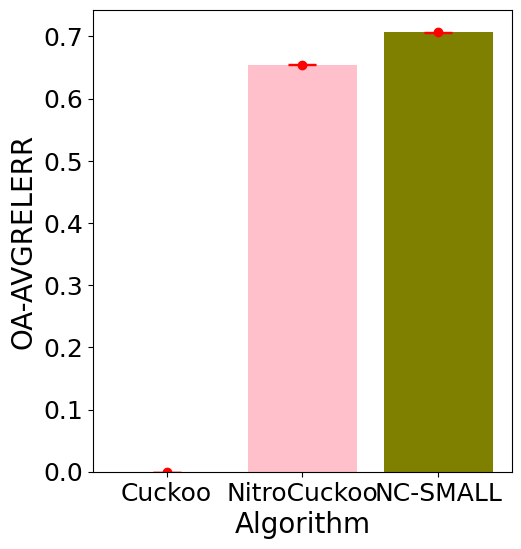}\label{fig:oa-avgrelerr:NITRO-Chicago15}}
	\subfloat[][Chic16Small]{\includegraphics[width=0.15\textwidth]{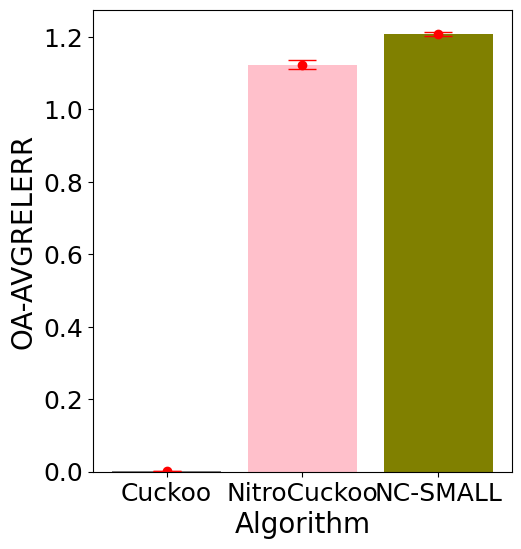}\label{fig:oa-avgrelerr:NITRO-Chicago16Small}}
	\subfloat[][Chic1610Mil]{\includegraphics[width=0.15\textwidth]{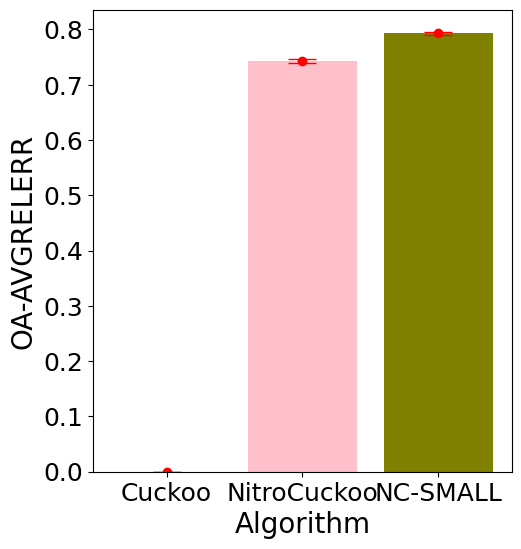}\label{fig:oa-avgrelerr:NITRO-Chicago1610Mil}}
	\subfloat[][Chicago16]{\includegraphics[width=0.15\textwidth]{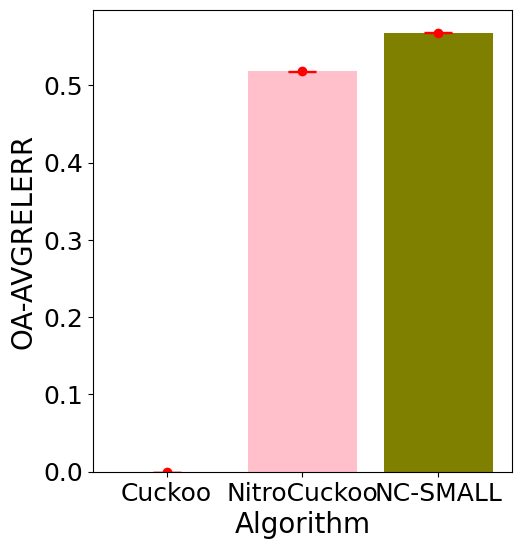}\label{fig:oa-avgrelerr:NITRO-Chicago16}}
	\caption{OA Arrival AVGRELERR with different Cockoo configurations}
	\label{fig:NITRO-oa-avgrelerr}
\end{figure*}

\begin{figure*}[t]
	\centering
	\subfloat[][ny19A]{\includegraphics[width=0.15\textwidth]{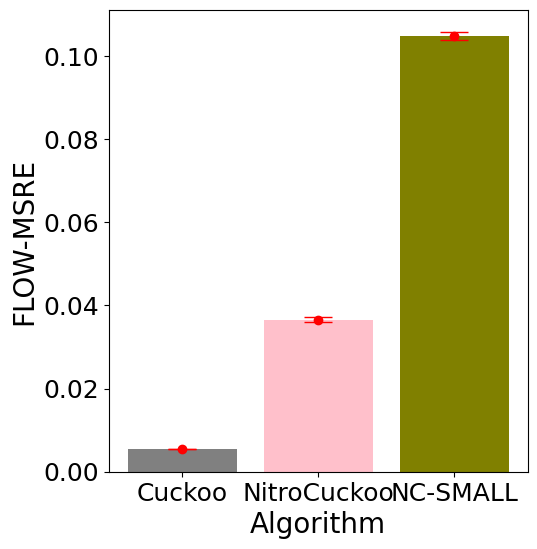}\label{fig:flow-msre:NITRO-ny19A}}
	\subfloat[][SJ14]{\includegraphics[width=0.15\textwidth]{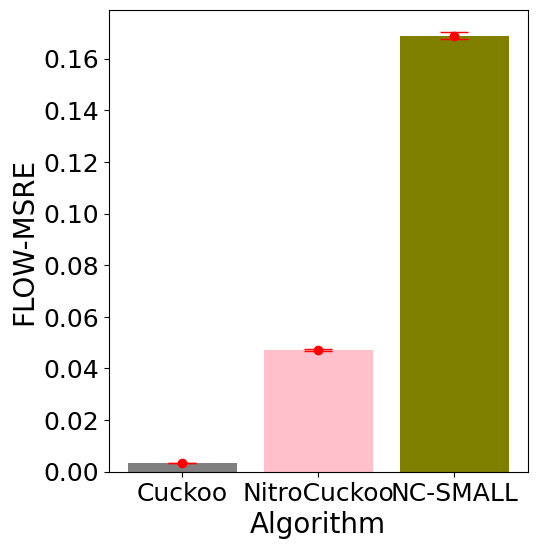}\label{fig:flow-msre:NITRO-SJ14}}
	\subfloat[][Chicago15]{\includegraphics[width=0.15\textwidth]{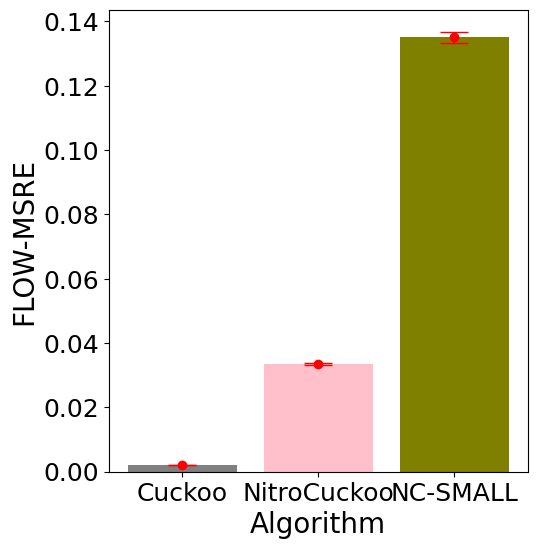}\label{fig:flow-msre:NITRO-Chicago15}}
	\subfloat[][Chic16Small]{\includegraphics[width=0.15\textwidth]{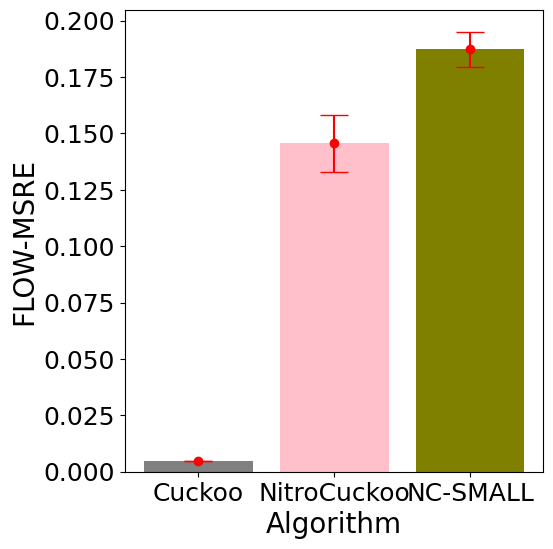}\label{fig:flow-msre:NITRO-Chicago16Small}}
	\subfloat[][Chic1610Mil]{\includegraphics[width=0.15\textwidth]{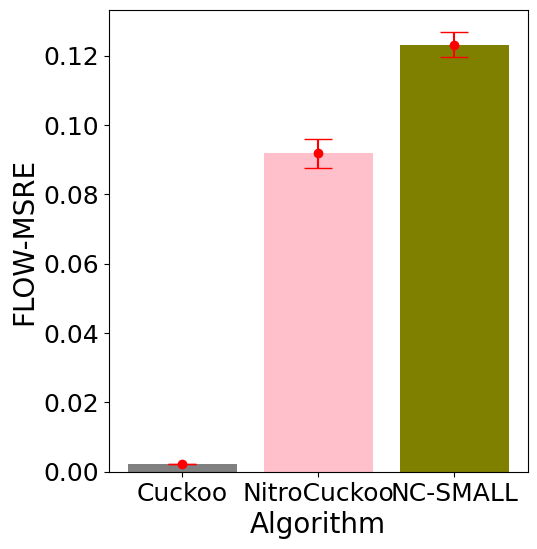}\label{fig:flow-msre:NITRO-Chicago1610Mil}}
	\subfloat[][Chicago16]{\includegraphics[width=0.15\textwidth]{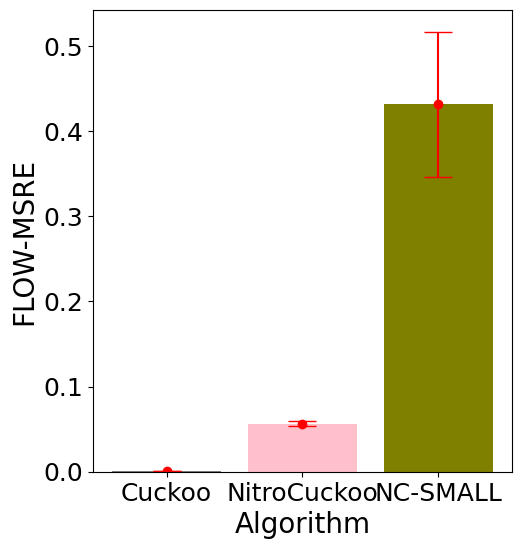}\label{fig:flow-msre:NITRO-Chicago16}}
	\caption{Per Flow MSRE with different Cockoo configurations}
	\label{fig:NITRO-flow-msre}
	%
	\subfloat[][ny19A]{\includegraphics[width=0.15\textwidth]{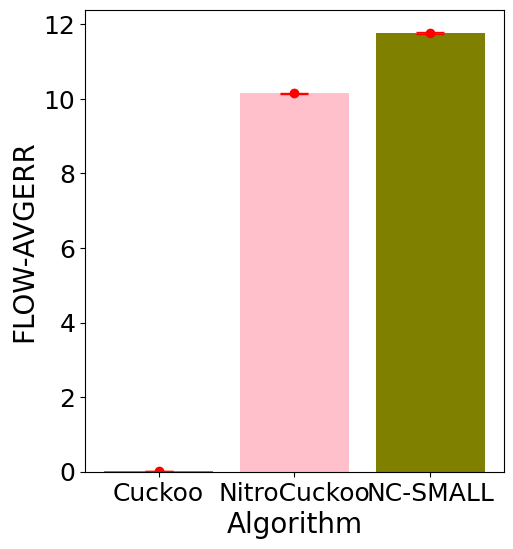}\label{fig:flow-avgerr:NITRO-ny19A}}
	\subfloat[][SJ14]{\includegraphics[width=0.15\textwidth]{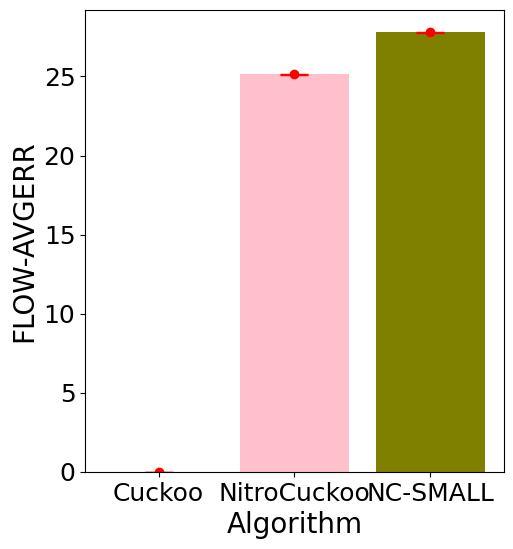}\label{fig:flow-avgerr:NITRO-SJ14}}
	\subfloat[][Chicago15]{\includegraphics[width=0.15\textwidth]{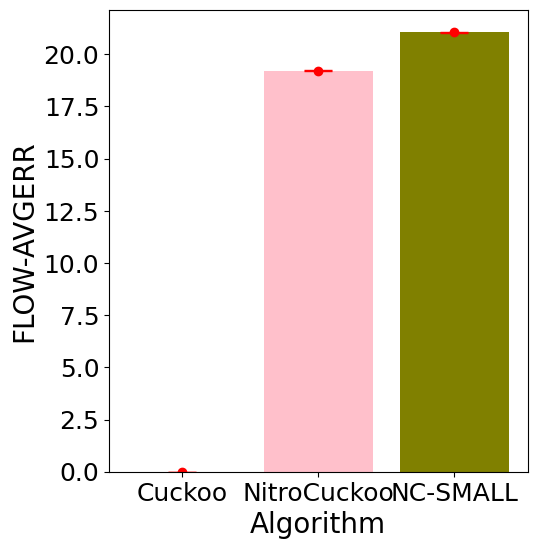}\label{fig:flow-avgerr:NITRO-Chicago15}}
	\subfloat[][Chic16Small]{\includegraphics[width=0.15\textwidth]{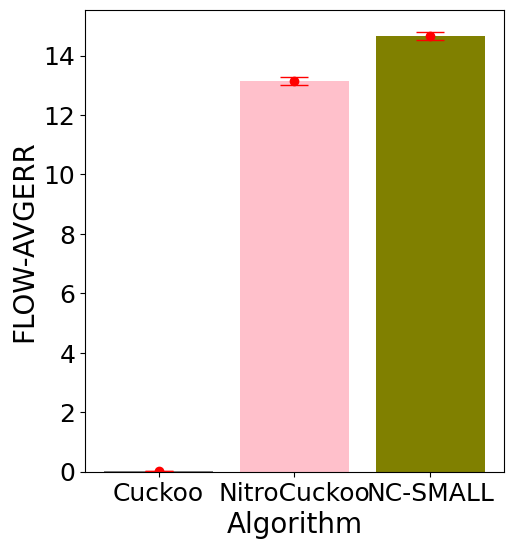}\label{fig:flow-avgerr:NITRO-Chicago16Small}}
	\subfloat[][Chic1610Mil]{\includegraphics[width=0.15\textwidth]{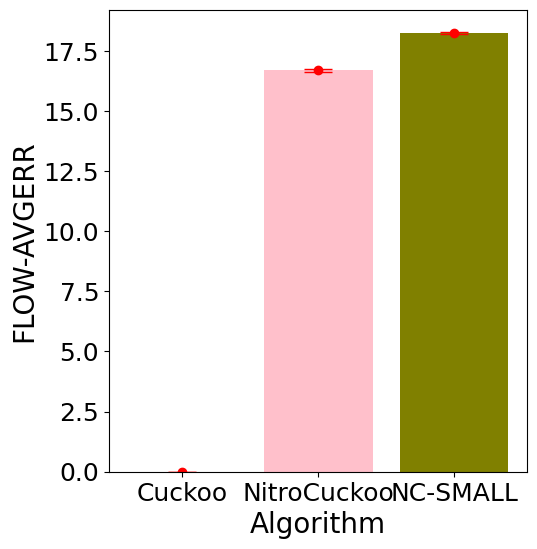}\label{fig:flow-avgerr:NITRO-Chicago1610Mil}}
	\subfloat[][Chicago16]{\includegraphics[width=0.15\textwidth]{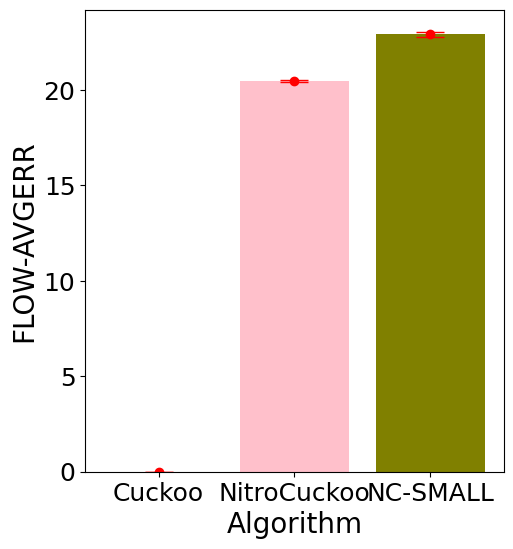}\label{fig:flow-avgerr:NITRO-Chicago16}}
	\caption{Per FLOW AVGERR with different Cockoo configurations}
	\label{fig:NITRO-flow-avgerr}
	%
	\subfloat[][ny19A]{\includegraphics[width=0.15\textwidth]{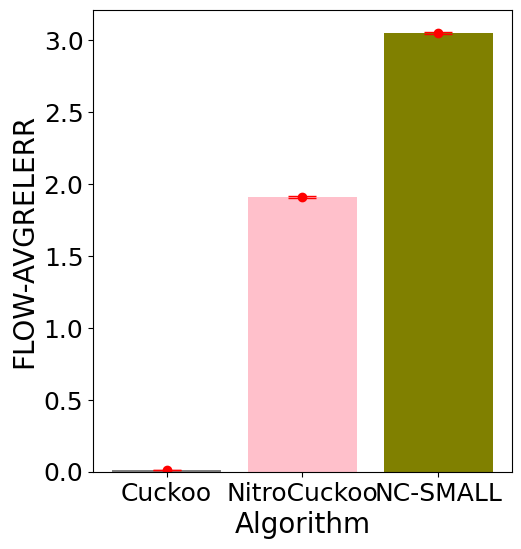}\label{fig:flow-avgrelerr:NITRO-ny19A}}
	\subfloat[][SJ14]{\includegraphics[width=0.15\textwidth]{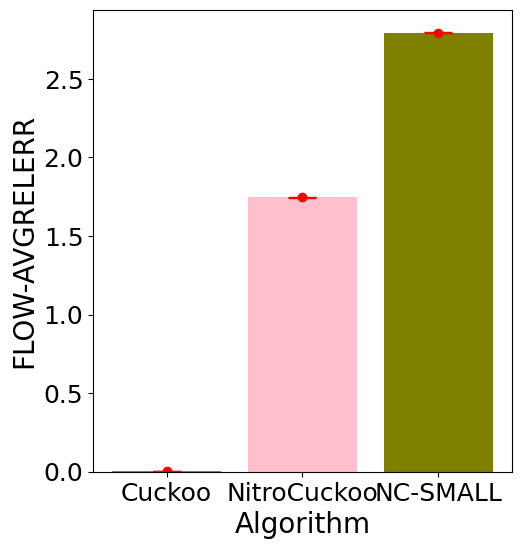}\label{fig:flow-avgrelerr:NITRO-SJ14}}
	\subfloat[][Chicago15]{\includegraphics[width=0.15\textwidth]{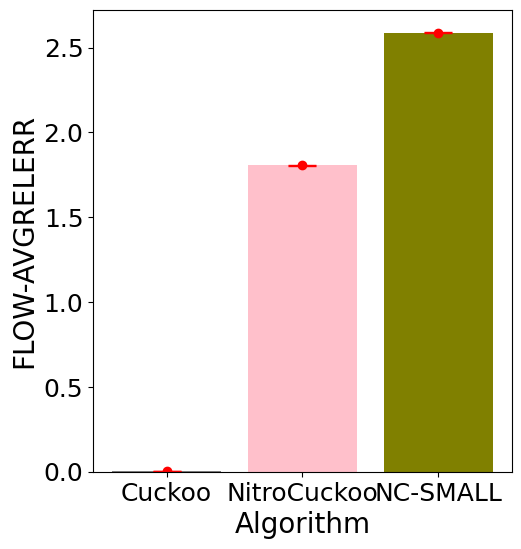}\label{fig:flow-avgrelerr:NITRO-Chicago15}}
	\subfloat[][Chic16Small]{\includegraphics[width=0.15\textwidth]{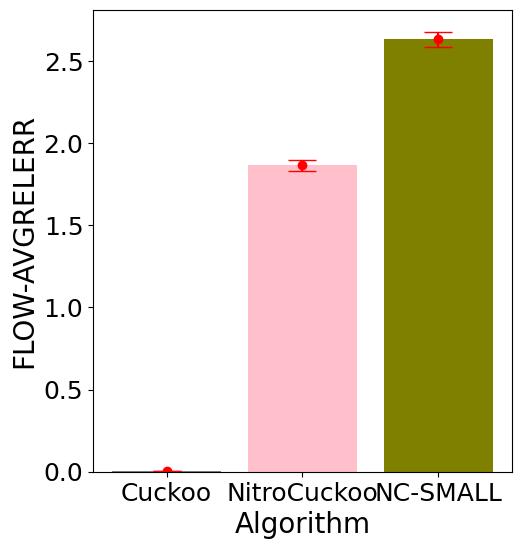}\label{fig:flow-avgrelerr:NITRO-Chicago16Small}}
	\subfloat[][Chic1610Mil]{\includegraphics[width=0.15\textwidth]{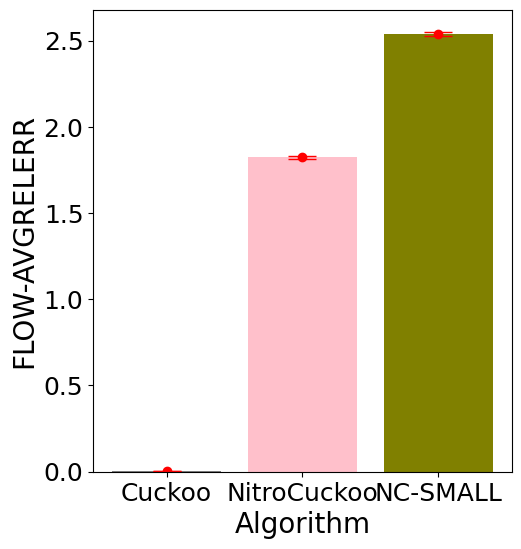}\label{fig:flow-avgrelerr:NITRO-Chicago1610Mil}}
	\subfloat[][Chicago16]{\includegraphics[width=0.15\textwidth]{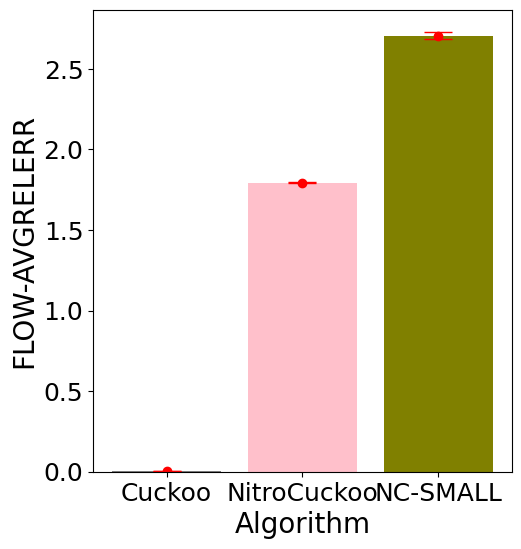}\label{fig:flow-avgrelerr:NITRO-Chicago16}}
	\caption{Per FLOW AVGRELERR with different Cockoo configurations}
	\label{fig:NITRO-flow-avgrelerr}
\end{figure*}

\begin{figure*}[t]
	\centering
	\subfloat[][ny19A]{\includegraphics[width=0.15\textwidth]{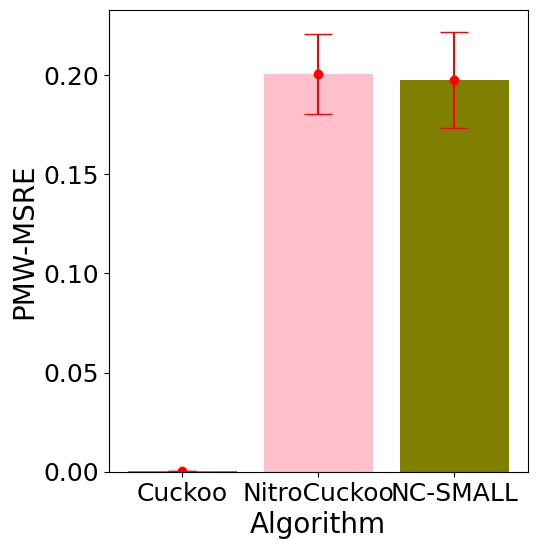}\label{fig:pm-msre:NITRO-ny19A}}
	\subfloat[][SJ14]{\includegraphics[width=0.15\textwidth]{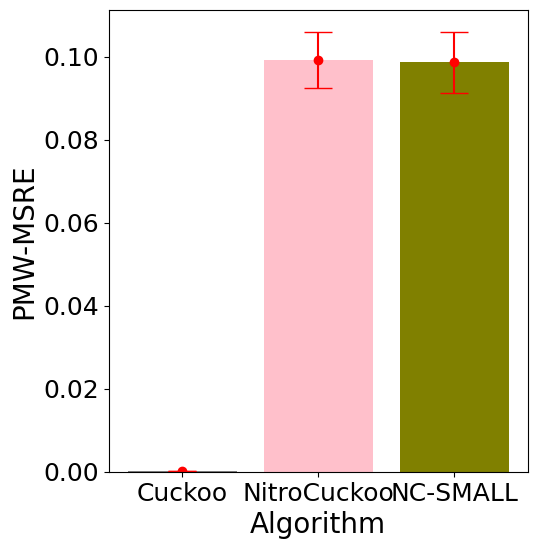}\label{fig:pm-msre:NITRO-SJ14}}
	\subfloat[][Chicago15]{\includegraphics[width=0.15\textwidth]{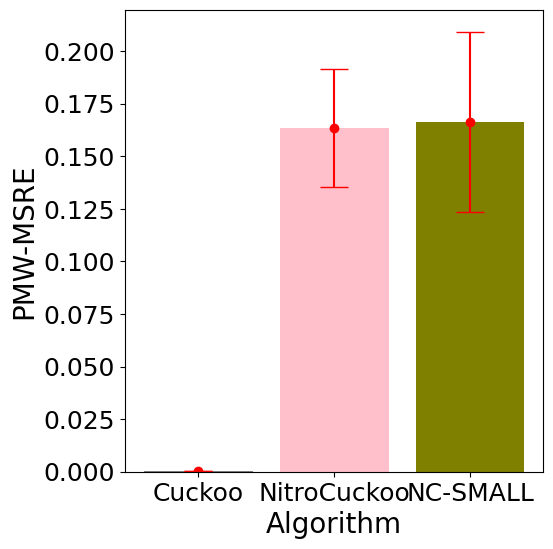}\label{fig:pm-msre:NITRO-Chicago15}}
	\subfloat[][Chic16Small]{\includegraphics[width=0.15\textwidth]{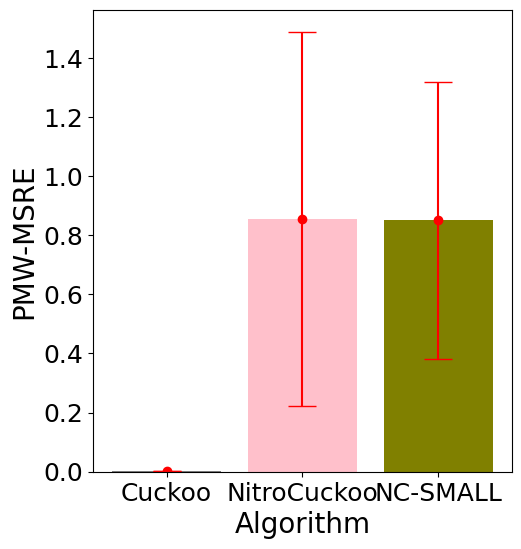}\label{fig:pm-msre:NITRO-Chicago16Small}}
	\subfloat[][Chic1610Mil]{\includegraphics[width=0.15\textwidth]{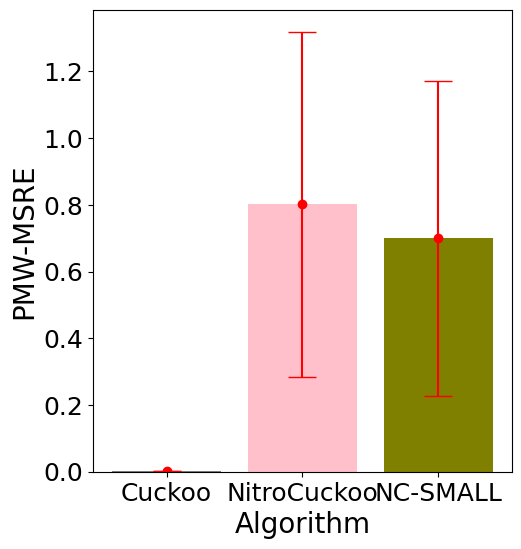}\label{fig:pm-msre:NITRO-Chicago1610Mil}}
	\subfloat[][Chicago16]{\includegraphics[width=0.15\textwidth]{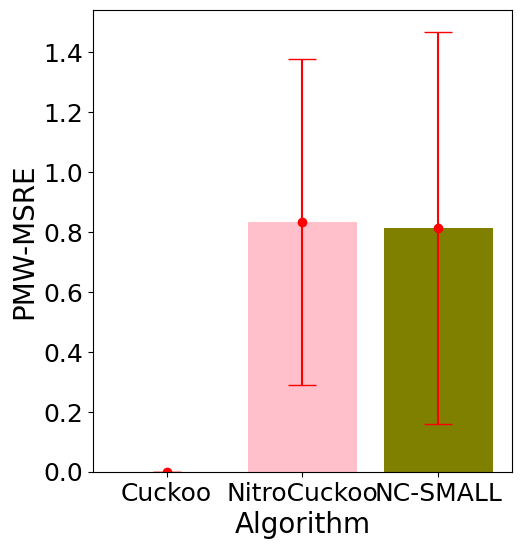}\label{fig:pm-msre:NITRO-Chicago16}}
	\caption{Postmortem MSRE with different Cockoo configurations}
	\label{fig:NITRO-pm-msre}
	%
	\subfloat[][ny19A]{\includegraphics[width=0.15\textwidth]{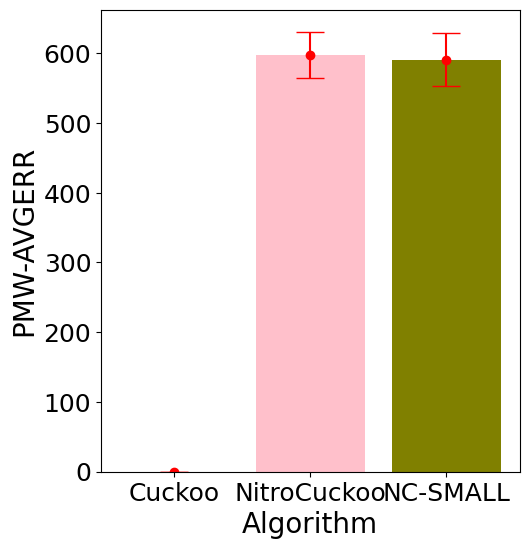}\label{fig:pm-avgerr:NITRO-ny19A}}
	\subfloat[][SJ14]{\includegraphics[width=0.15\textwidth]{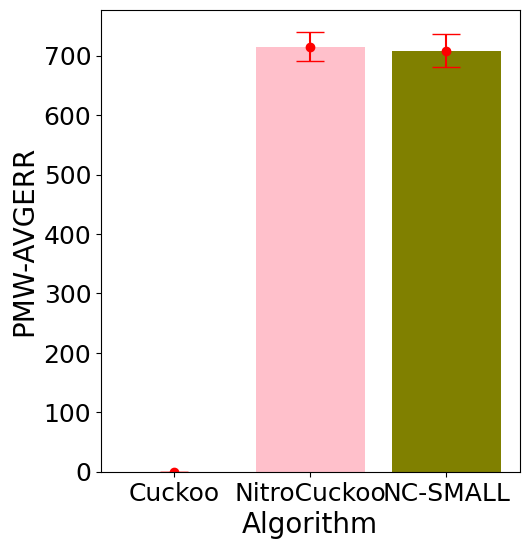}\label{fig:pm-avgerr:NITRO-SJ14}}
	\subfloat[][Chicago15]{\includegraphics[width=0.15\textwidth]{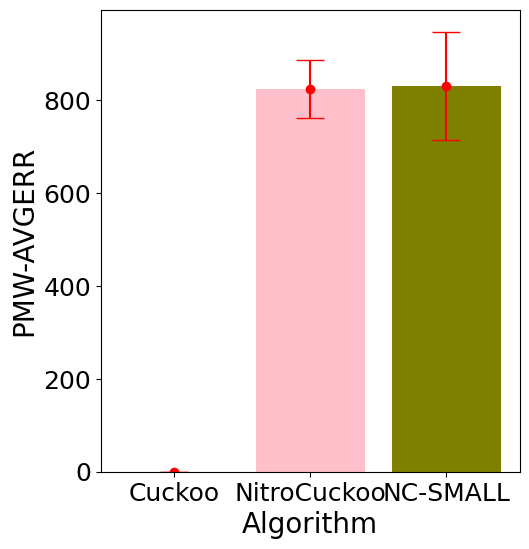}\label{fig:pm-avgerr:NITRO-Chicago15}}
	\subfloat[][Chic16Small]{\includegraphics[width=0.15\textwidth]{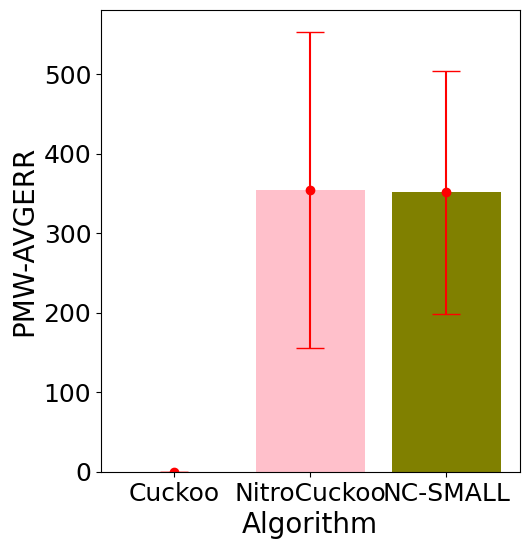}\label{fig:pm-avgerr:NITRO-Chicago16Small}}
	\subfloat[][Chic1610Mil]{\includegraphics[width=0.15\textwidth]{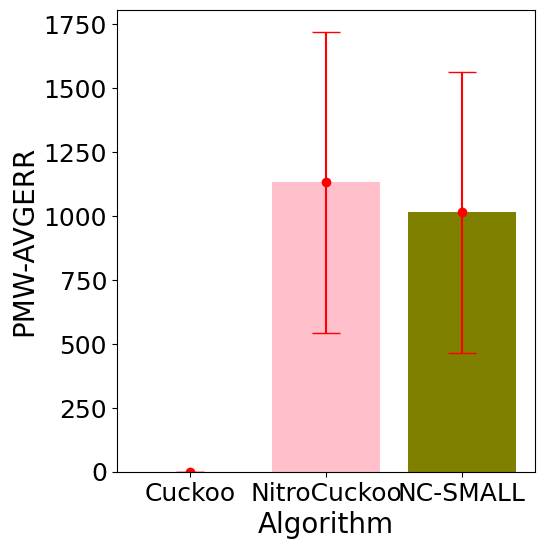}\label{fig:pm-avgerr:NITRO-Chicago1610Mil}}
	\subfloat[][Chicago16]{\includegraphics[width=0.15\textwidth]{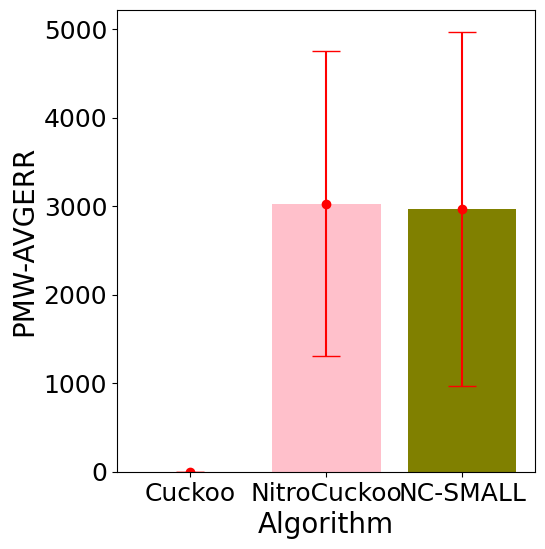}\label{fig:pm-avgerr:NITRO-Chicago16}}
	\caption{Postmortem AVGERR with different Cockoo configurations}
	\label{fig:NITRO-pm-avgerr}
	%
	\subfloat[][ny19A]{\includegraphics[width=0.15\textwidth]{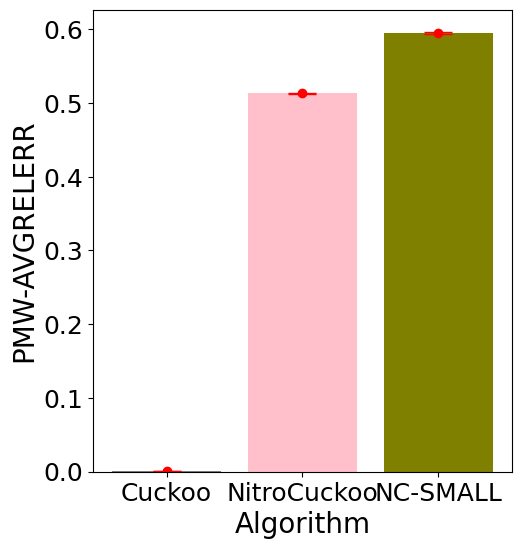}\label{fig:pm-avgrelerr:NITRO-ny19A}}
	\subfloat[][SJ14]{\includegraphics[width=0.15\textwidth]{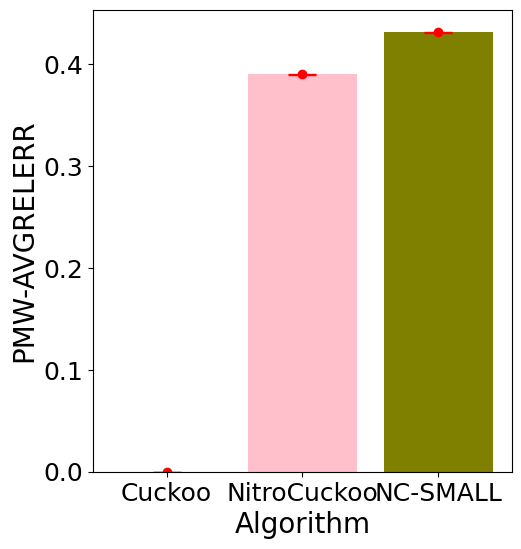}\label{fig:pm-avgrelerr:NITRO-SJ14}}
	\subfloat[][Chicago15]{\includegraphics[width=0.15\textwidth]{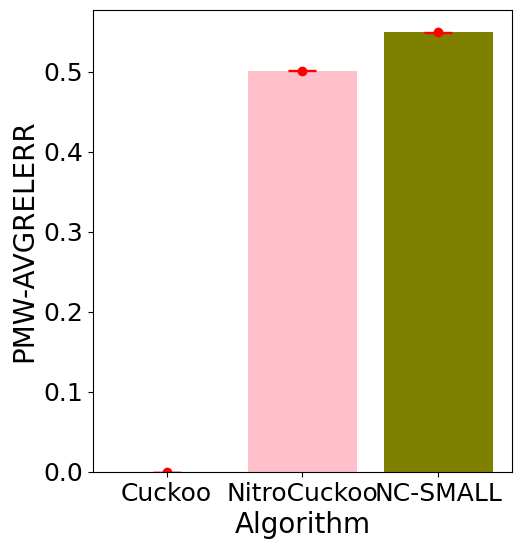}\label{fig:pm-avgrelerr:NITRO-Chicago15}}
	\subfloat[][Chic16Small]{\includegraphics[width=0.15\textwidth]{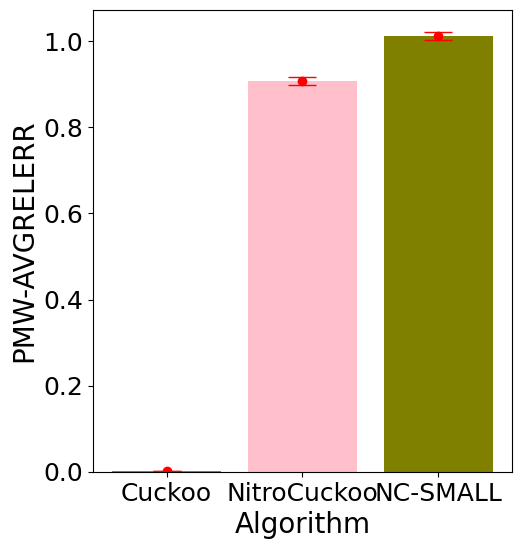}\label{fig:pm-avgrelerr:NITRO-Chicago16Small}}
	\subfloat[][Chic1610Mil]{\includegraphics[width=0.15\textwidth]{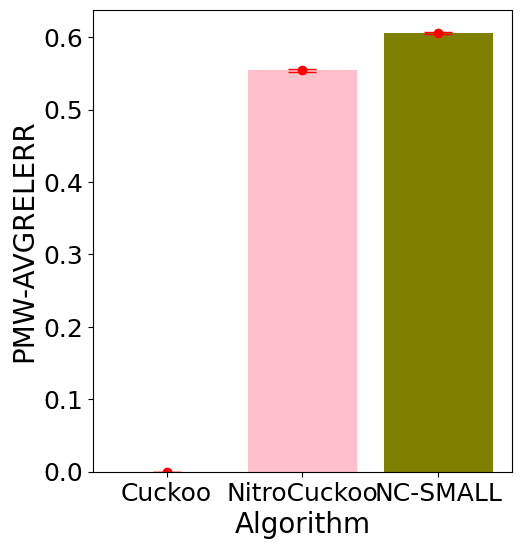}\label{fig:pm-avgrelerr:NITRO-Chicago1610Mil}}
	\subfloat[][Chicago16]{\includegraphics[width=0.15\textwidth]{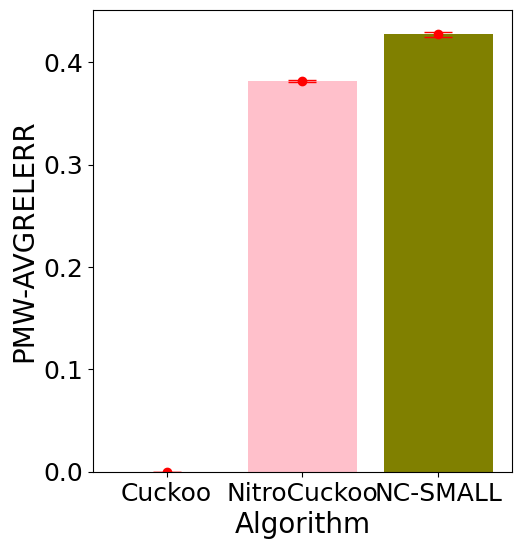}\label{fig:pm-avgrelerr:NITRO-Chicago16}}
	\caption{Postmortem AVGRELERR  with different Cockoo configurations}
	\label{fig:NITRO-pm-avgrelerr}
\end{figure*}

The impact of the smaller table allocation with NC-SMALL on the error metrics is shown in Figures~\ref{fig:NITRO-oa-msre}--\ref{fig:NITRO-pm-avgrelerr}.
In the case of the on arrival and postmortem metrics, the results are inconclusive, with slight tendency to worsening the error.
NC-SMALL fares much worse in the per flow metrics, since here small flows, whose relative error tends to be higher, have a much larger impact on the error calculation than in the on arrival and postmortem cases.
Yet, the errors with any of theses algorithms, Cuckoo, NitroCuckoo, and NC-SMALL, are much lower than many of the other alternatives we explored in this paper.
Hence, overall, we conclude that NC-SMALL is one of the best approaches.

\section{Conclusion}
\label{sec:discussion}

In this work, we have compared several Rust implementations of frequency estimation algorithms in terms of speed (throughput), estimation error under several metrics, and space requirements.
Specifically, we have realized and compared the following algorithms: HASH (serves as a baseline), NitroHash, Cuckoo (representing fingerprint based hash tables), NitroCuckoo, NC-SMALL, CMS (representing shared counter arrays sketches), Nitro CMS, CMS-NOMI, Space Saving, and Space Saving with the RAP optimization.
The measurements were carried over several real world network traffic monitoring traces from major backbone routers.

In terms of throughput, HASH was the fastest without Nitro, and NitroHash the fastest overall, both in the case of write only, write-read, and read only benchmarks.
In general, the Nitro optimization helps a lot for the write only benchmark, but its impact in other cases is inconsistent.
While it always improves performance for HASH and CMS, for read-only, NitroCuckoo is slower than Cuckoo.
The reason is that querying Cuckoo takes on average more time for non-exiting items, and with Nitro the table is under-loaded.
We speculate that this problem persists in many other fingerprint based hash tables, including TinyTable~\cite{TinyTable}, Counting Quotient Filter~\cite{CountingQuotientFilter}, and DASH~\cite{DASH}.
The results of the NC-SMALL method showed that it is best to configure NitroCuckoo with a much smaller table, which fits better in the hardware cache.
This more than offsets the impact of the longer lookup process in the read only and write-read tests, and improved throughput even in the write only case.

Our SpaceSaving implementations (with and without RAP) were the slowest, motivating trying a more optimized implementation along the lines of~\cite{ben2016heavy}.
Also, we speculate that software implementations of CMS would be better off without the minimum increment (or conservative update) optimization, de to its computational cost.

Interestingly, all error metrics yielded qualitatively similarly looking behavior, just at different scale.
The only minor exception is NitroCuckoo and NitroHash over short traces.
This is because Hash is accurate and Cuckoo has a marginal error, while the sampling approach of Nitro takes some time to converge~\cite{NitroSketch}.
Still, even on the relatively short traces, NitroHash and NitroCuckoo were more accurate than SpaceSaving, CMS, and NitroCMS.
Similarly to the findings of~\cite{RAP}, RAP improved the estimation error of SpaceSaving by orders of magnitude.

Space wise, CMS and SpaceSaving (plus SpaceSaving-RAP) are significantly more efficient than HASH and Cuckoo.
HASH exhibits the benefit of elasticity, as it can adjust its size to the number of unique items, which is orders of magnitudes lower than the total trace size.
Here, NitroHash seems to be a very compelling approach, since its memory requirements are very reasonable for any type of modern hardware, it is the fastest, and its estimation error is very low compared to the alternatives we tested.
NC-SMALL is also a very promising approach, almost as fast as NitroHash in the write only case, the second fastest in the write-read and read only case, its error is still relatively small, and its memory consumption is~manageable.

We note that CMS or SpaceSaving are useful for situations where the space requirement of the filter must be known and fixed.
Also, their significantly lower space requirements can be beneficial, e.g., if the filters need to be sent over a limited bandwidth network, in large multi-tenancy situations, or when holding a single filter per relation in a data base that stores a very large number of relations.
SpaceSaving is more compact when the identifiers are relatively small (or when fingerprints can be used instead of full identifiers), and otherwise CMS should be the preferred choice.

\paragraph*{Acknowledgments:} I would like to thank Eytan Singher for helping me getting up to speed with Rust programming, and to Alec Mocatta for helping me understand how to use the Amadeus CMS implementation. Thanks also to Ran Ben Basat, Gil Einziger, and Rana Shahout for insightful comments that greatly improved the presentation in this paper. This work was partially funded by the Israeli Science Foundation grant \#3119/21.

	
\bibliographystyle{plain}
\bibliography{refs}
	
	
	

\end{document}